\documentclass{aa}  
\usepackage{lineno}
\usepackage{float}
\usepackage[tableposition=top]{caption}
\usepackage{printlen}
\usepackage{color}
\usepackage[breaklinks=true]{hyperref}
\usepackage{caption}
\usepackage{subcaption}
\usepackage{multirow}
\usepackage[export]{adjustbox}
\usepackage{multirow}
\usepackage{natbib,twoopt}
\usepackage{tabu}
\usepackage{xcolor}
\usepackage{graphicx,colortbl}
\usepackage{txfonts}
\usepackage{amsmath} 
\bibpunct{(}{)}{;}{a}{}{,} 
\makeatletter
  \newcommandtwoopt{\citeads}[3][][]{\href{http://adsabs.harvard.edu/abs/#3}%
    {\def\hyper@linkstart##1##2{}%
     \let\hyper@linkend\@empty\citealp[#1][#2]{#3}}}
  \newcommandtwoopt{\citepads}[3][][]{\href{http://adsabs.harvard.edu/abs/#3}%
    {\def\hyper@linkstart##1##2{}%
     \let\hyper@linkend\@empty\citep[#1][#2]{#3}}}
  \newcommandtwoopt{\citetads}[3][][]{\href{http://adsabs.harvard.edu/abs/#3}%
    {\def\hyper@linkstart##1##2{}%
     \let\hyper@linkend\@empty\citet[#1][#2]{#3}}}
  \newcommandtwoopt{\citeyearads}[3][][]%
    {\href{http://adsabs.harvard.edu/abs/#3}
    {\def\hyper@linkstart##1##2{}%
     \let\hyper@linkend\@empty\citeyear[#1][#2]{#3}}}
\makeatother

\titlerunning{}
\authorrunning{Siu-Tapia et al.}
\begin{document}

   \title{Superstrong photospheric magnetic fields in sunspot penumbrae}

   \subtitle{}

   \author{A. Siu-Tapia
          \inst{1,}\inst{2}
           \and
          A. Lagg\inst{1} \and
          M. van Noort \inst{1} \and
          M. Rempel \inst {3}\and
          S. K. Solanki\inst{1,}\inst{4} 
          }

   \institute{Max-Planck-Institut f\"ur Sonnensystemforschung, Justus-von-Liebig-Weg 3,
37077 G\"ottingen, Germany
\and
Instituto de Astrof\'isica de Andaluc\'ia (IAA-CSIC), Apdo. 3004, E-18080 Granada, Spain\\
              \email{siu@iaa.es}
         \and
High Altitude Observatory, NCAR, P. O. Box 3000, Boulder, CO 80307, USA
       \and
School of Space Research, Kyung Hee University, Yongin, 446-701 Gyeonggi, Republic of Korea
\\
             }


 
  \abstract
   {Recently, there have been some reports of unusually strong photospheric magnetic fields (which can reach values of over 7 kG) inferred from Hinode SOT/SP sunspot observations within penumbral regions. These superstrong penumbral fields are even larger than the strongest umbral fields on record and appear to be associated with supersonic downflows. The finding of such fields has been controversial since they seem to show up only when spatially coupled inversions are performed.}
   {Here, we investigate and discuss the reliability of those findings by studying in detail observed spectra associated with particularly strong magnetic fields at the inner edge of the penumbra of active region 10930. }
   {We apply  classical diagnostic methods and various inversions with different model atmospheres to the observed Stokes profiles in two selected pixels with superstrong magnetic fields, and compare the results with a magnetohydrodynamic simulation of a sunspot whose penumbra contains localized regions with strong fields (nearly 5 kG at $\tau=1$) associated with supersonic downflows.}
   {The different inversions provide different results: while the SPINOR 2D inversions consider a height-dependent single-component model and return B$>$7 kG and supersonic positive $v_{LOS}$ (corresponding to a counter-Evershed flow), height-dependent 2-component inversions suggest the presence of an umbral component (almost at rest) with field strengths $\sim 4-4.2$ kG and a penumbral component with $v_{LOS}\sim16-18$ km s$^{-1}$ and field strengths up to $\sim5.8$ kG. Likewise, height-independent 2-component inversions find a solution for an umbral component and a strongly redshifted ($v_{LOS}\sim15-17$ km s$^{-1}$) penumbral component  with B$\sim4$ kG. According to a Bayesian Information Criterion, the inversions providing a better balance between the quality of the fits and the number of free parameters considered by the models are the height -independent 2-component inversions, but they lie only slightly above the SPINOR 2D inversions. Since it is expected that the physical parameters all display considerable gradients with height, as supported by MHD sunspot simulations, the SPINOR 2D inversions are the preferred ones.
} 
   { 
According to the MHD sunspot simulation analyzed here, the presence of counter-Evershed flows in the photospheric penumbra can lead to the necessary conditions for the observation of $\sim$ 5 kG fields at the inner penumbra.
Although a definite conclusion about the potential existence of fields in excess of 7 kG cannot be given, their nature could be explained (based on the simulation results) such as the consequence of the extreme dynamical effects introduced by highly supersonic counter-Evershed flows ($v_{LOS}>10$ km s$^{-1}$ and up to $\sim$30 km s$^{-1}$ according to SPINOR 2D), which would be much faster and more 
compressive downflows than those found in the MHD simulations and therefore could lead to  the field intensification up to considerably stronger fields. Also, a lower gas density would lead to a deeper depression of the $\tau=1$ surface, making possible the observation of deeper-lying stronger fields. The superstrong magnetic fields are expected to be nearly force-free, so that they can attain much larger strengths than expected when considering only balance between magnetic pressure and the local gas pressure.}


   \maketitle
%

\section{Introduction}

Sunspots are the largest concentrations of magnetic flux on the solar surface and, due to the significant suppression of the convective motions caused by the strong fields, they are seen as dark regions on the photosphere. The various brightness levels observed in sunspots indicate differences in temperature, and therefore different magnetic regimes. Such a relation between brightness/temperature and magnetic field strength in the photosphere has been extensively studied \citep[e.g.,][]{Lites1990,Solanki1993,Keppens1996, Mathew2003,Mathew2004,Tiwari2015}.

\begin{figure*}[htp!]
 \centering
\begin{subfigure}[b]{0.5\textwidth}
           \centering
  \includegraphics[width=\textwidth]{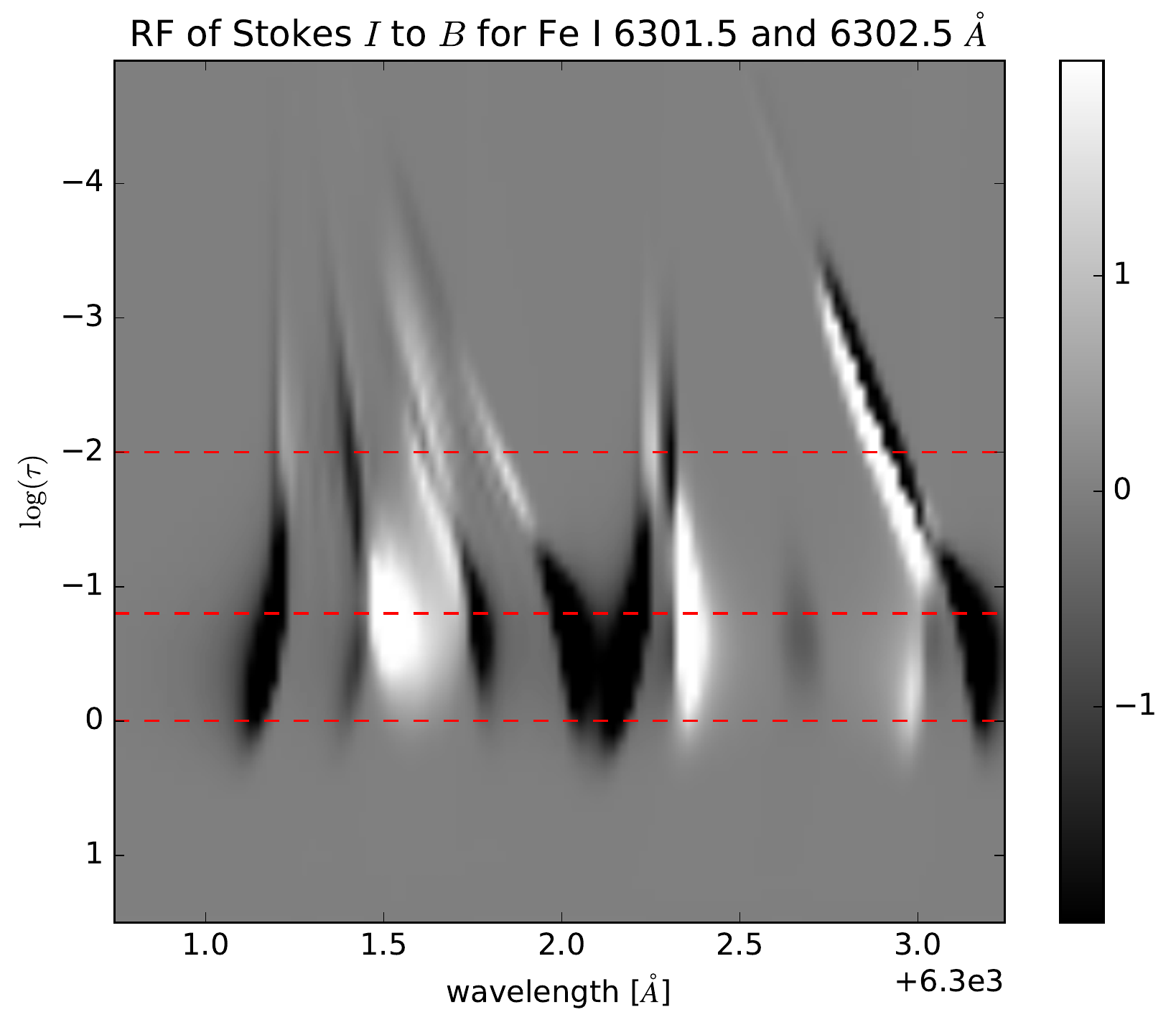}
	\label{fig:RFa}
    \end{subfigure}%
 ~   
    \begin{subfigure}[b]{0.5\textwidth}
\centering
     \includegraphics[width=\textwidth]{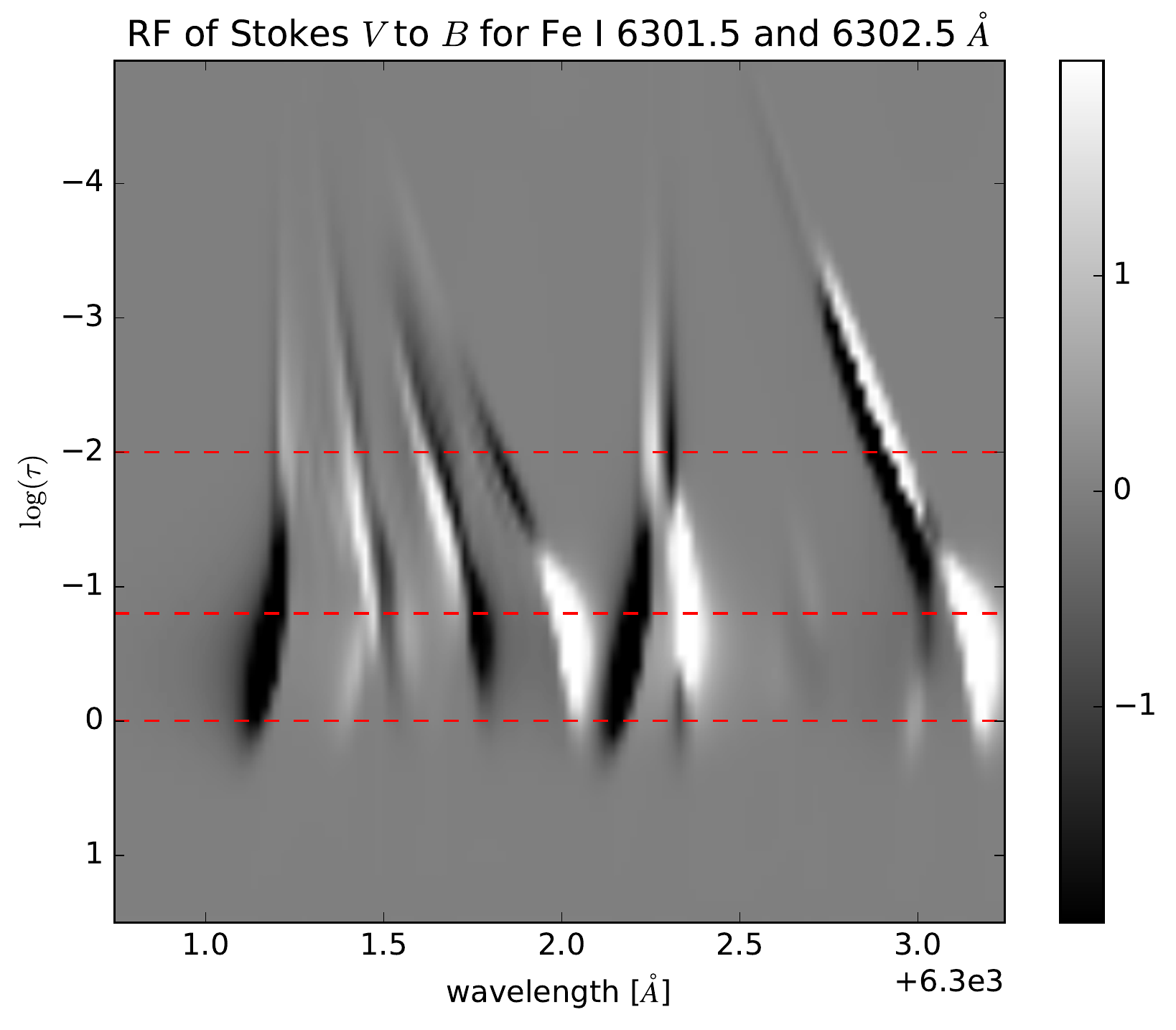}    
	\label{fig:RFb}
    \end{subfigure}

     \caption[]{Normalized response function of Stokes $I$ (left) and Stokes $V$ (right) to the magnetic field $B$, multiplied by $1/\Delta \tau_C [10^{-5}$ G$^{-1}]$, for the pair of Fe I absorption lines at 6301.5 and 6302.5 $\AA$ in the atmosphere inferred by the SPINOR 2D inversions for the penumbral pixel marked with a blue `+' in Fig. \ref{fig:LFPa}. Some physical parameters for such an atmosphere are presented in Table \ref{tab:1}. The horizontal dashed lines indicate the selected nodes for the SPINOR 2D inversion: $\log(\tau)=0,-0.8$ and $-2$.}
\label{fig:RF}
\end{figure*}

\begin{figure*}[htp!]
 \centering
\begin{subfigure}[b]{0.46\textwidth}
           \centering
  \includegraphics[width=\textwidth]{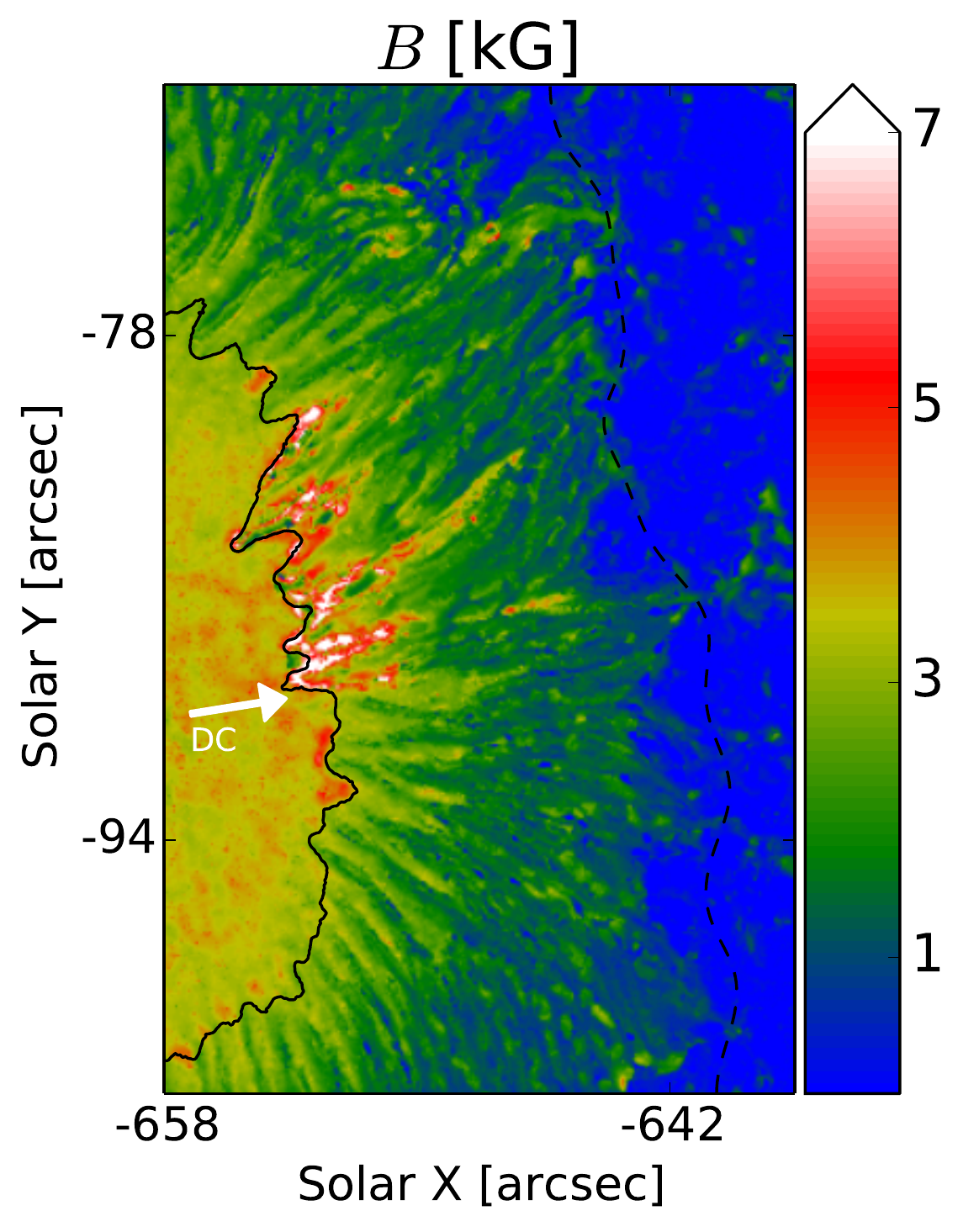}
 \caption{}
	\label{fig:LFP_Bmap}
    \end{subfigure}%
    ~ 
    \begin{subfigure}[b]{0.5\textwidth}
\centering
     \includegraphics[width=\textwidth]{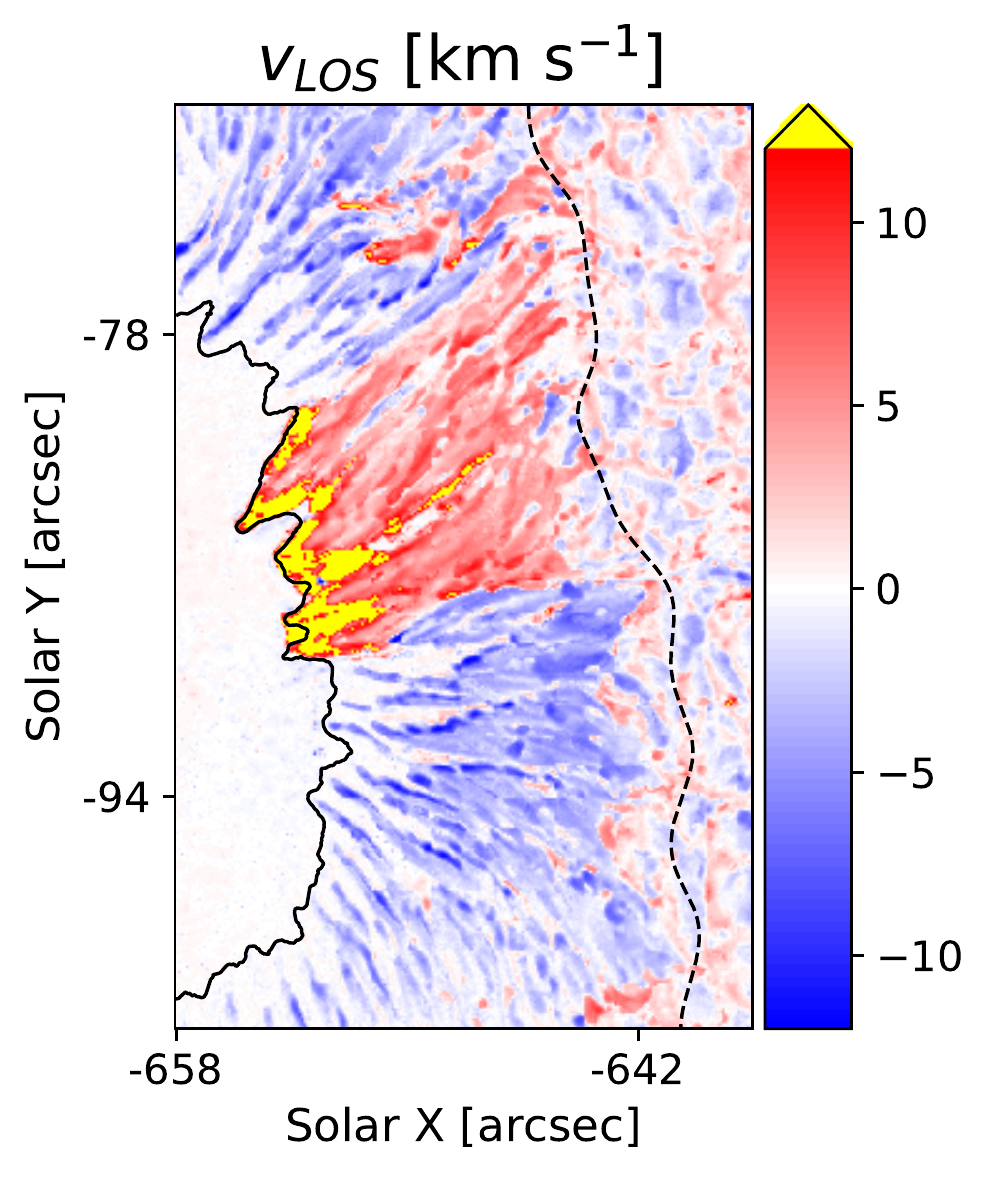}    
\caption{}
	\label{fig:LFPb}
    \end{subfigure}

     \caption[]{ A portion of the  (center-side) penumbra of the main sunspot in AR 10930  (cf. Fig. 5 in \citet{siu2017}). (a) Map of the magnetic field strength at $\log(\tau)=0$, as inferred with SPINOR 2D inversions. The arrow points towards the solar disk-center. (b) Line-of-sight velocity as inferred with SPINOR 2D inversions at $\log(\tau)=0$. Positive values (red-to-yellow colors) indicate plasma motions away from the observer, while negative values (blue colors) indicate motions towards the observer.   Solid and dashed black contour lines were placed at $I_c/I_{QS}=0.26$ and $I_c/I_{QS}=0.94$, respectively. The large penumbral sector where red-to-yellow colors dominate harbors a fast counter-EF which supersonically sinks (with speeds larger than 12 km s$^{-1}$) near the inner penumbral boundary \citep{siu2017}. Notice that the color-bar scales have been saturated.}
\label{fig:LFP}
\end{figure*}

Thus, the strongest magnetic fields in sunspots are generally found within their central dark regions or \textit{umbrae}, where the field is closely vertical
with typical strengths between 2.5-4 kG \citep[e.g.,][]{Livingston2002}. The largest field strength ever recorded within an umbral region is nearly 6.1 kG \citep{Livingston2006}, but an even larger value (above 6.2 kG) was recently reported by \citet{Okamoto2018}, and was observed in a light bridge.
Moreover,  field strengths in excess of 7 kG have been found within sunspot \textit{penumbrae} \citep{vannoort2013,siu2017}. Penumbrae are  the less dark regions partially or completely surrounding umbrae and where  the magnetic field is filamentary, strongly inclined, and generally weaker than in the umbrae (typically the field varies from about 1.5-2.5 kG at the inner penumbral boundary to 0.5-1 kG towards the outer boundary) \citep[e.g.,][]{Lites1990,Skumanich1994,Westendorp2001b}.

Sunspot penumbrae are additionally characterized by a gas outflow with speeds of several kilometers per second, the so-called Evershed flow \citep[EF;][]{Evershed1909}, which is directed along the penumbral filaments, i.e. along  the bright elongated channels with more inclined fields within the penumbra itself. 
\citet{Tiwari2013}, in a detailed analysis of the penumbral fine structure, found that the EF has  its sources in hot upflows that occur at the inner endpoints of the penumbral filaments (heads)  and part of it sinks in concentrated cooler dowflows that occur at the outer endpoints (tails) where the magnetic field  bends over vertically and displays a strengthening of about 1.5-2.5 kG on average, so that it can reach up to 3.5 kG in individual tails.

The superstrong penumbral fields reported by \citet{vannoort2013} (reaching up to 7.5 kG) were observed in some tails of complex penumbral filaments, i.e. those with a single tail but with more than one head, near the penumbral periphery in supersonic downflow areas (estimated line-of-sight velocities of up to 22 km s$^{-1}$). Such values
 were inferred by using a sophisticated inversion technique that takes into account the spatial coupling between the pixels of the observed image when considering the instrumental effects that cause the image degradation \citep[SPINOR 2D,][]{vannoort2012,vannoort2013}.

In contrast, the superstrong fields reported by \citet{siu2017} ($>$ 7 kG based on SPINOR 2D inversions) were observed in the tails of inverted penumbral filaments carrying a counter-EF (CEF), i.e. a gas inflow towards the sunspot umbra at photospheric heights, near the inner penumbral boundary. Unlike in \citet{vannoort2013} and \citet{Tiwari2013}, \citet{siu2017} considered  the outer endpoint of the CEF-carrying filaments to be heads, since at such places the sources of the CEF were found; and the inner endpoints of the CEF-carrying filaments to be tails, where the sinks of such flow were observed. Therefore, they found  on average field strength of $\sim$4.5 kG (when excluding all those fields in excess of 7 kG) at the tails of the CEF-carrying filaments. Such superstrong penumbral fields were also found to be associated with the supersonic downflows occurring at the sinks of the CEF.

\begin{figure}[htp!]
 \centering
\begin{subfigure}[b]{0.5\textwidth}
           \centering
  \includegraphics[width=\textwidth]{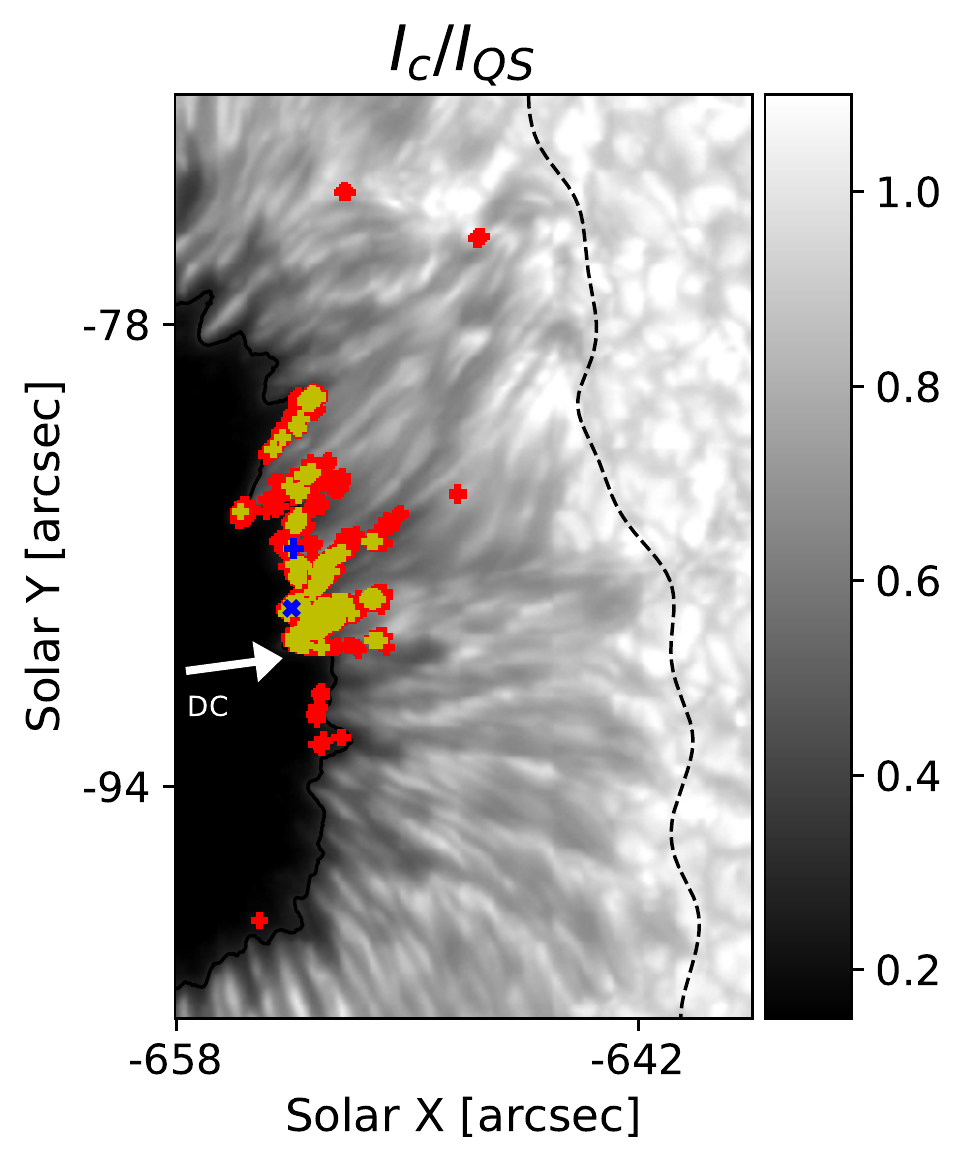}
 \caption{}
	\label{fig:LFPa}
    \end{subfigure}%
    
    \begin{subfigure}[b]{0.5\textwidth}
\centering
     \includegraphics[width=\textwidth]{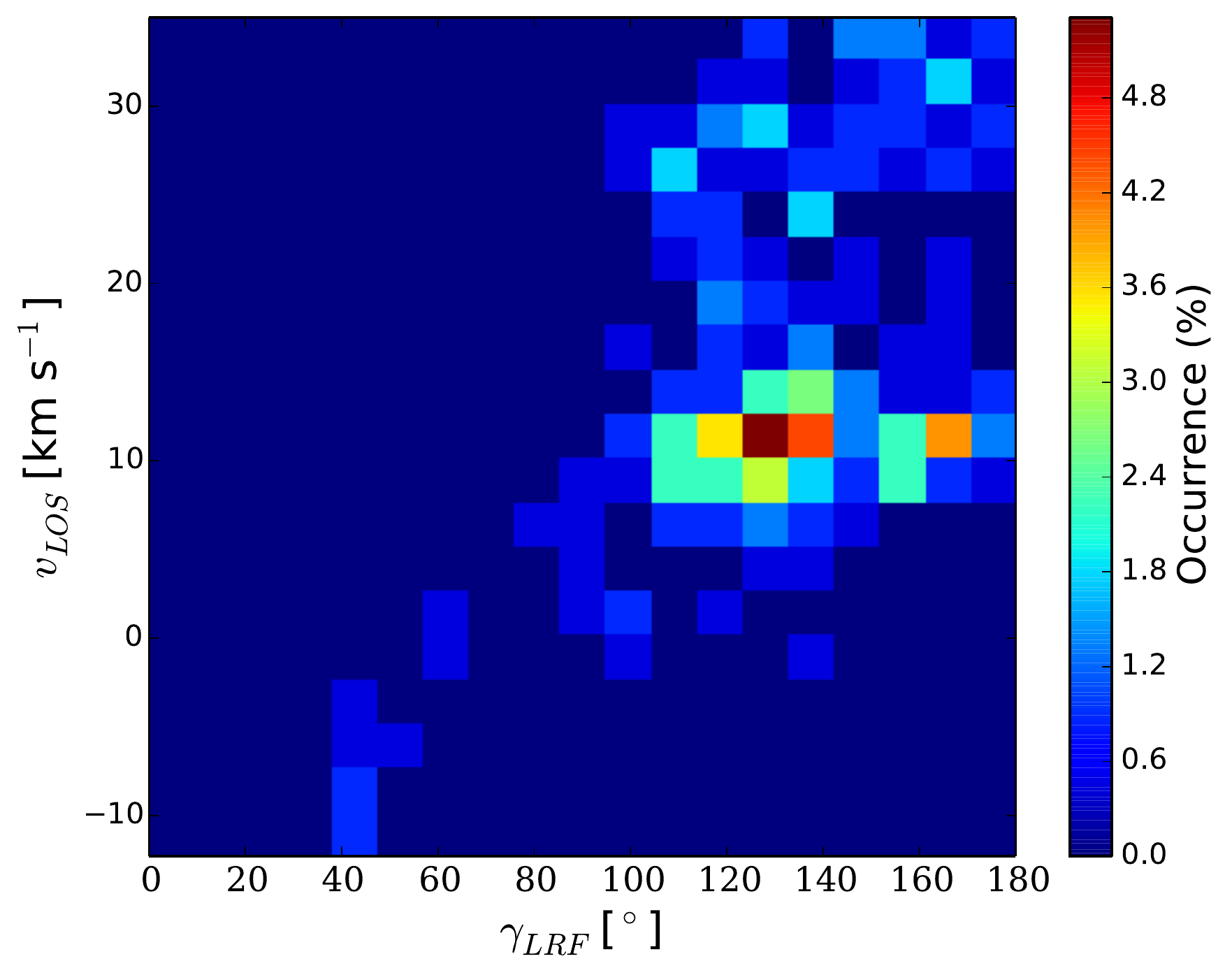}    
\caption{}
	\label{fig:LFP2}
    \end{subfigure}

     \caption[]{(a) Normalized continuum intensity map as observed by the Hinode SOT/SP  in AR 10930. The red markers indicate the  location of 845 pixels where the SPINOR 2D inversions return $B\geq5$ kG at $\log(\tau)=0$, of which 226 harbor fields larger than $7$ kG at $\log(\tau)=0$ (yellow  markers, also named in the text such as large-field-pixels or LFPs). The blue markers  highlight the location of two  LFPs (`+': LFP 1, `x': LFP 2) selected to test the robustness of our inversion solutions (see text). The arrow points towards the solar disk-center. (b) Scatter plot of the magnetic field inclination in the local-reference-frame  $\gamma_{LRF}$ (after removal of the field azimuth ambiguity through the NPFC method) versus the line-of-sight velocity $v_{LOS}$, from SPINOR 2D at $\log(\tau)=0$, for the 226 LFPs. }
\label{fig:LFP}
\end{figure}

The superstrong penumbral fields reported by both \citet{vannoort2013} and \citet{siu2017} are unusual and  are even stronger than the strongest umbral fields found by \citet{Livingston2006} of $\sim6.1$ kG. Moreover, in both studies, spatially coupled inversions were employed, and therefore, the reality of these extraordinary field strengths needs to be confirmed with other techniques. The existence of field strengths in excess of 7 kG possibly has an impact on the approximations made by the inversion code, which can lead to errors in the density stratifications if the non-vertical field components are significant, and therefore in the atmospheric stratifications of the physical parameters inferred by the code. Hence, the reliability of the inversion results for such peculiar pixels needs to be tested.

This work is aimed at discussing the reliability of those penumbral fields in excess of 7 kG reported by \cite{siu2017}  concentrated at the inner boundary of the penumbra of the main sunspot in active region (AR) 10930. 
Here we examine the reality of such findings by applying some classical diagnostic methods and various inversion techniques with different model atmospheres to the Stokes profiles observed with the spectropolarimeter (SP) of the Solar Optical Telescope (SOT) on board Hinode. 
Finally, since the mere possibility of such superstrong fields in sunspots leads to the question about the necessary physical conditions for their appearance, we analyze the physical structure of a sunspot MHD simulation with CEFs which was reported by \citet{Rempel2015} \citep[see also][]{Siu2018}.

This work is organized as follows: In section 2 we describe
our data and inversion technique. In section 3, some classical diagnostic methods are applied to peculiar pixels with extreme spectra. In section 4, various inversions with different model atmospheres are considered. The results are presented in section 5. In section 6, we compare the observations with  MHD sunspot simulations and study the mechanisms that can amplify the magnetic field in the penumbra. Finally, in section 7 we discuss our results and
draw our conclusions.

\section{Observations and spatially coupled inversions}
For this study we use the spectropolarimetric observations of  AR NOAA 10930 from the Hinode SOT/SP instrument \citep{Lites2001,Lites2013}, which simultaneously measures the full Stokes profiles of a pair of absorption lines of Fe I that are formed in the lower solar photosphere with central wavelengths at 6301.5 and 6302.5 $\AA$ (having effective Land\'e factors $g_{L}$=1.67 and 2.49, respectively).

The SP instrument scanned the main sunspot in AR 10930 (whose umbra displayed a negative magnetic polarity) on Dec 8 2006 at an heliocentric angle of $\sim$47$^{\circ}$ while operating in normal mode, i.e., using an exposure time of 4.8 seconds per slit position and a spatial sampling of 0.16$^{\prime\prime}$.
These observations were corrected for dark current,
flat field, orbital drift and instrumental cross-talk by reducing the row data with IDL
routines of the Solar-Soft package \citep{Lites2013b}.

As described by \citet{siu2017}, the atmospheric properties of the main sunspot in AR 10930 
were derived by inverting the observed Stokes profiles with the SPINOR 2D inversion code \citep{vannoort2012,vannoort2013}, which is the spatially coupled version of the SPINOR inversion code \citep{Frutiger2000}, based on the STOPRO routines \citep{Solanki1987b} that solve the radiative transfer equations for polarized light.

The method is able to invert 2-D maps of spectropolarimetric data that have been degraded spatially in a known way (information that is contained in the point-spread function or PSF of the telescopes). Then, the code is able to use the information contained in the spectral dimension and the known spatial degradation properties to constrain a parameterization of the atmosphere over the whole FOV.
The  image degradation is applied to the solution rather than by deconvolving the original data themselves, which is the clasical approach, while the code performs a coupled inversion of all the pixels simultaneously.

According to \citet{vannoort2012}, the spatially coupled inversion method is stable to oversampled data and produces an inversion result with a resolution up to the resolution limit of the telescope. To be able to reach the diffraction limit of Hinode SOT, we oversampled all Stokes maps by a factor two, to 0.08$^{\prime\prime}$,  following \citet{vannoort2013}. 
Thus, by considering a spatial grid that is denser than  the original data and by employing a single-component atmospheric model per pixel, the inverted atmospheric parameters returned by the code correspond to the best fits
to the Stokes profiles once the blurring effect of the telescope's PSF has been
taken into account. This significantly improves the
spatial resolution, allowing structures at the diffraction limit of the telescope to be properly resolved.

\begin{figure*}
\centering
   \includegraphics[width=0.49\textwidth]{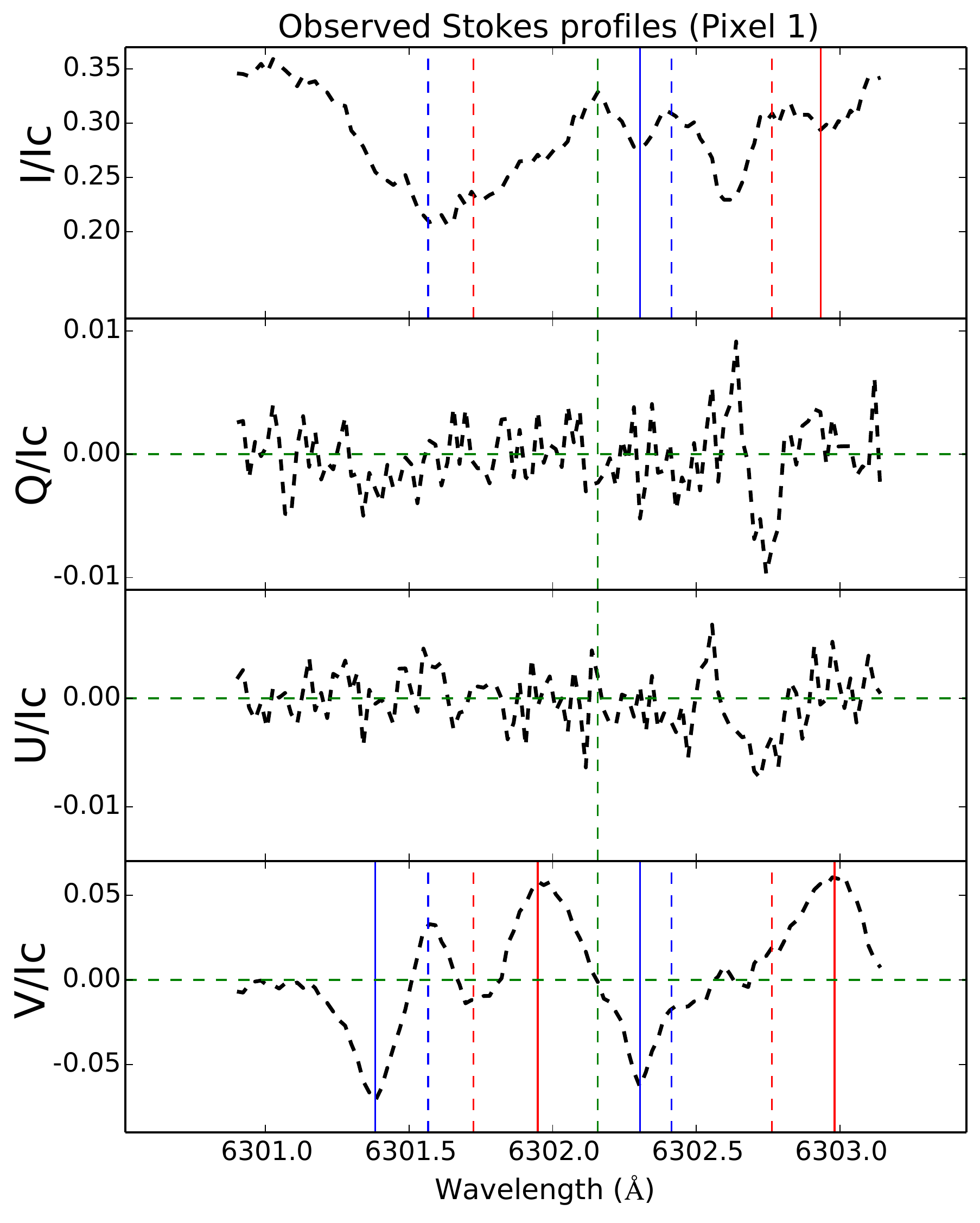}
     \includegraphics[width=0.49\textwidth]{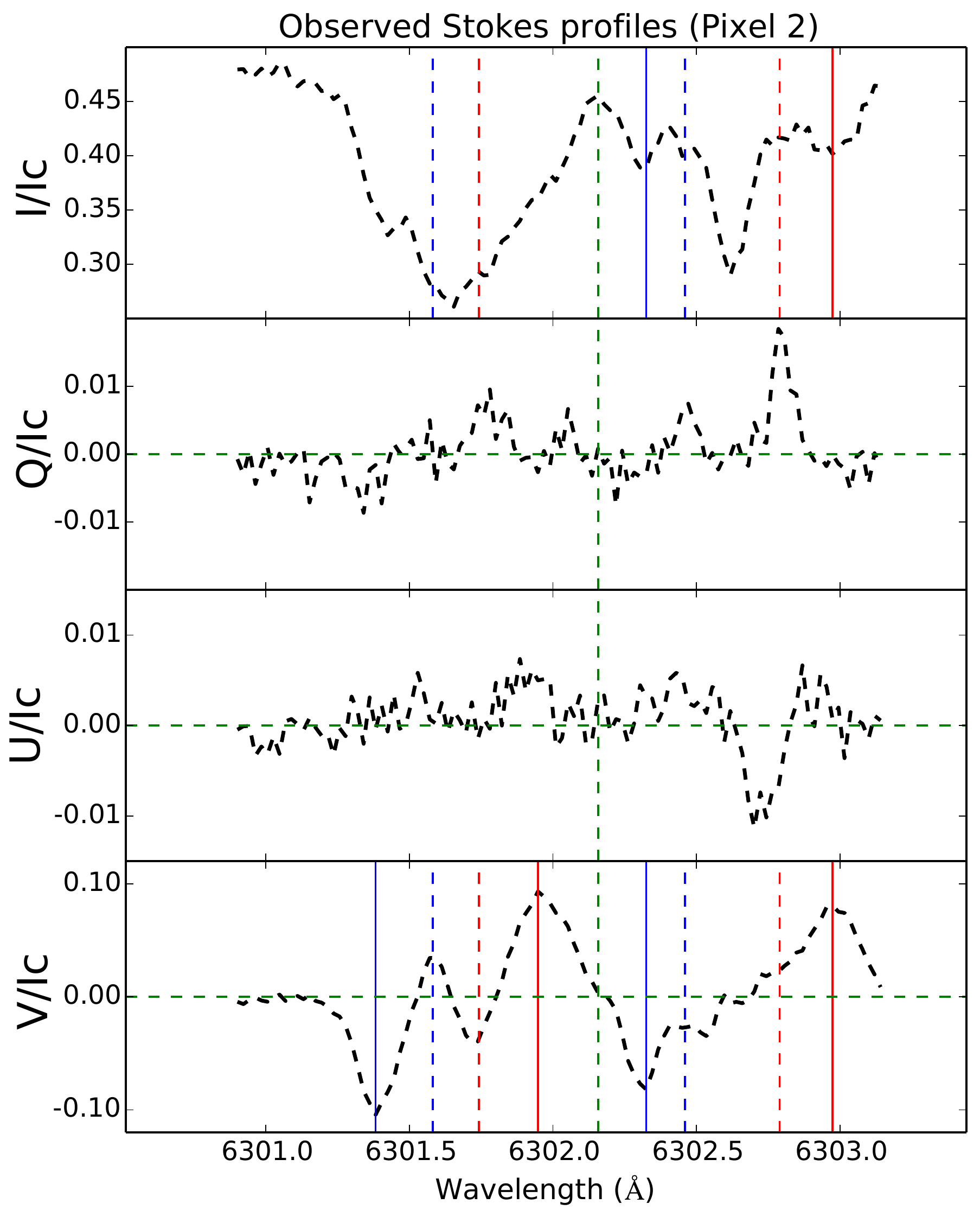}

\caption[Observed Stokes profiles in two LFPs]{ Observed Stokes profiles (dashed curves) in  two  pixels (\textit{left:} pixel 1, \textit{right:} pixel 2) located at the inner penumbra of the CEF region (see blue crosses on maps displayed in Fig. \ref{fig:LFPa}). From top to bottom: Stokes $I$, $Q$, $U$ and $V$. 
 The vertical lines in Stokes $I$ and $V$ panels indicate the position of the $\lambda^-$ (blue) and $\lambda^+$ (red) used to calculate the magnetic field strength with three different methods (see the main text for a description of the methods): the solid lines in the Stokes $V$ panels show the splitting derived from method 1; the solid lines in the Stokes $I$ panels correspond to the line splitting obtained with method 2; and, the dashed lines in both $I$ and $V$ panels indicate the splitting derived from method 3. The vertical dashed green line shows the wavelength position used to separate the two Fe I lines.}
\label{fig:3}
   \end{figure*}

For the SPINOR 2D inversions used in this work \citep[cf.][]{siu2017} all the atmospheric free parameters were initially defined at three optical depth nodes, placed at $\log(\tau)=-2$, $-0.8$, and $0$. This set of optical depth nodes was shown to work well for the inversion of Hinode/SP observations of sunspots by \citet{siu2017}.  
At each of the three chosen nodes the temperature $T$, magnetic field strength $B$, field inclination relative to the line-of-sight $\gamma_{LOS}$, field azimuth $\phi$, line-of-sight velocity $v_{LOS}$, and a microturbulent velocity $v_{MIC}$, were fitted, leading to 18 free parameters in total.

In Figure \ref{fig:RF}, the response functions (RFs) of Stokes $I$ and $V$ to the magnetic field $B$ are plotted as functions of wavelength and optical depth for the pair of Fe I absorption lines at 6301.5 and 6302.5 $\AA$. To compute these RFs we have used the atmospheric parameters retrieved by SPINOR 2D inversions (see Table \ref{tab:1}) in a penumbral pixel with superstrong field (labeled as LFP 1 and marked with a blue `+' in Fig. \ref{fig:LFPa}). These plots show that the selected nodes at $\log(\tau)=0,-0.8$ and $-2$ for performing our inversions all formally lie within the formation region of the lines. In particular, the lowest node at $\log(\tau)=0$ is also sensitive to the magnetic field changes, which means that information about the magnetic field at such node is available for the analyzed wavelength range.

\subsection{Sunspot's magnetic and velocity field as inferred by SPINOR 2D }

As reported by \citet{siu2017}, the SPINOR 2D inversions return very large magnetic field strengths, $B\geq5$ kG at all three height nodes ($\log(\tau)=-2.0,-0.8$ and 0) in the inner center-side penumbra of the main sunspot in AR 10930 (see field strength map in Fig. \ref{fig:LFP_Bmap} for the height node at $\log(\tau)=0$ only). Such strong magnetic fields are located at  places that coincide with supersonic sinks of the counter-EF (CEF, see Fig. \ref{fig:LFPb} and Figs. 3 and 5 in \citet{siu2017}). These are among the largest field strengths ever observed in penumbral environments \citep[see also][]{vannoort2013}. 

In particular, field strengths above $7$ kG (whiter regions near the inner penumbral boundary in Fig. \ref{fig:LFP_Bmap}) are even larger than the largest umbral fields found by \citet{Livingston2006} of $\sim6.1$ kG, who also found that only a very small fraction of sunspots (around the $0.2\%$ in a 9-decade record of $\sim32 000$ sunspots)  have umbral fields stronger than 4 kG.

The scatter-plot in Figure \ref{fig:LFP2} shows the SPINOR 2D inversion results for the magnetic field inclination angle in the local-reference-frame (LRF) after the azimuthal disambiguation was resolved with the Non-Potential Magnetic Field Computation
method \citep[NPFC,][]{Georgoulis2005} versus the line-of-sight (LOS) velocity at $\log(\tau)=0$, for all the 226  pixels where the SPINOR 2D inversions return $B\geq7$ kG (hereafter large field pixels or LFPs,  yellow markers in Fig. \ref{fig:LFPa}). 

Most of the LFPs 
(see main  population in Fig. \ref{fig:LFP2}) appear  associated with supersonic LOS velocities, i.e. around $v_{LOS}=10$ km s$^{-1}$ \citep[the sound speed is around $C_s \sim 5-8$ km s$^{-1}$ in penumbrae, e.g.,][]{vannoort2013} and even larger than $20$ km s$^{-1}$ in some pixels.  The peak of the LFPs distribution occurs near $\gamma_{LRF}=140^{\circ}$. 

Therefore, according to the SPINOR 2D inversions, the LFPs are of umbral polarity (which is negative) with a large longitudinal component and, under the assumption of field-aligned flows, they contain supersonic downflows at $\log(\tau)=0$. See \citet{siu2017} for more details on their brightness and thermal structure.

\section{Analysis}
\subsection{Zeeman splitting and center-of-gravity methods}

Most of the LFPs inferred by the SPINOR 2D inversions in AR10930 are located at or close to the umbral/penumbral boundary of the penumbral region hosting a CEF (cf. Figs. 3 and 5 in \citet{siu2017}),  and display very complex Stokes profiles.  
Figure \ref{fig:3} shows the observed Stokes profiles (black dashed curves) from two selected LFPs. 
The locations of the selected LFPs 
are marked with blue crosses in Figure \ref{fig:LFPa} (`+': pixel 1, `x': pixel 2).
All Stokes parameters in these LFPs 
exhibit large asymmetries and the Stokes $V$ profiles display more than two lobes.
Their best fits from SPINOR 2D  are not nearly as good as in most of the penumbral pixels (see Fig. 2 in \citet{siu2017}).   It is noticeable that the continuum intensity and the Stokes $V$ profiles are only imperfectly reproduced in both examples (see also, orange curves in Fig. \ref{fig:3**}).

\begin{table*}
\caption[Results of direct measurement of Zeeman splitting (Methods 1 and 2) and center-of-gravity (COG) methods in two LFPs]{Results of direct measurement of Zeeman splitting (Methods 1 and 2) and center-of-gravity (COG) method in two LFPs.
}
\begin{center}
    \begin{tabular}{c c c c}
    \hline
 \multirow{2}{*}{    method}  &  \multirow{2}{*}{ pixel}& $B$ [kG]  &  $B$ [kG] \\ 
  & & (6301.5 $\AA$)&  (6302.5 $\AA$)\\ \hline
 \multirow{2}{*}{1. Method 1 (Stokes $V$)}&1&9.4&7.9\\ 
&  2&  9.4&7.2\\ \hline
\multirow{2}{*}{2. Method 2 (Stokes $I$)}&  1&&7.1\\ 
&  2&&6.9\\ \hline
\multirow{2}{*}{3. Method 3 (COG)}& 1 &   2.6&3.9\\ 
 & 2& 2.7&3.6\\ \hline
    \end{tabular}
\end{center}

  \label{tab:1*}
\end{table*}

As a very simple alternative estimation of the field strength in the selected LFPs, 
we compute the 
magnetic field strength directly from the splitting of the two observed Fe I line profiles at $6301.5$  $\AA$ (line 1) and $6302.5$ $\AA$ (line 2), using the Zeeman splitting formula:

\begin{equation}
B=\frac{\lambda^{+}-\lambda^{-}}{2.0} \frac{4 \pi m c}{e g_L \lambda_0^2},
\label{eq:1a}
\end{equation}

\noindent where $\lambda_0$ is the central wavelength of the line, $\lambda^{\pm}$ are the centroids of the right and left circularly polarized line components ($\sigma$-components), $m$ and $e$ are the electron mass and charge, $g_L$ is the effective Land\'e factor of the transition ($g_L=1.67$ for 
line 1 and $g_L=2.5$ for line 2), 
and $c$ is the speed of light. 
Given the huge field strengths inferred by the SPINOR 2D inversions, the Zeeman splitting should be complete, so that 
 this approach is expected to prove the inversion results in case the super-strong fields are real. However, the Zeeman splitting approach can be misleading if there are strong gradients with height or multiple components in the resolution element. A strong magnetic field near optical depth unity which rapidly decreases with height might not display a complete Zeeman splitting in the profiles but very broad wings of
the Stokes $I$ and $V$ profiles.

The critical point about using Equation \ref{eq:1a} is determining $\lambda^-$ and $\lambda^+$ due to the large asymmetries observed in all four Stokes profiles from the LPF.  We use three ways: \textit{Method 1)}  Selecting $\lambda^-$ where Stokes $V$ takes on the largest negative value in the blue wing of the corresponding line, and $\lambda^+$ where Stokes $V$ is largest in its red wing. \textit{Method 2)} Placing $\lambda^-$ and $\lambda^+$   where the two deepest minima of Stokes $I$  away from the central wavelength are found (using 
line 2 only, since line 1 
is insufficiently split). \textit{Method 3)} Applying the center of gravity method or COG method \citep{Semel1967,Semel1970,Rees1979,Cauzzi1993}, in which $\lambda^{\pm}=\lambda_{COG}^{\pm}$, where $\lambda_{COG}^{\pm}$ are the center of gravity wavelengths of the centroids of the right and left circularly polarized components, respectively, of the corresponding line, i.e.:

\begin{equation}
\lambda_{COG}^{\pm}=\frac{\int{\lambda(I_{c}-(I \pm V)) d\lambda}}{\int{(I_{c}-(I \pm V)) d\lambda}},
\label{eq:3a}
\end{equation}

The position of $\lambda_{COG}^{\pm}$ and the wavelengths delimiting the integration intervals in Equation \ref{eq:3a} for each of the Fe I lines, are indicated in  Figure \ref{fig:3} for the observed spectra in the two selected LFPs (dashed vertical lines). The resultant field strengths obtained with the methods described above are listed in Table \ref{tab:1*}, for both LFPs. 

The  $B$ values computed with methods 1 and 2 are indeed very large, lying between $\sim 7$ kG (obtained from line 2) 
and $\sim9.4$ kG (from line 1) 
in both LFPs. 
On the contrary, the COG method provides considerably smaller field strength values in both LFPs: $B\sim2.6$ kG from  line 1 
and $3.6-3.9$ kG from line 2. 
The difference between the COG and direct splitting methods likely stems from the fact that the profiles are not simple Gaussians, but show complex shapes indicating a range of field strengths (of which methods 1 and 2 sample only the largest), but is also partly due to the non-longitudinal direction of the field (methods 1 and 2 determine field strength, while method 3 only gives the LOS component).

\begin{figure*}
   \resizebox{\hsize}{!}
            {

\includegraphics[width=0.5\textwidth]{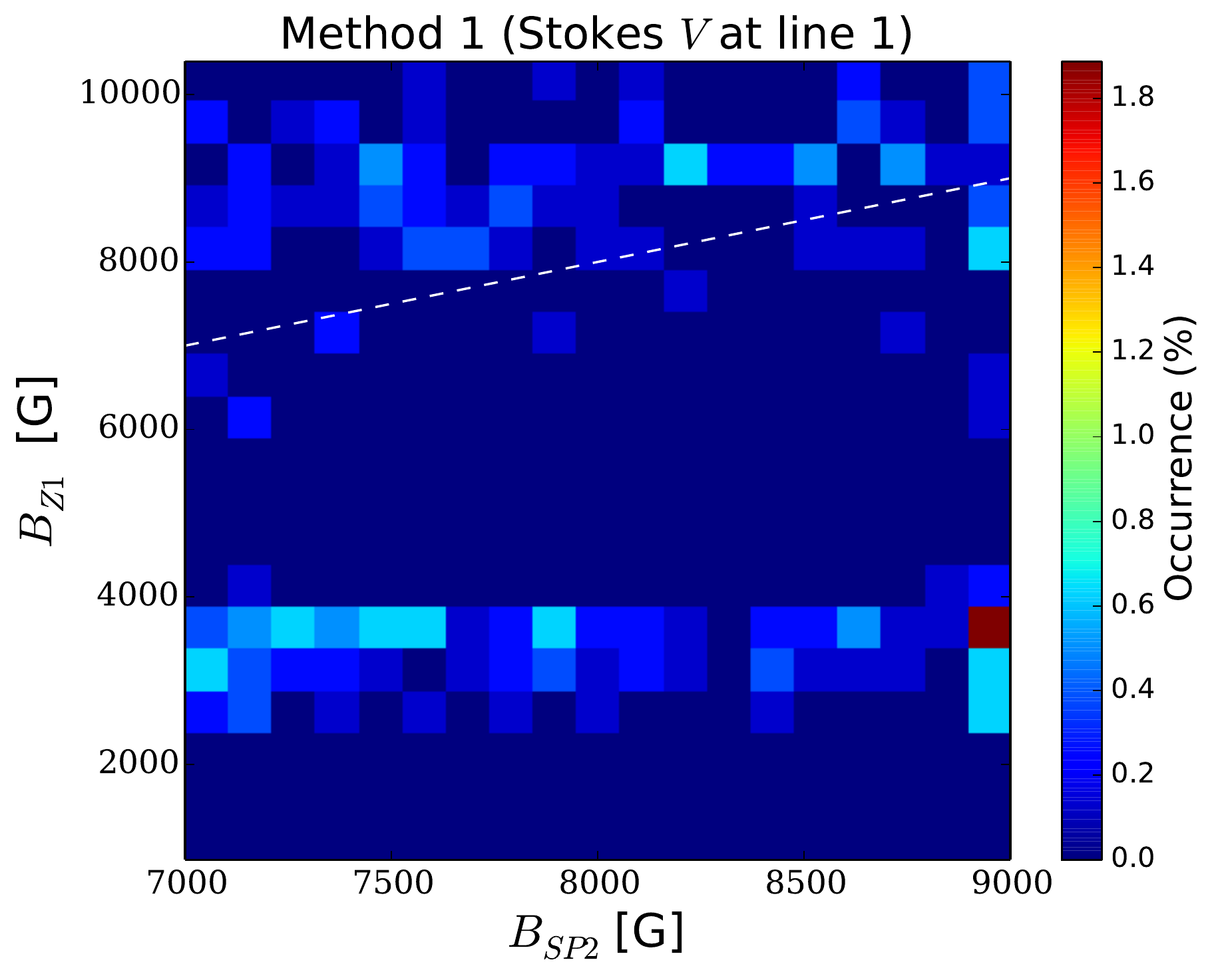}
     \includegraphics[width=0.5\textwidth]{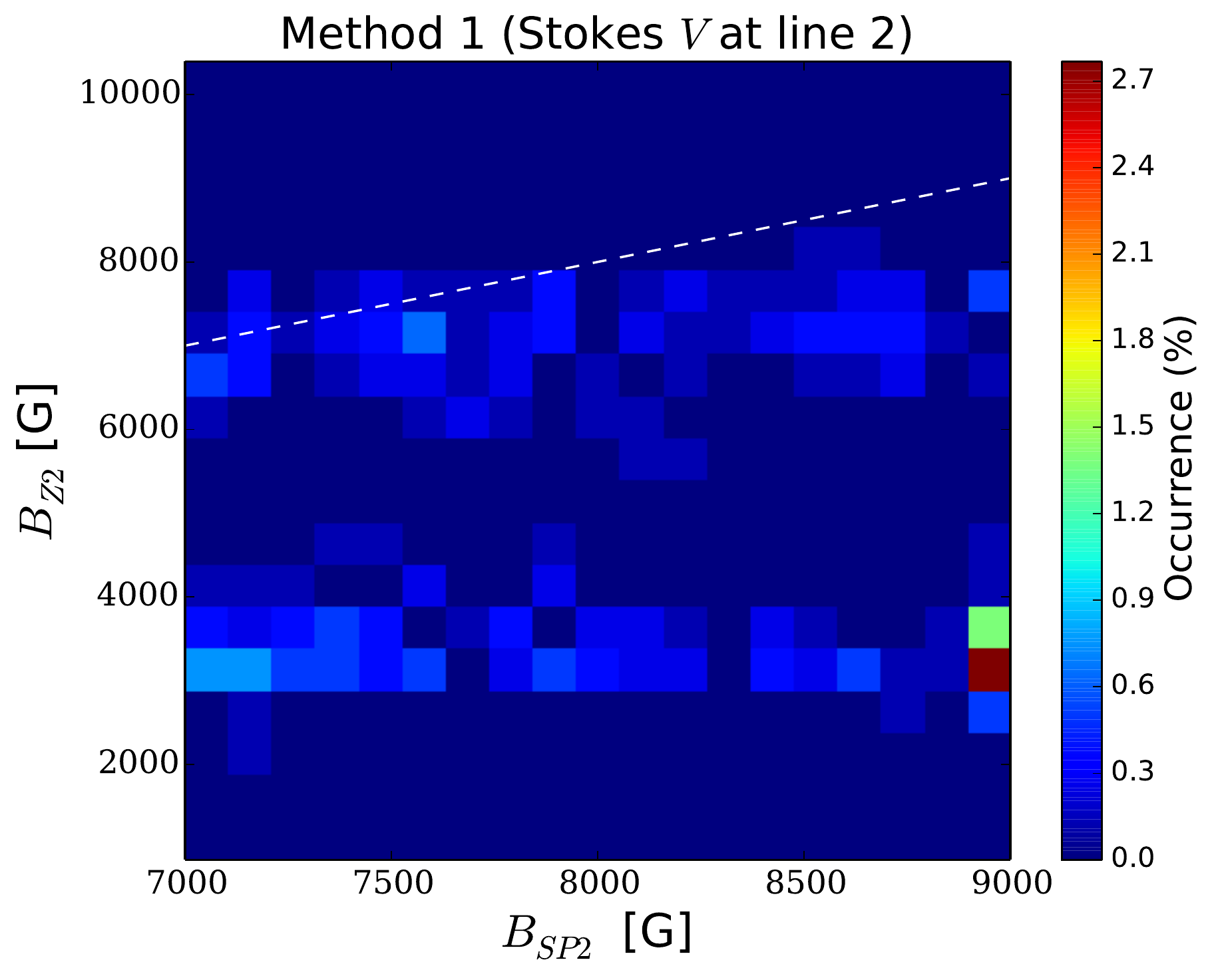}}

\hfill \resizebox{0.5\hsize}{!}
            {  
\includegraphics[width=\textwidth]{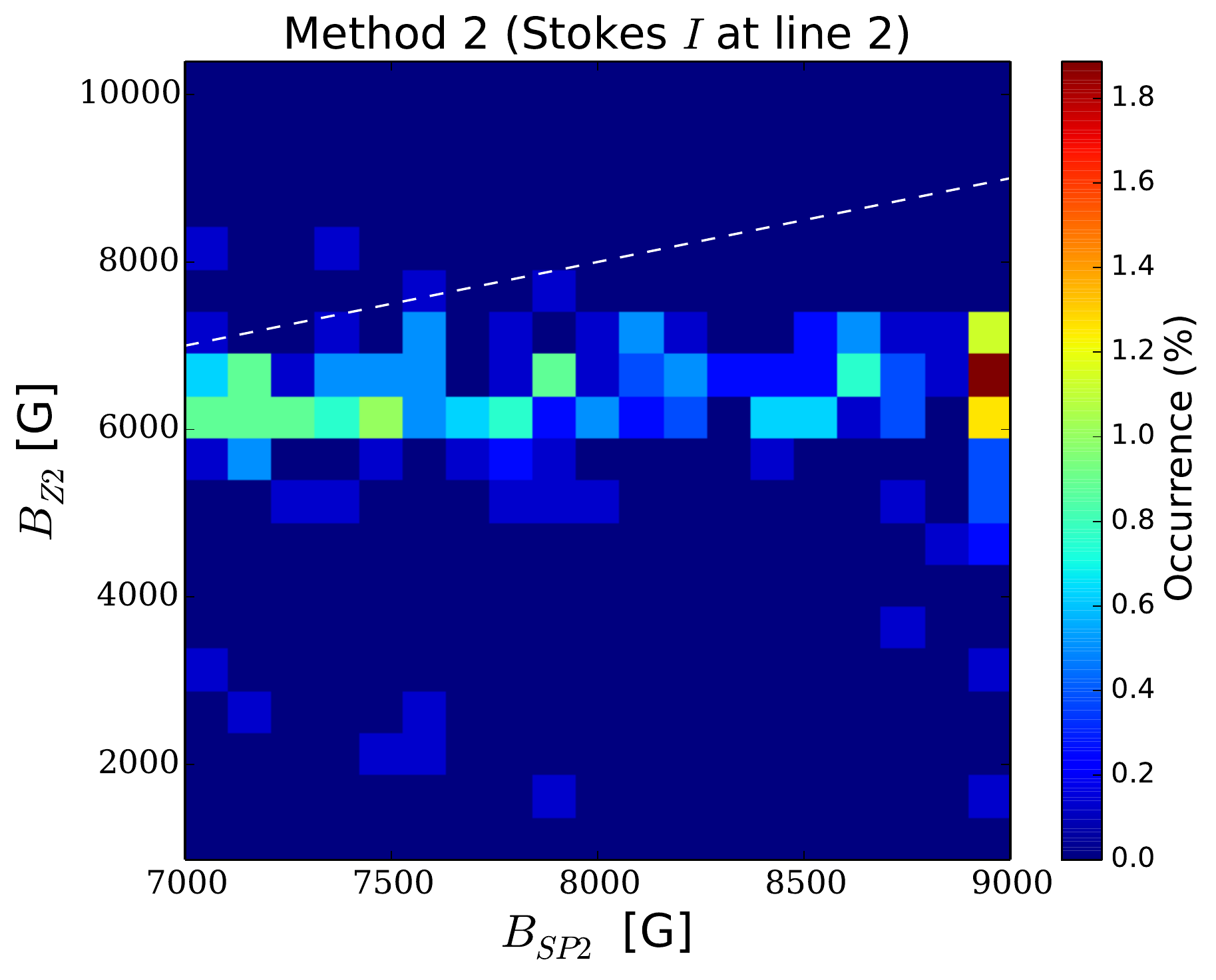}
}

\resizebox{\hsize}{!}
            {
\includegraphics[width=0.5\textwidth]{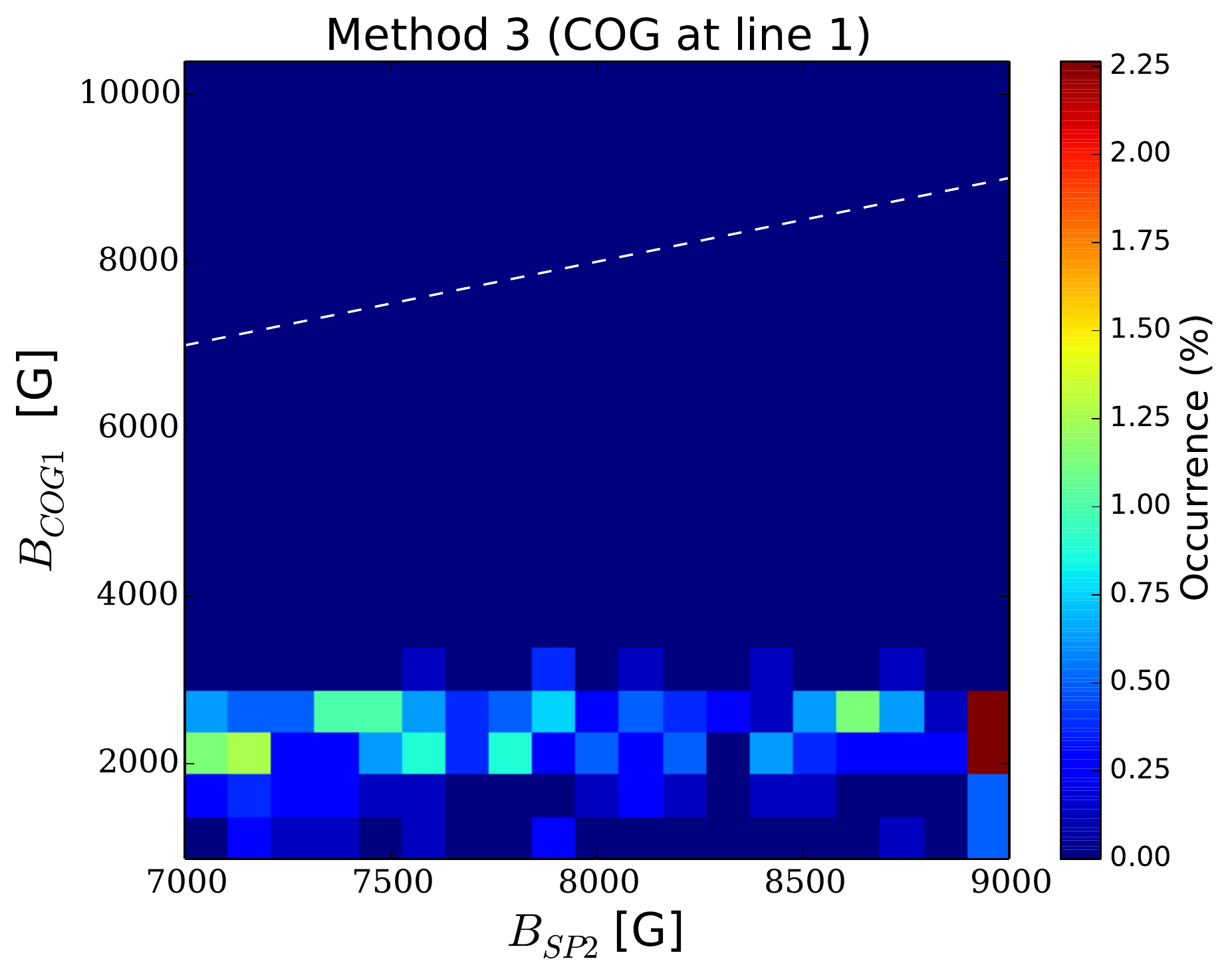}
     \includegraphics[width=0.5\textwidth]{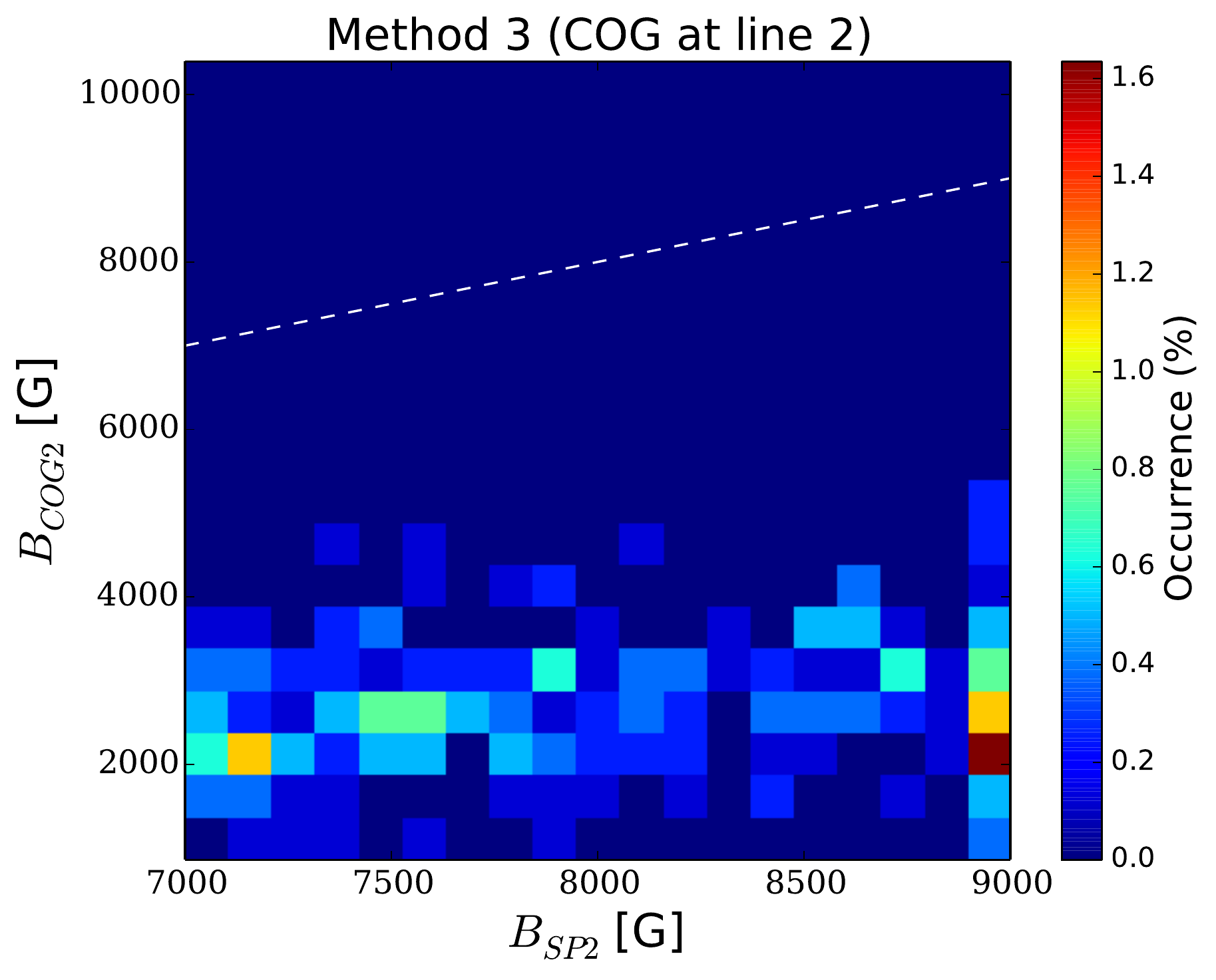}}

\caption[Scatter-plots of Zeeman and COG methods vs SPINOR 2D inversions in the LFPs]{Scatter plots of the 226 LFPs where SPINOR 2D returns $B>7$ kG at $\log(\tau)=0$ ($B_{SP2}$). The y axis indicates magnetic field values obtained with each of the 3 alternative methods (direct Zeeman splitting and COG methods, $B_Z$ and $B_{COG}$ respectively) as described in the text and displayed in Table \ref{tab:1*}. From top to bottom: Methods 1, 2, and 3 for line 1 at  $\lambda=6301.5 \AA$ (subscript 1, left plots)  and  for line 2 at $\lambda=6302.5 \AA$ (subscript 2, right plots). Dashed lines represent expectation values if both methods give identical results (white).}
\label{fig:3_scatter}
   \end{figure*}

\begin{figure*}
 \begin{subfigure}[b]{0.5\textwidth}
\centering
\includegraphics[width=\textwidth]{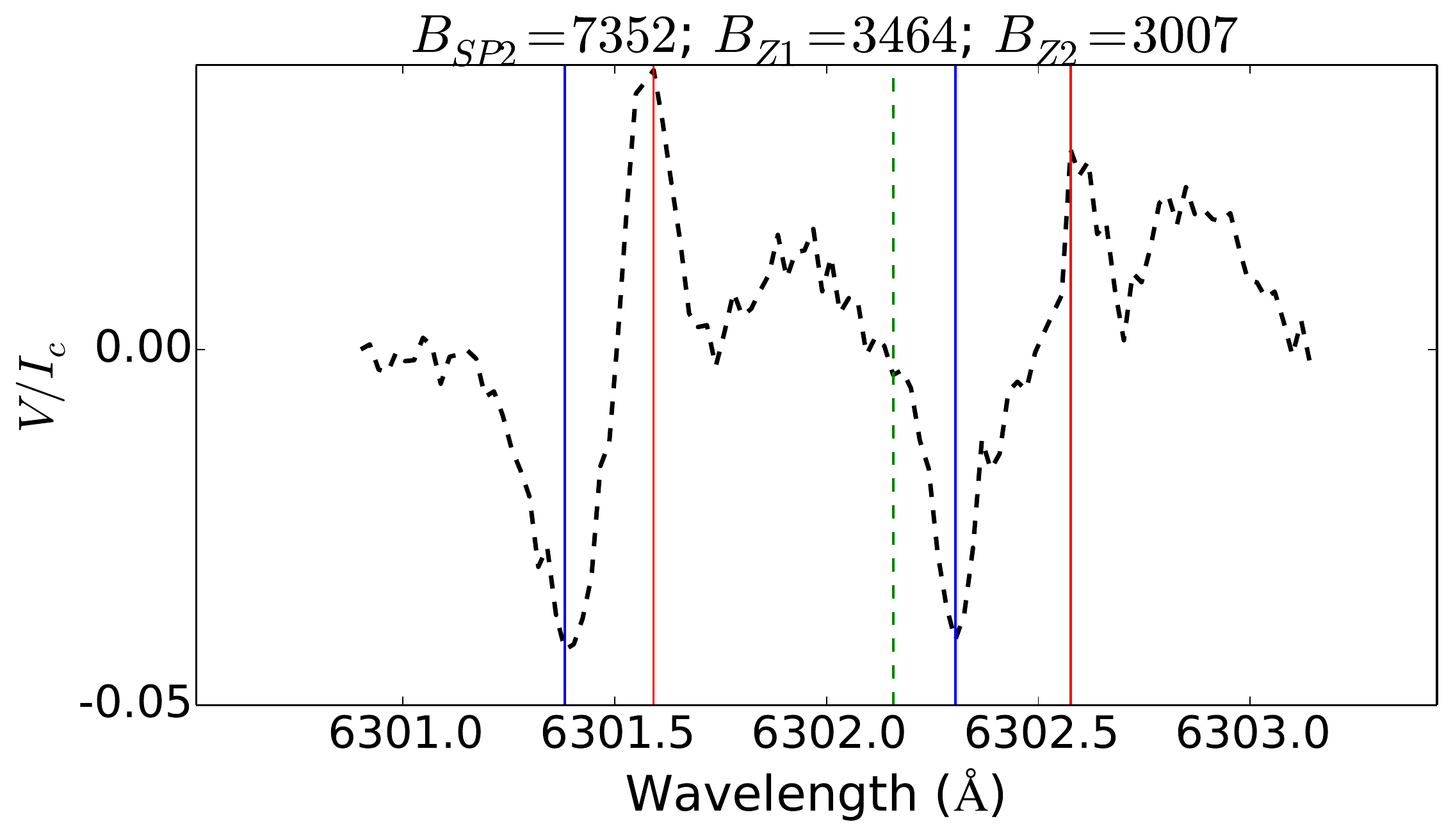}
        \caption{}
	\label{fig:3_IVa}
     \end{subfigure}%
    ~
    \begin{subfigure}[b]{0.5\textwidth}
        \centering
\includegraphics[width=\textwidth]{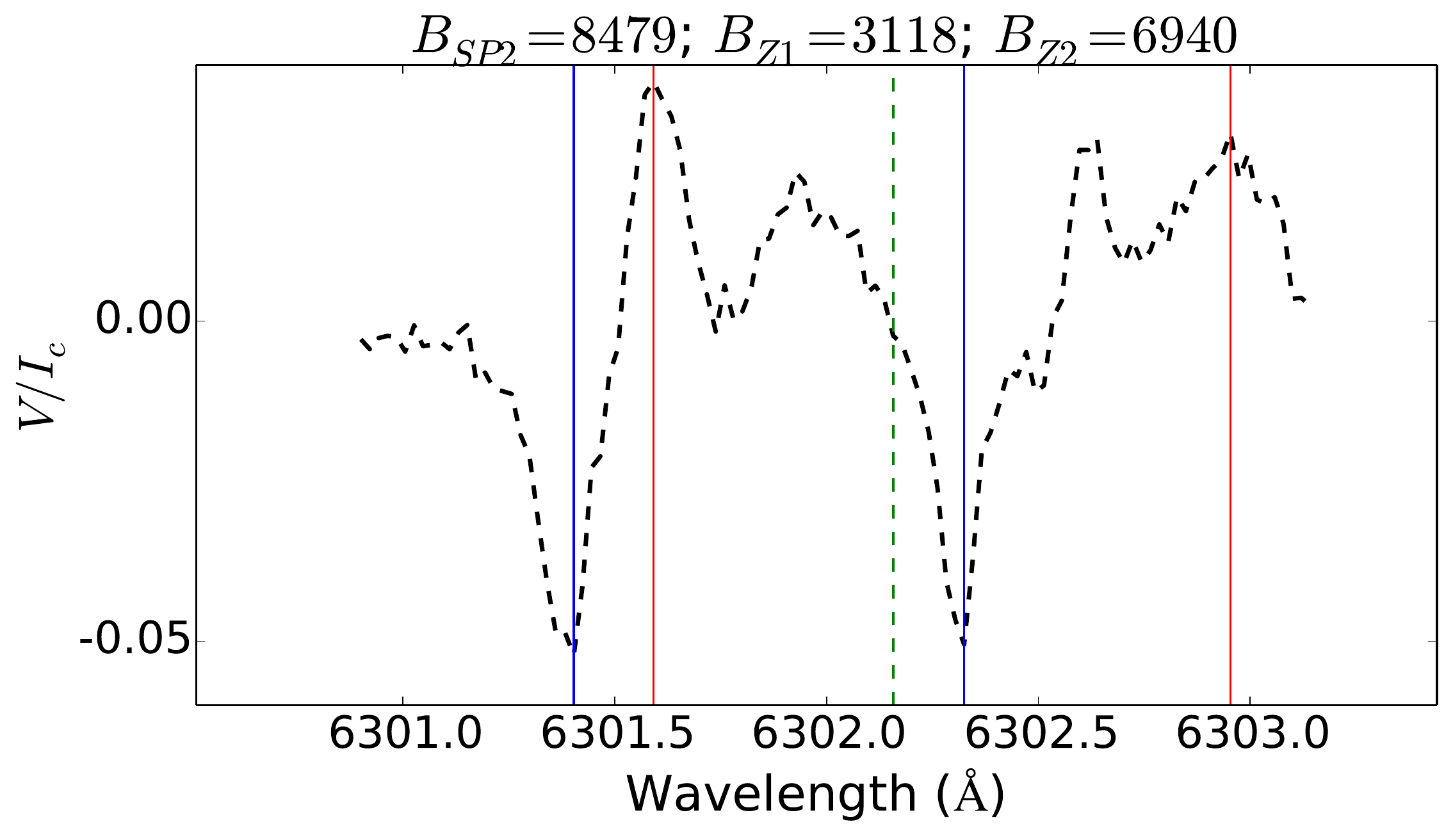}
        \caption{}
	\label{fig:3_IVb}
     \end{subfigure}%

  \begin{subfigure}[b]{0.5\textwidth}
        \centering
\includegraphics[width=\textwidth]{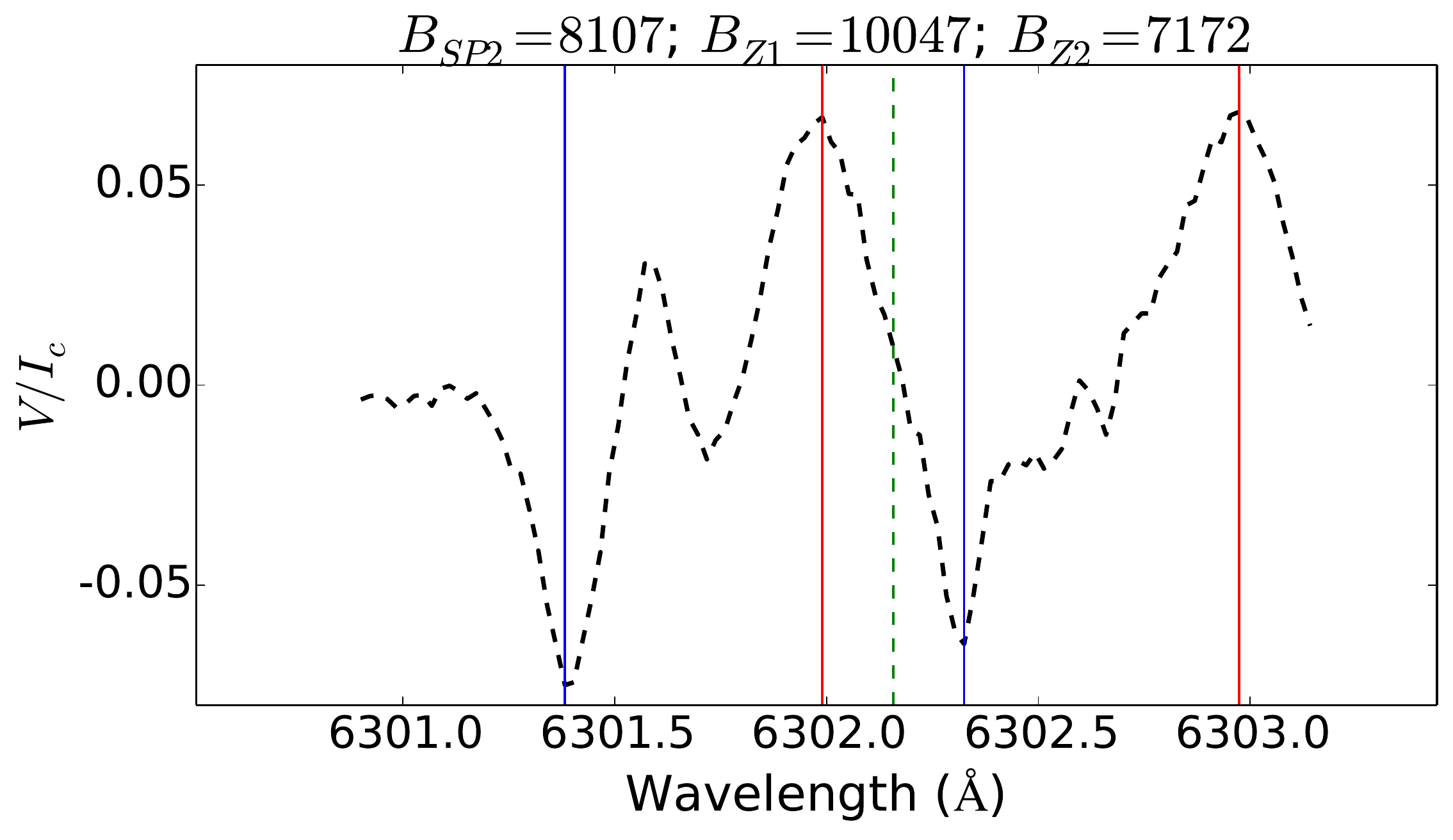}
\caption{}
	\label{fig:3_IVc}
     \end{subfigure}%
 ~
    \begin{subfigure}[b]{0.5\textwidth}
        \centering
\includegraphics[width=\textwidth]{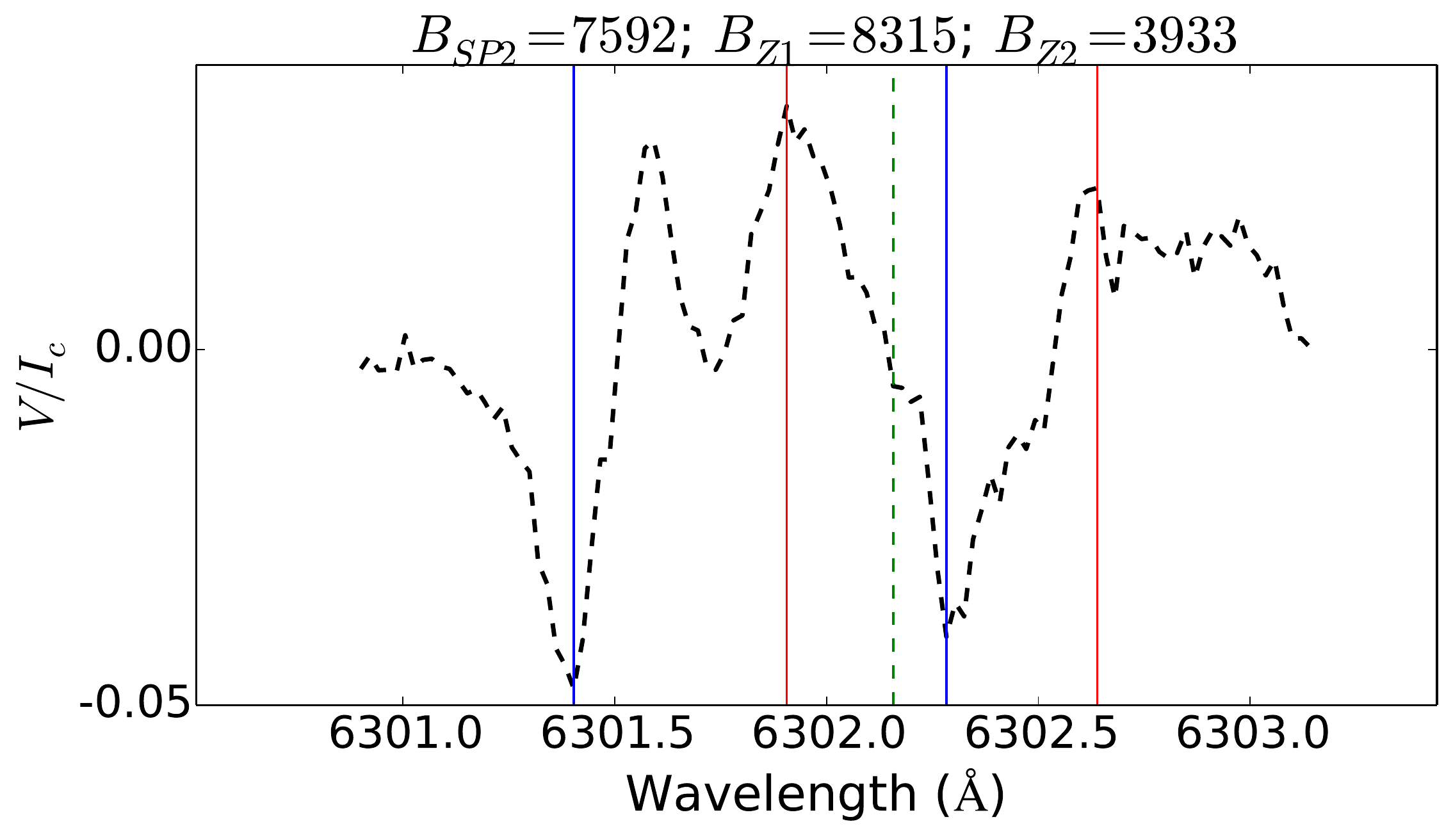}
\caption{}
	\label{fig:3_IVd}
     \end{subfigure}%
\caption[Four examples of observed Stokes $V$ profiles in the LFPs]{Four examples of observed Stokes $V$ profiles in the LFPs. The vertical lines indicate the position of the $\lambda^-$ (blue) and $\lambda^+$ (red) used to calculate the magnetic field strength with method 1. The headers of the plots indicate the magnetic field strengths as derived from: SPINOR 2D inversions ($B_{SP2}$),   method 1 applied to the Fe  I line at $\lambda=6301.5 \AA$ ($B_{Z1}$), and method 1 applied to the Fe I line at $\lambda=6302.5 \AA$ ($B_{Z2}$). }
\label{fig:3_IV}
   \end{figure*}

Scatter plots in Figure \ref{fig:3_scatter} show the results of all three methods applied to all 226 LFPs found in the penumbra. 
Method 1 (top panels) returns  mainly two different solutions in the two Fe I lines, one of which is centered at $\sim3.5$ kG in both lines and a second solution  that is centered at $\sim9$ kG for line 1 
and at $\sim7$ kG for line 2. 
The fact that method 1 senses two very different field strengths in both cases 
 is a consequence of the complex shape of the multi-lobed Stokes $V$ profiles in the observed spectra, see e.g. Figure \ref{fig:3_IV} which displays four different shapes of Stokes $V$ profiles that prevail among the 226 LFPs. For profiles like the one shown in  Figure \ref{fig:3_IVa}, method 1 selects the innermost lobes in both lines (given that in this kind of profiles the inner lobes on the red wing of the lines satisfy the criterion used by  method 1 to select $\lambda^+$), therefore returning field strengths that are almost  half  of the SPINOR 2D inversion result ($B_{SP2}\sim7.3$ kG). Similar situations occur in profiles with shapes as shown in Figures \ref{fig:3_IVb} and \ref{fig:3_IVd}, where the inner lobes on the red wings for line 1 and line 2, respectively, are the largest. In contrast, in profiles like those shown in Figures \ref{fig:3_IVb} (line 2),  \ref{fig:3_IVc} 
(both lines), and \ref{fig:3_IVd} (line 1), the method selects the most external lobes, therefore computing very large  field strengths as displayed in Figure \ref{fig:3_scatter} (top panels).

In contrast, method 2 (middle panel in Fig. \ref{fig:3_scatter}) finds a field strength distribution centered at $\sim7$ kG; whilst method 3 (bottom panels) mainly senses field strengths of the order of 2.5 kG for line 1 and around 3 kG for line 2.

The correlation between the field strengths from the methods and the results from SPINOR 2D are very low, particularly for method 3. However, there is some level of correlation between SPINOR 2D and the cloud of pixels displaying the strongest field solution in method 1. Likewise, there is a good correlation between the clouds of pixels displaying the weaker field solution in method 1 with the solutions in method 3. 

The large discrepancy between the methods has to do with the fact that only the LFPs have been plotted. Ideally, to check if the methods are consistent at lower field values but depart from each other only at strong fields, scatter-plots with a larger sample of pixels covering a broader range of magnetic field values in the x axis are needed.
However, it is not possible to apply the direct Zeeman splitting method to profiles that do not show distinguishable sigma components, i. e. to most of the penumbral pixels, except for the LFPs.

For the COG method, differences are expected due to velocity gradients, vertical field gradients, temperature, etc., that are not taken into account by the method (the gradients in $B$ and $v_{LOS}$ are not taken into account in Eqs. \ref{eq:1a} and \ref{eq:3a}).
  The presence of such gradients is indicated by the asymmetries of the Stokes profiles.  Moreover, if there are spatially unresolved areas containing field inhomogeneities (e.g. different field strengths and/or different field polarities located next to each other within the same resolution element), then the resultant line profiles from such pixels will be the summation of all the different magnetic components, so that  the number of lobes in the observed Stokes $V$ profiles will depend on the number of  unresolved components \citep[e.g.,][]{Stenflo1993}. Most Stokes $V$ profiles from the 226 LFPs  display more than two lobes, suggesting the presence of  different magnetic field components in each of those pixels. Nonetheless, according to methods 1 and 2,
 the very large wavelength separation between the most external lobes 
observed in the Stokes $V$ profiles 
could still reflect that one of the field components is particularly strong. 
In such a scenario, an important aspect to consider is if the Paschen-Back effect would play an important role under the presence of such strong fields, i.e. if the splitting of atomic levels caused by such strong external magnetic fields would dominate over the $LS$-coupling.

\begin{figure*}
   \resizebox{\hsize}{!}
            {\includegraphics[width=0.5\textwidth]{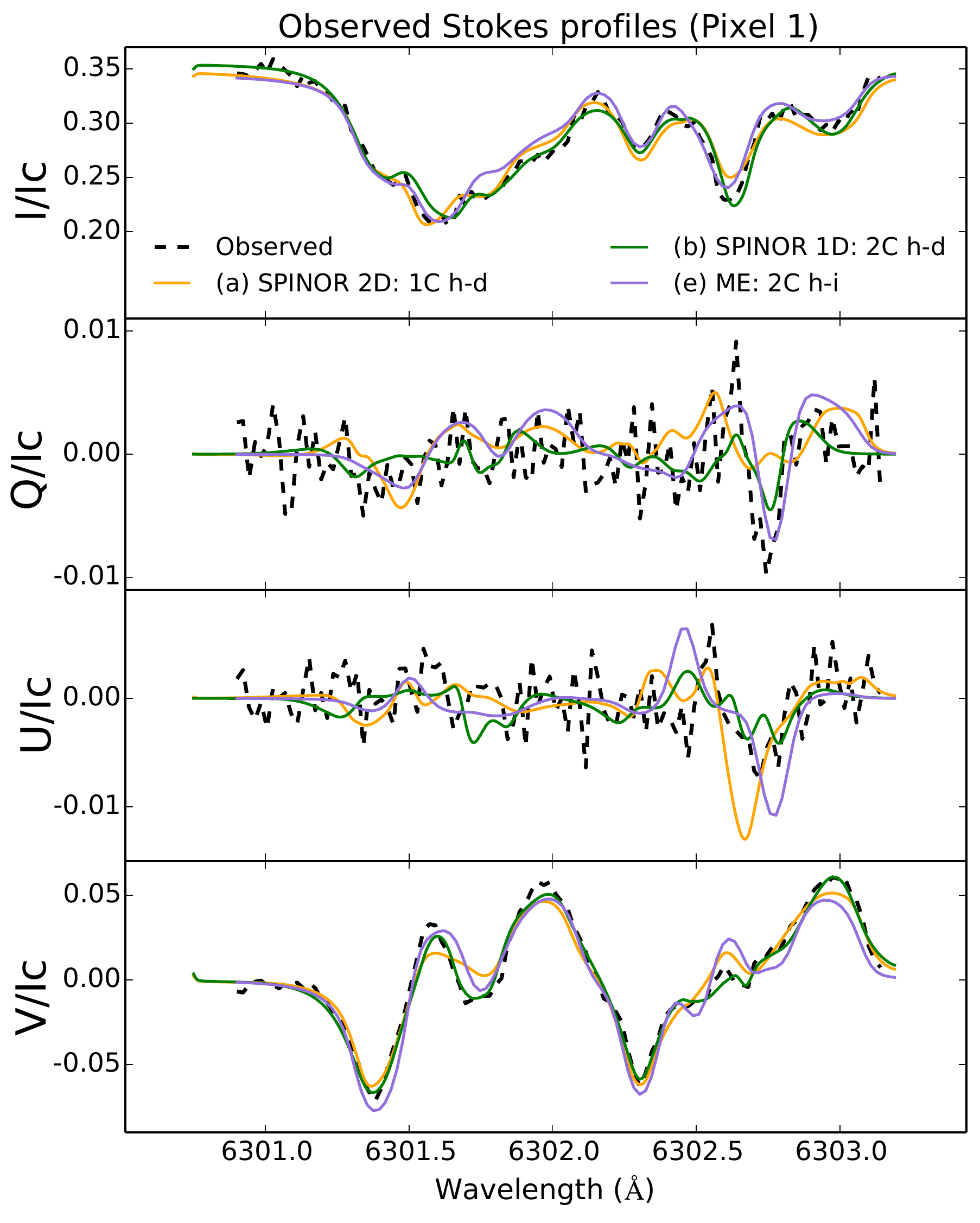}
        \includegraphics[width=0.5\textwidth]{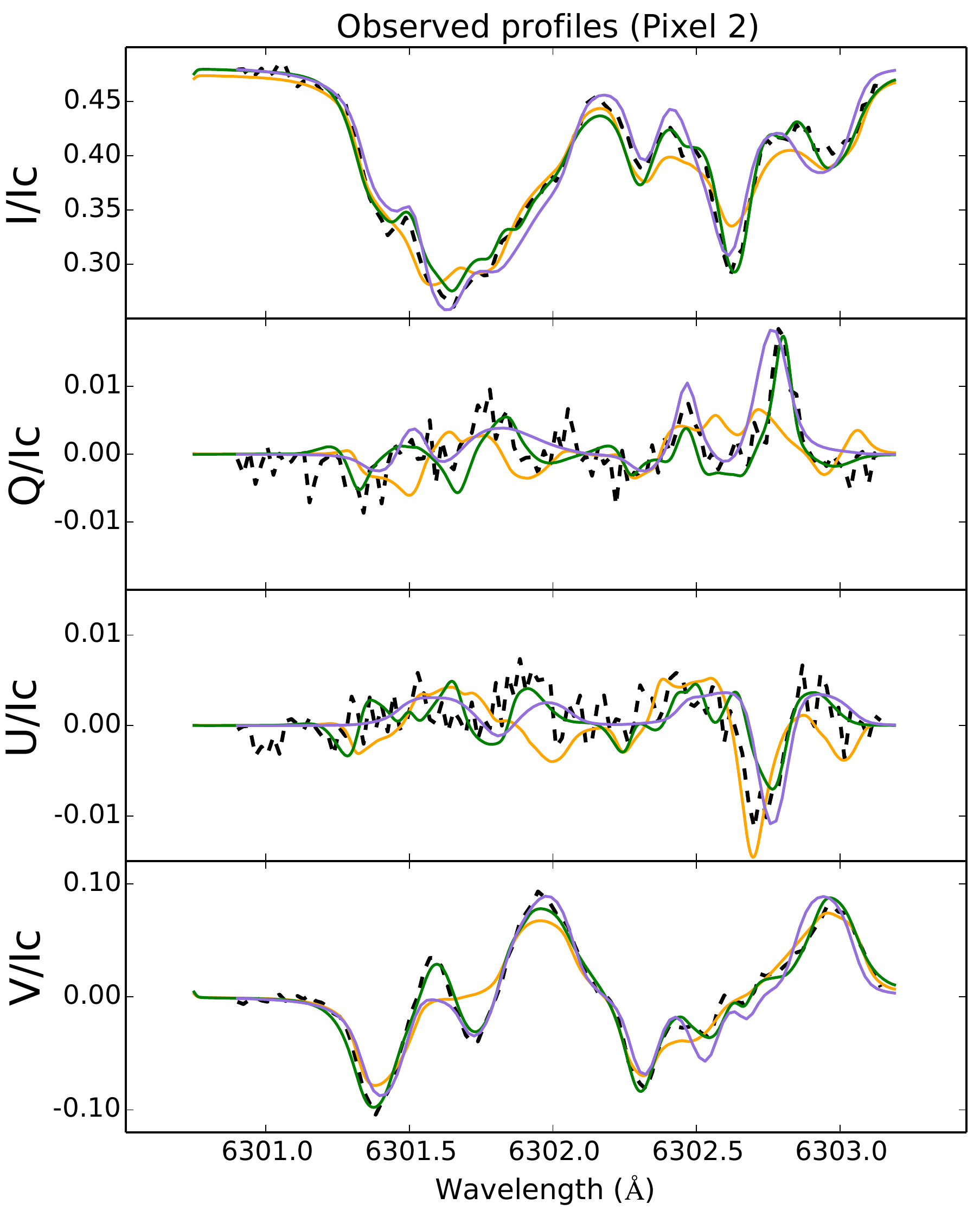}
}

\caption[Observed Stokes profiles in two selected LFPs and their best fits returned from different inversions ]{ Observed Stokes profiles (dashed curves), and their best fits returned by SPINOR 2D (orange curves),  SPINOR 1D 2-component height-dependent inversions (green curves) and Milne-Eddington 2-component inversion (purple curves)  in the two selected LFPs.  
Plots are in the same format as plots in Figure \ref{fig:3}.}
\label{fig:3**}
   \end{figure*}

The Paschen-Back regime occurs when $\Delta E_{ik}\gg \mu_B g_L MB$, where  $\Delta E_{ik}$ is the energy difference in the atom between the terms of the multiplet structure, $\mu_B$ is the Bohr magneton, $g_L$ is the Land\'e factor, $M$ is the magnetic quantum number, and $B$ is the magnetic field strength.
The Fe I $^{5}P_2$$-$ $^{5}D_2$ $\lambda=6301.5 \AA$  and the Fe I $^{5}P_1$$-$ $^{5}D_0$ $\lambda=6302.5 \AA$ lines belong  to the same multiplet No. 816 \citep{Moore1945}. The minimum energy difference of the lower levels of these lines is $\Delta E_{ik}\approx 0.032$ eV and leads to a magnetic field `threshold value' for the Paschen-Back effect of the order of $10^3$ kG, a value that is well above the field strength values inferred by methods 1, 2, and the SPINOR 2D inversions.
Nonetheless, according with laboratory experiments the Paschen-Back effect actually takes place under the presence of magnetic fields with strengths $10-100$ kG  when the multiplet splitting energy difference is so small that the two adjacent lines of the same multiplet are separated by a distance of less than 1 $\AA$ \citep{Frisch1963}. 
In the present case, since the analyzed pair of Fe I lines are separated by around 1$\AA$ and the Zeeman splittings being discussed are seemingly very large (corresponding to field strengths approaching 10 kG in some LFPs according to method 1), it is possible that the Paschen-Back effect starts playing some role in the magnetic splitting for those cases. However, even within the Paschen-Back regime,  the magnetic field could still be measured with sufficient accuracy 
according to the laboratory results of \citet{Moore1945}.

\begin{figure*}
   \resizebox{\hsize}{!}
            {\includegraphics[width=0.5\textwidth]{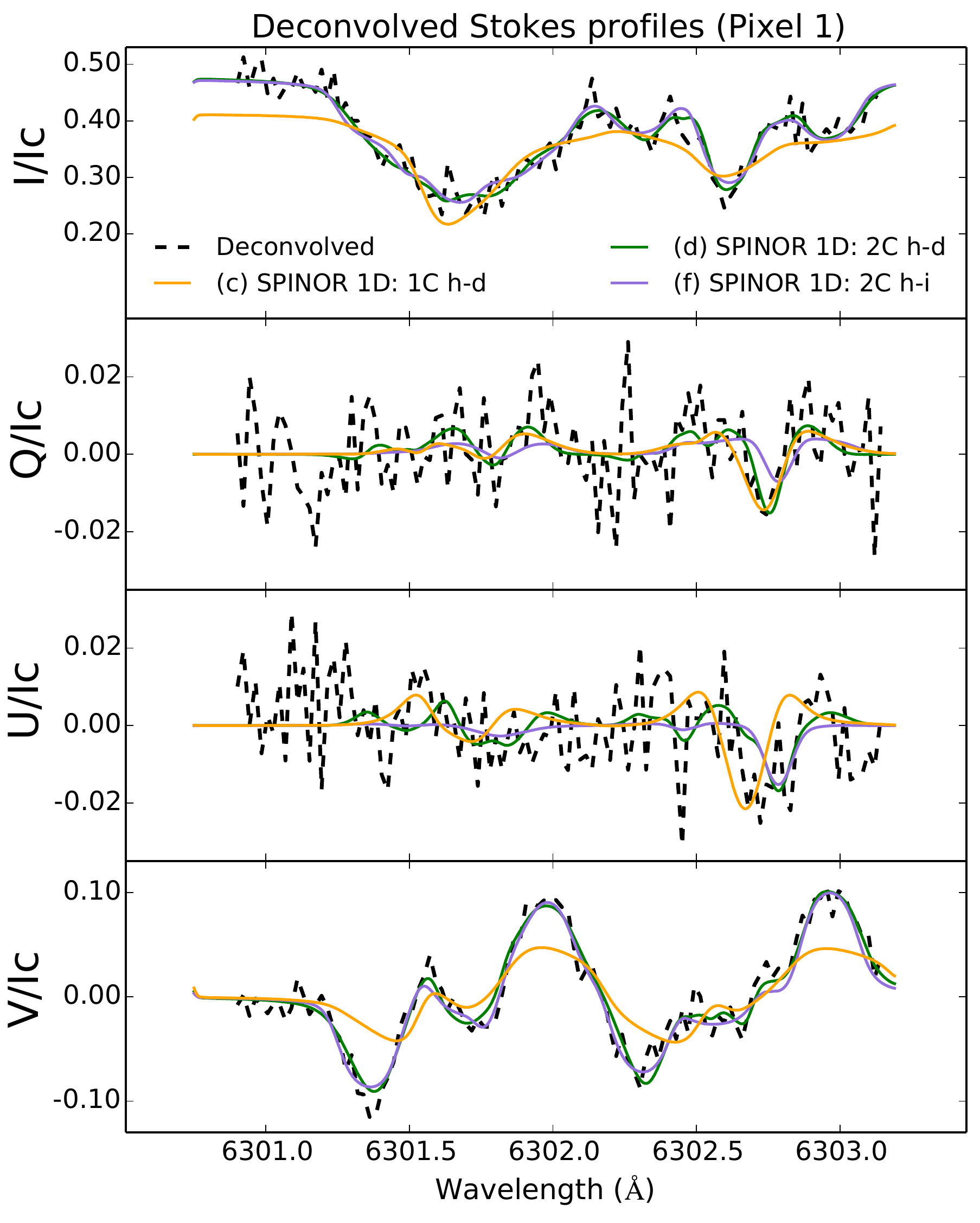}
        \includegraphics[width=0.5\textwidth]{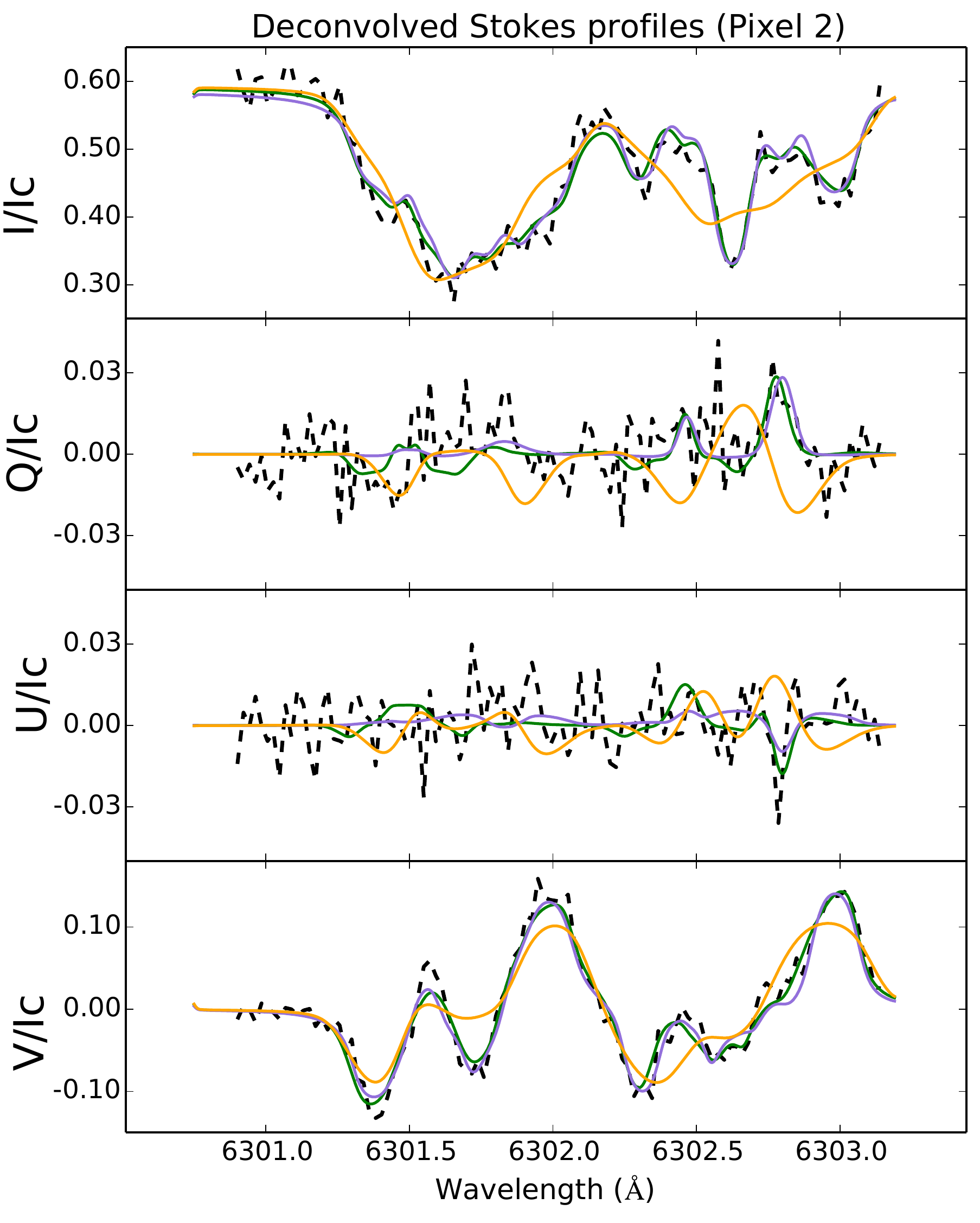}
}

\caption[Deconvolved Stokes profiles in two LPFs and their best fits returned from different inversions ]{ Deconvolved Stokes profiles (dashed curve), and their best fits returned by SPINOR 1D  inversions, using  a single-component height dependent model atmosphere  (orange curves), a 2-component height-dependent model (green) and a 2-component  height independent model atmosphere (purple curves)  for the two selected  LFPs. Plots are in the same format as plots in Figure \ref{fig:3}.}
\label{fig:3***}
   \end{figure*}

Another plausible scenario is that the observed multi-lobed Stokes $V$ profiles are produced by  two unresolved atmospheric components that
display large differences in their Doppler velocity
\citep[e.g.,][]{Solanki1993b,Martinez2000,Borrero2002,Schlichenmaier2002b,Bellot2004}. One of the components could  be associated with the umbral magnetic field in the sunspot (i.e. nearly at rest) and the second one with the filamentary CEF penumbra (strongly redshifted). This possibility is considered in the following section based on the fact that all the LFPs appear to be mostly located near, or at the umbral/penumbral edge (see Figs. \ref{fig:LFP_Bmap} and \ref{fig:LFPa}).

\section{Inversions}

In order to gain more insight into the reliability of the large magnetic field strengths returned by the SPINOR 2D inversion code, we now apply 5 additional inversions, considering two atmospheric components in some of them. It is noteworthy that the inclusion of a second atmospheric component causes the number of free parameters, $n$, to increase to almost twice that in a single component model used in the SPINOR 2D inversions. While this can and should lead to a much better fit for a complex Stokes profile (lower $\chi^2$), it also increases the risk of obtaining artificial (unphysical) results. 

The merit functions, $\chi^2$, are not computed in the same way by the different inversion codes; and hence, they are in principle not comparable among each other. To perform a valid comparison between the different fits, in the following, the minimum $\chi^2$ is chosen to be the sum over the squared differences between the observed and the synthetic profiles resulting from the best fits in each case.

In Tables \ref{tab:1} and \ref{tab:2}, we show some of the parameters corresponding to the  best fits of the Stokes profiles in the two selected LFPs, obtained from all 6 different inversions. These inversions can be classified into two categories:

\subsection{Height-dependent inversions\label{ch31:hd}}

 The first category includes \textit{height dependent inversions} (Table \ref{tab:1}), i.e. inversions in which all the parameters are allowed to vary with optical depth (with nodes being set at $\log(\tau)=-2.0,-0.8$ and $0$), such as (a) SPINOR 2D inversions, with 18 free parameters; (b) SPINOR 1D \citep{Frutiger2000b} 2-component inversions, applied to the observed Stokes profiles (best fits are shown by green curves in Fig. \ref{fig:3**}), with 37 free parameters; (c) SPINOR 1D single-component, applied to the deconvolved Stokes profiles, with 18 free parameters and  (d) SPINOR 1D 2-component inversions, applied to the deconvolved Stokes profiles, with 37 free parameters.

Inversions (c) and (d) are intended to resemble the SPINOR 2D technique as far as accounting of the instrumental effects over the observed Stokes profiles is concerned, although the treatment is not as consistent as that by SPINOR 2D.  We retrieve the deconvolved Stokes profiles from the spatially degraded observed Stokes profiles by using an effective point-spread function (PSF)  \citep{Danilovic2008} constructed from the pupil function of the 50-cm \textit{Hinode} SOT \citep{Suematsu2008} and applying the Richardson-Lucy deconvolution method (see \citet{Richardson1972} and \citet{Lucy1974} for details). The resultant deconvolved Stokes profiles in the selected pixels are displayed in Figure \ref{fig:3***} (black dashed curves) together with their best fits obtained by SPINOR 1D (c) and (d) (orange and green curves in Fig.  \ref{fig:3***}, respectively).

\subsection{Height-independent inversions\label{ch31:id}}
The second category corresponds to \textit{height independent inversions} (Table \ref{tab:2}), i.e., we assume no variation with optical depth of atmospheric parameters by using (e) a 2-components Milne-Eddington inversion \citep{Skumanich1987,Lagg2004,Lagg2009,Borrero2011b} applied to the observed Stokes profiles (best fits are shown by purple curves in Fig. \ref{fig:3**}), with 15 free parameters; and (f)  2-component 1-node SPINOR 1D inversions applied to the deconvolved Stokes profiles  (best fits are shown by purple curves in Fig. \ref{fig:3***}), with 17 free parameters.

\section{Results}

The SPINOR 2D  best-fits give $B\sim8.3$ kG at $\log(\tau)=0$ for the profiles in pixel 1, while $B=8$ kG is obtained for both pixels at $\log(\tau)=-0.8$ (see Table \ref{tab:1}), with $\chi^2=14$ and $35$ for each fit, respectively. Note that these fits  do not succeed in perfectly reproducing the reversed central lobes of the Stokes $V$ profiles observed in both pixels. The deconvolved spectra displayed in Figure \ref{fig:3***} also show the reversed central lobes in Stokes $V$, which suggests that they are not a result of the mixing of signals due to instrumental effects. Nonetheless, inversions (c) also fail in reproducing the central reversed lobes, providing a much poorer fit than SPINOR 2D ($\chi^2=44$ and $47$ for each pixel, respectively, when taking into account an increased noise of $\sim5\sigma$ in the deconvolved Stokes profiles caused by the deconvolution itself) and featuring $B\sim7$ kG at $\log(\tau)=0$ in both pixels. Thus, in both cases, the 1-component inversions cannot fit these four-lobe profiles to high precision. At the same time, both inversions  return very large field values. Such extreme field values are likely the result of providing a good fit mainly to the external lobes of Stokes $V$ to reproduce the large wavelength separation in terms of the Zeeman splitting.

\begin{figure*}[htp!]
 \centering

\begin{subfigure}[b]{0.47\textwidth}
\includegraphics[width=\textwidth]{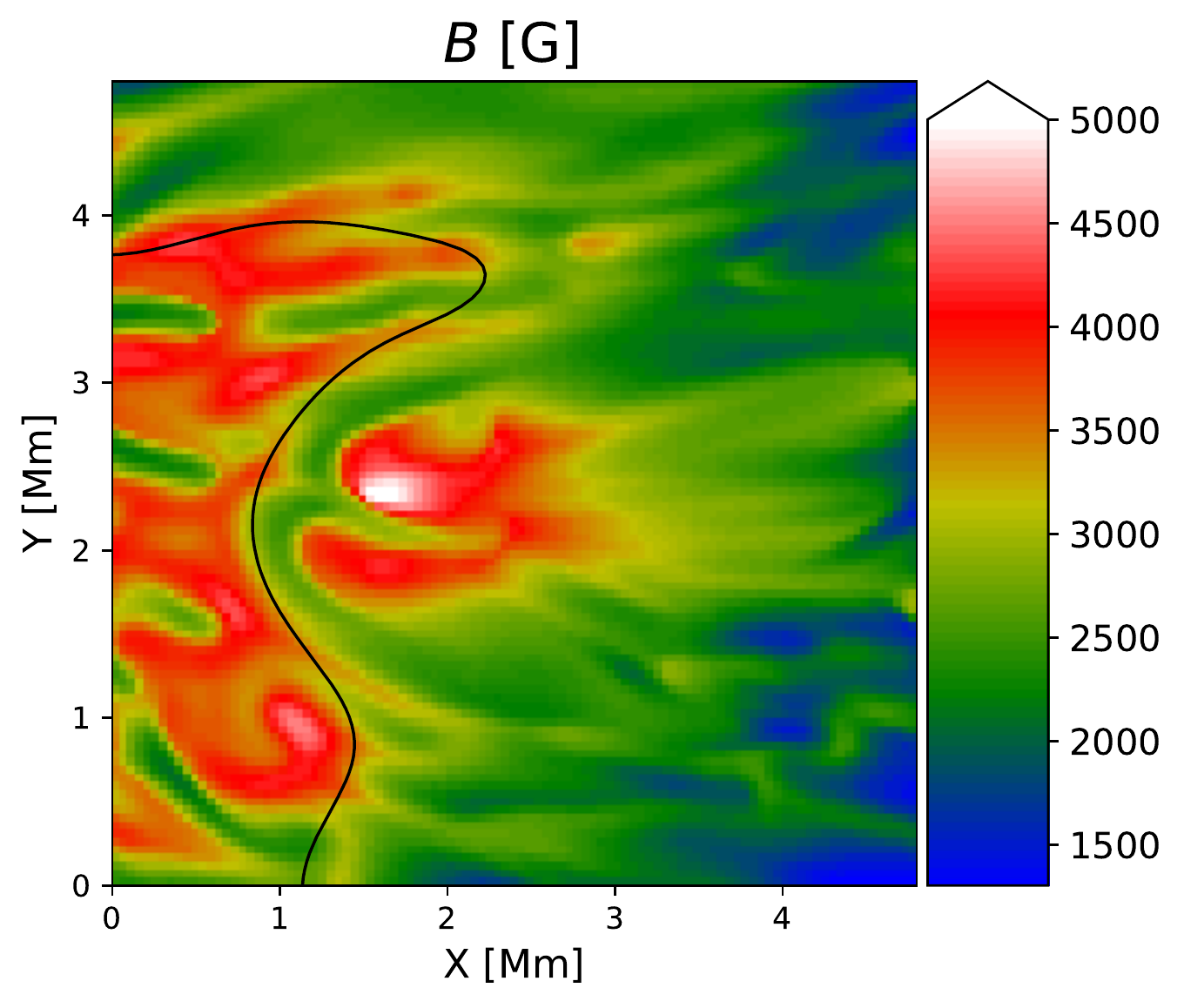}
 \caption{}
\label{fig:mhd_para}
 \end{subfigure}
~
\begin{subfigure}[b]{0.47\textwidth}
\includegraphics[width=\textwidth]{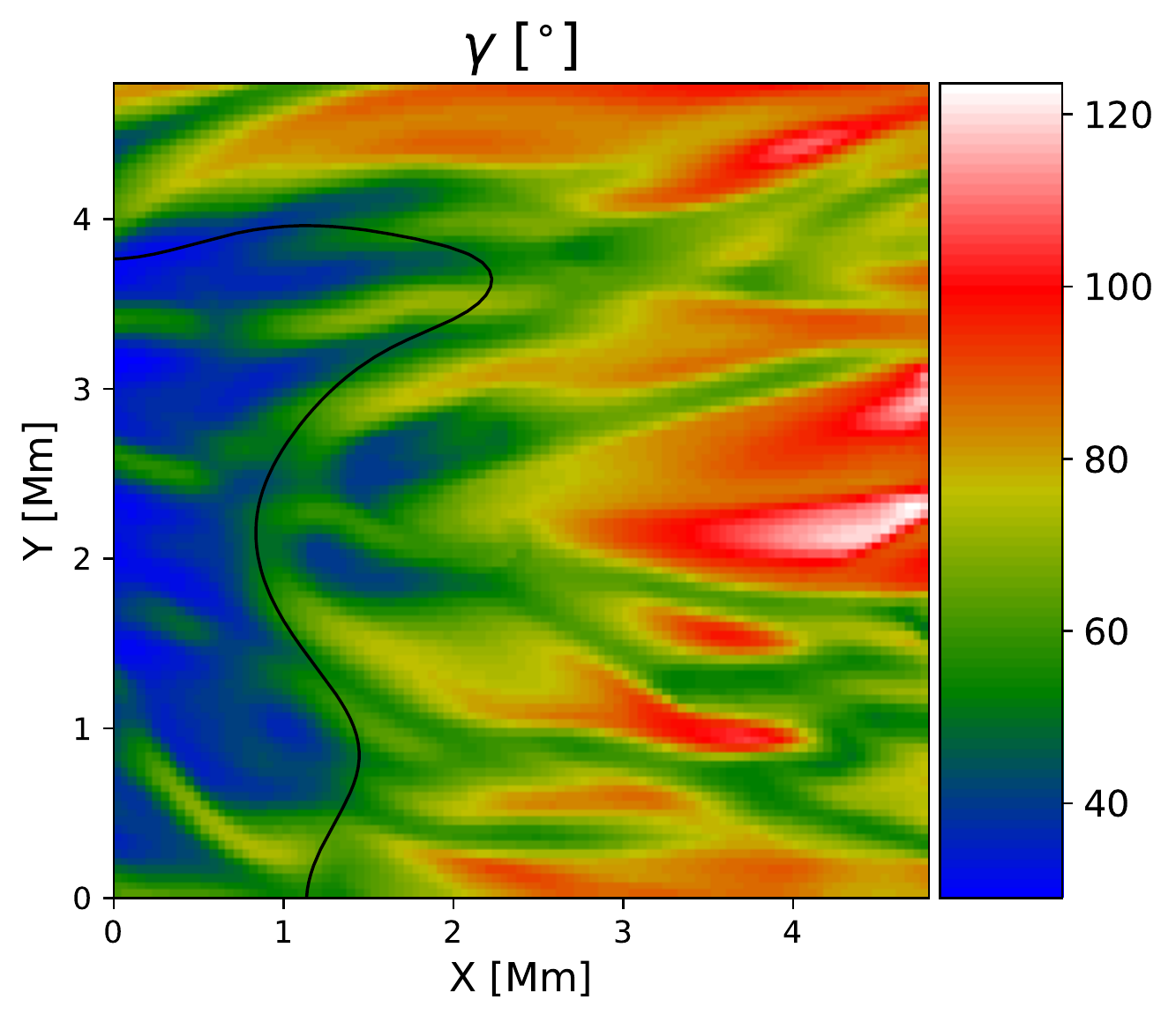}
\caption{}
\label{fig:mhd_parb}
 \end{subfigure}

\centering
\begin{subfigure}[b]{0.47\textwidth}
\includegraphics[width=\textwidth]{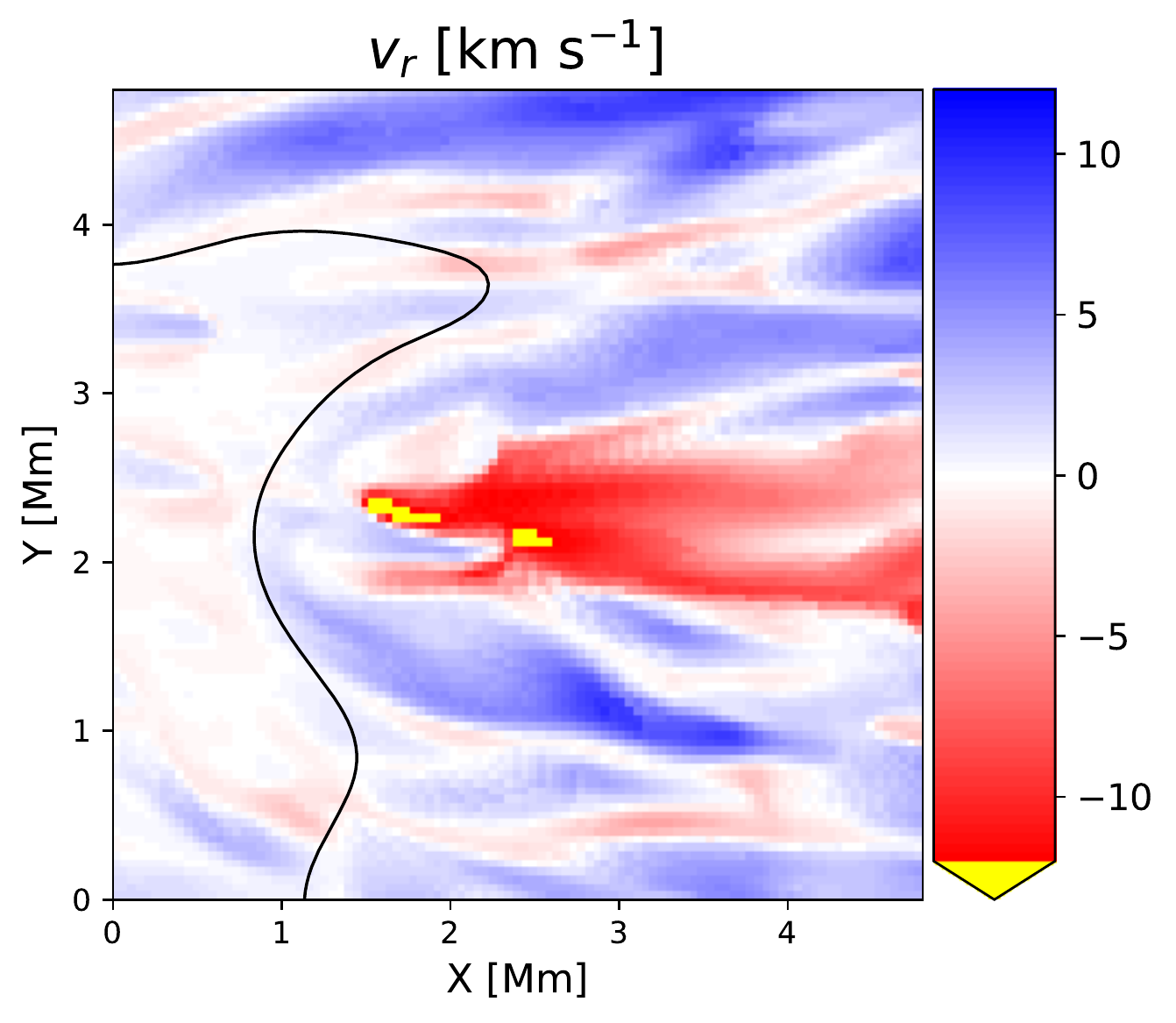}
\caption{}
\label{fig:mhd_parc}
 \end{subfigure}
~
\begin{subfigure}[b]{0.47\textwidth}
\includegraphics[width=\textwidth]{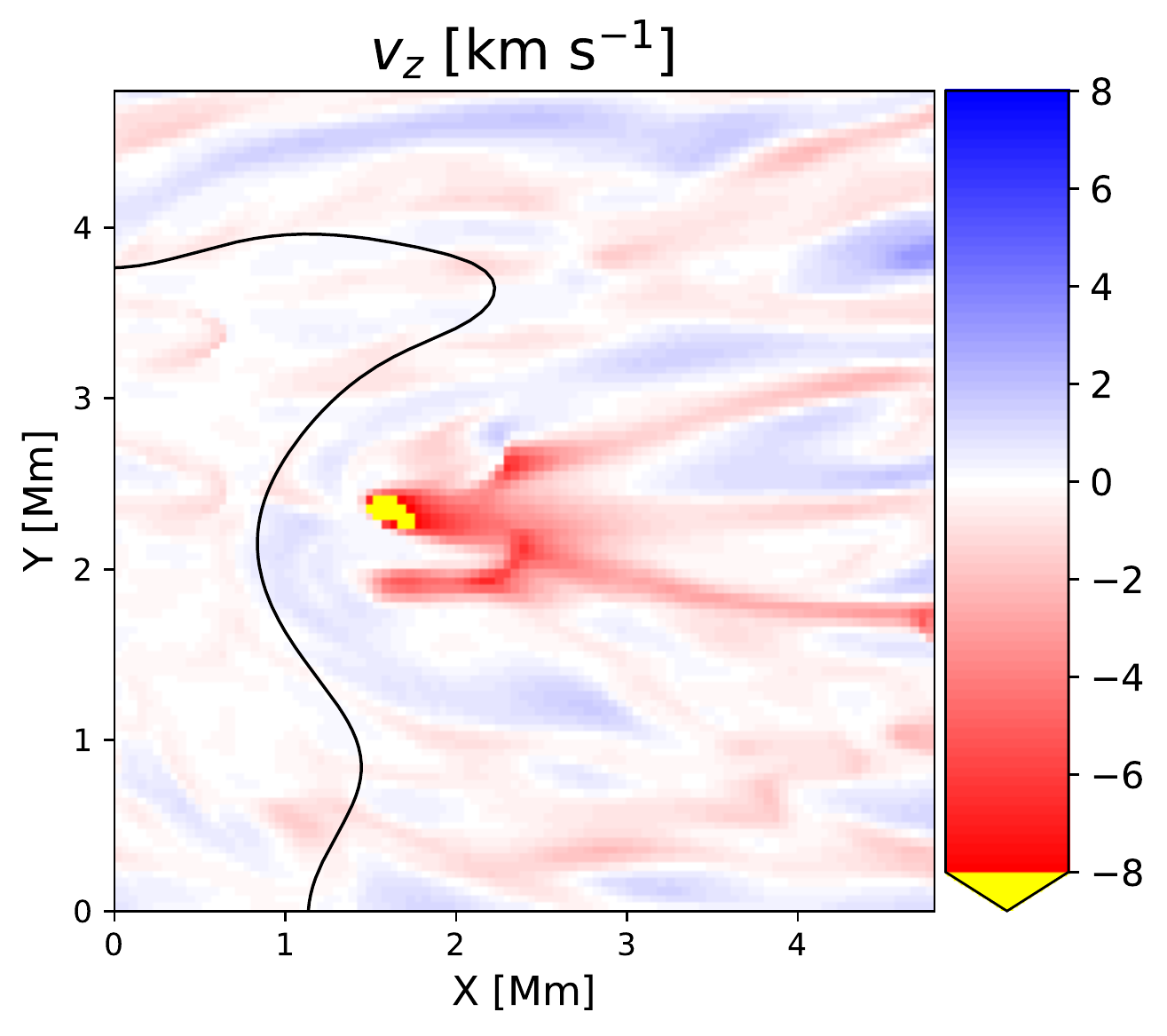}
\caption{}
\label{fig:mhd_pard}
 \end{subfigure}

    \caption[]{A portion of the inner penumbra in the  MURaM sunspot simulation by \citet{Rempel2015} with some filaments hosting a counter-EF \citep[see also][]{Siu2018}. The maps show: (a) the magnetic field strength $B$ [G]; (b) the field inclination with respect to the vertical $\gamma$ [$^{\circ}$], i.e. $\gamma$=0$^{\circ}$ represents a vertical field of  umbral polarity, $\gamma$=90$^{\circ}$ a horizontal field, and $\gamma$=180$^{\circ}$  a vertical field of opposite polarity to the umbra; (c) radial flow velocity $v_r$ [km s$^{-1}$]; and (d) the vertical flow velocity $v_z$  [km s$^{-1}$].
Negative $v_r$ and $v_z$ values (red-to-yellow colors) indicate inflows and downflows, respectively. This sign convention differs from the
one used in observational studies, where negative values denote flows moving towards the observer along the line-of-sight. The black contour lines where placed at $I_c/I_{QS}<0.45$ near the umbra(left)-penumbra(right) boundary. 
All maps show the corresponding physical parameters at $\log(\tau)=0$.}
\label{fig:mhd_par}
\end{figure*}

While the height-dependent 2-component inversions produce better fits to the four-lobe profiles than the 1-component inversions, $\chi^2=8$ and $12$ for each pixel respectively in inversions (b), and $\chi^2=14$ and $18$ respectively in inversions (d),  they also use a much larger number of free parameters than the 1-component inversions so that a comparison is not straightforward.
The components 1 in inversions (b) and (d) give $B\sim4-4.2$ kG and $v_{LOS}\sim 1$ km s$^{-1}$ at $\log(\tau)=0$ in both pixels, roughly consistent with the umbral environment in the vicinity of those pixels  ($B\sim4.2$ kG at $\log(\tau)=0$, according with SPINOR 2D). Furthermore, the  $v_{LOS}$ in the components 1  are relatively small at all three atmospheric layers, as expected for umbral environments.
In contrast, the components 2 in inversions (b) and (d) give $B\sim4.4$ and $\sim 4.9$ kG, respectively, for pixel 1 at $\log(\tau)=0$; and $B\sim5$ and $5.8$ kG, respectively, for pixel 2; with $v_{LOS} \gtrsim
 16$ km s$^{-1}$ and with the filling factors $\alpha$, i.e. the mixing ratio between the 2 components, being slightly larger for the second components.  
The second component in both pixels could then correspond to  the tails of the penumbral filaments harboring the CEF, with a large redshift and stronger fields than in the surrounding  umbra; which is qualitatively compatible with the results from SPINOR 2D,  but  quantitatively suggests lower penumbral field strengths of the order of $4-5$ kG (although approaching $6$ kG in  one pixel).

The SPINOR 1D inversion code can find a solution involving 2 components, one of which is strongly wavelength shifted to mimic the seemingly very strongly split spectral line. This nearly halves the field strength, although even in this case, we get $B$ values reaching up to nearly $6$ kG. 
These are still very large field strengths and are atypical for penumbral environments. They are close to the record measurement of $6.1$ kG in sunspot umbrae \citep{Livingston2006}.
However, due to the large number of free parameters involved, 
it is difficult to judge if the results they provide are more reliable than those from  SPINOR 2D inversions.

A formal approach to compare the different results obtained from  inversions that consider different model assumptions would be through the comparison of their error bars. Unfortunately, specifying the uncertainties in the fitted atmospheric parameters that take into account the possible degeneracies between parameters is an intrinsic difficulty facing inversions. Especially in the case of the spatially coupled inversions, the changes in the parameters of a single pixel severely affect the result, and therefore the uncertainties of the parameters, of the neighboring pixels. This fact makes the computation of formal errors for a single pixel principally impossible, and therefore the SPINOR 2D inversion code  does not provide errors.

As a simple proxy for the quality of the fits, we compare the models by using the Bayesian Information Criterion \citep[BIC;][]{Schwarz1978}, which is based on the crude approximation of Gaussianity of the posterior with respect to the model parameters:
\begin{equation}\label{eq:bic}
BIC=\chi_{min}^2+n\ln N,
\end{equation}
\noindent 
where $\chi_{min}^2$ is the merit function of the best-fits to the Stokes profiles in each model, $n$ is the number of free parameters and $N$ is the number of observed points. The computed values of the BIC for each  fit are shown in Tables \ref{tab:1} and \ref{tab:2}. The model with the smallest value of the BIC is the preferred one.
The height-dependent model preferred by the BIC is SPINOR 2D in both pixels. However, one of the fundamental problems of this criterion is that it penalizes all parameters equally, not taking into account situations in which data do not constrain some parameters \citep[see e.g.][]{Asensio2012}.

\begin{figure*}[htp!]
 \centering

\begin{subfigure}[b]{0.45\textwidth}
\includegraphics[width=\textwidth]{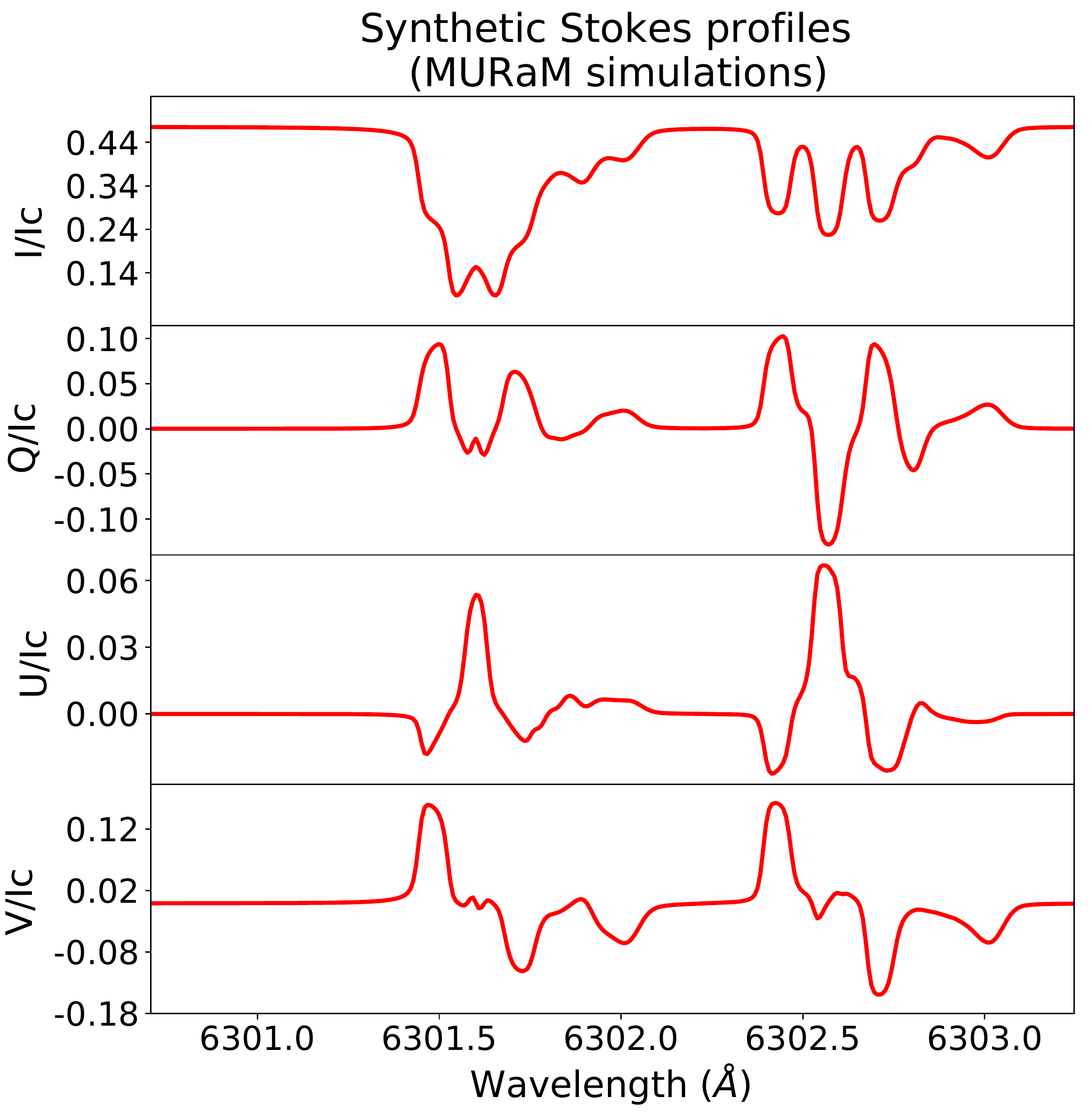}
 \caption{}
\label{fig:mhd_stokesa}
 \end{subfigure}
~
\begin{subfigure}[b]{0.45\textwidth}
\includegraphics[width=\textwidth]{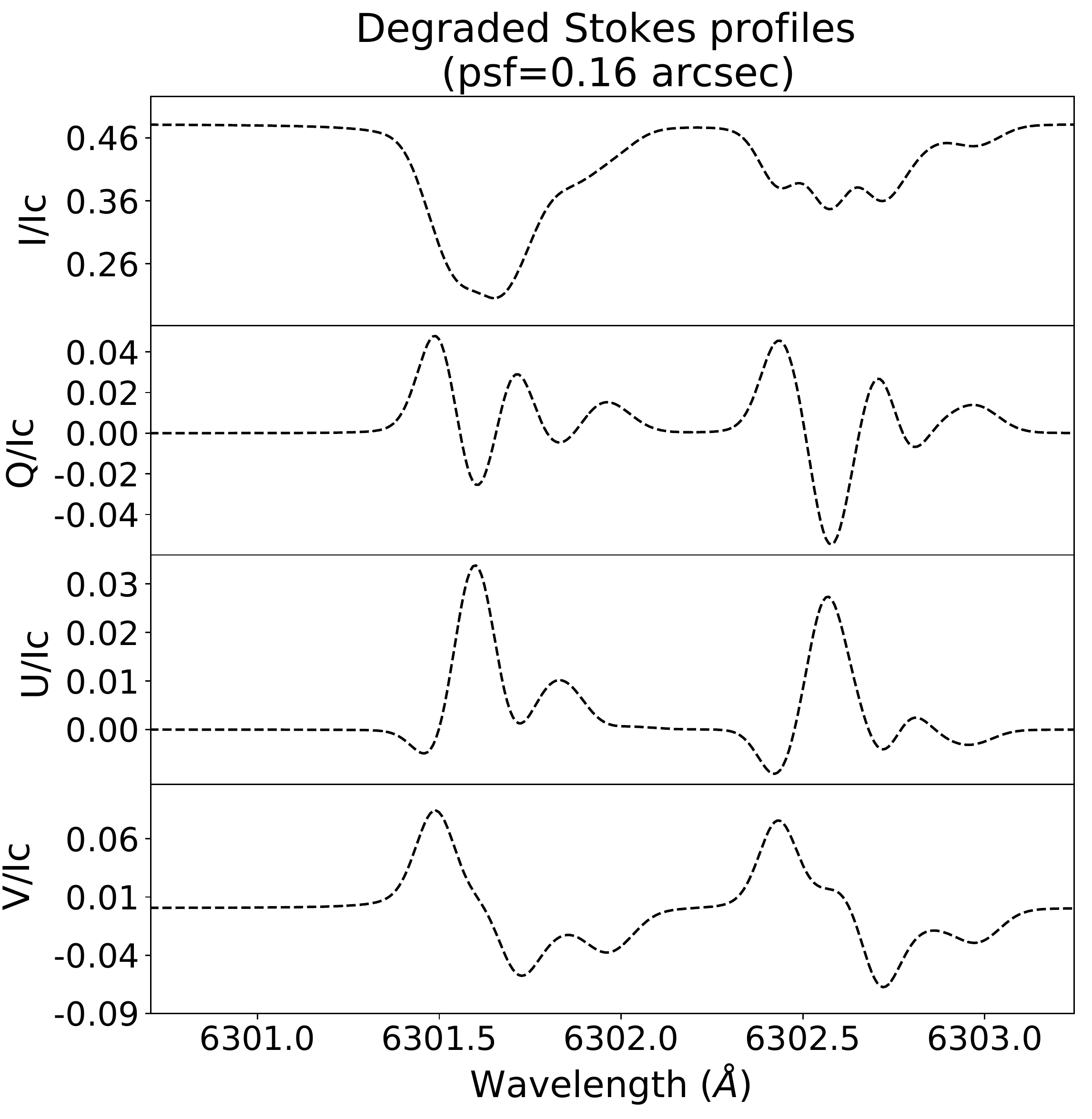}
\caption{}
\label{fig:mhd_stokesb}
 \end{subfigure}

    \caption[A set of emergent synthetic Stokes profiles in the MURaM  sunspot simulation at the location of a supersonic downflow in the tail of a CEF-carrying filament]{(a) A set of emergent synthetic Stokes profiles in the MURaM  sunspot simulation at the location of a supersonic downflow in the tail of a CEF-carrying filament. The spectra were synthesized by using the SPINOR code (STOPRO routines). (b) Degraded Stokes profiles obtained from the convolution of the synthetic spectra with a point-spread-function (PSF) of 0.16 arcseconds, similar to the case of the Hinode telescope.}
\label{fig:mhd_stokes1}
\end{figure*}

The height-independent 2-component inversions (e) and (f) give nearly identical results to each other  in both pixels, with lower field strengths of around $3.5$ kG in component 1, which is almost at rest ($v_{LOS} < 1$ km s$^{-1}$) compared to component 2 in which $B\sim4$ kG and $v_{LOS} \sim 16$ km s$^{-1}$. 
However, the resultant values of $B$ and $v_{LOS}$ given by the two height-independent inversions  (e) and (f)  generally resemble  the results from the height-dependent 2-component inversions (b) and (d) at $\log(\tau)=-0.8$.
This is not surprising since the sensitivity to $v_{LOS}$ and $B$ perturbations is higher at $\log(\tau)=-0.8$ than at $\log(\tau)=0$ for the Fe I 6302.5 $\AA$ line (see, e.g.,  response functions computed by \citet{Cabrera2005} and Fig. \ref{fig:RF}).
Even if the absorption line is not formed at a single depth, the height-independent inversions mainly provide information on the physical conditions prevailing at depths at which the line is more sensitive. 
Thus, if stronger magnetic fields are present in deeper  layers of the solar atmosphere (e.g. at $\log(\tau)=0$), as suggested by the results of inversions (a),  (b),  (c) and (d), they cannot be retrieved by inverting the Stokes profiles of the current wavelength range with a height-independent inversion technique only. 

The BIC values obtained for inversions (e) and (f)
are only slightly better than the ones obtained for SPINOR 2D in both pixels. Even if they both succeed in capturing many relevant aspects of all four Stokes profiles (despite the increased noise in the deconvolved spectra), it is very unlikely that the physical parameters in the pixels of interest display no gradients with height. Generally in sunspots, one would rather expect large gradients  with height of the physical parameters (particularly of the magnetic field) as supported by MHD simulations. In such cases, height-dependent inversions provide a more appropriate model atmosphere. The SPINOR 2D inversions are the most reliable model in this sense, since they take into account the height stratification of the physical parameters while keeping a good balance between the quality of the fit and the number of free parameters in the model, according to the obtained BIC values in the two studied pixels with peculiar spectra.

\section{Strong photospheric penumbral fields in a MURaM MHD simulation}
We now use the 3D high-resolution sunspot simulation by \citet{Rempel2015} \citep[see also][]{Siu2018} with a pixel resolution of 48 km in the horizontal direction and 24 km in the vertical direction to investigate the possible origin of super-strong penumbral magnetic fields associated with supersonic downflows in CEF-carrying filaments, and to compare  their synthetic spectra with the observed Stokes profiles reported in the previous sections.

As reported by \citet{Rempel2015} and \citet{Siu2018}, the sunspot simulation covers a time-span of 100 solar hours and after $t\sim 50$ hours, it displays radially aligned penumbral filaments with fast Evershed ouflows along them; but in some regions of the penumbra, the filaments carry instead a CEF, i.e. radial inflows directed toward the sunspot umbra and strong downflows ($v_z$<$-$8 km s$^{-1}$) at the end of such filaments, i.e. in their inner endpoints where the magnetic field is noticeably enhanced, up to values of around 5 kG and $\gamma\sim40^{\circ}$ near the local $\tau=1$ level (see for example Fig. \ref{fig:mhd_par}, cf. Fig. 2 in \citet{Siu2018}).

In a rather simple and quick attempt to quantify the effect of the 3D atmospheric structure and magnetic field on the profiles of the analyzed spectral lines, we employed the forward part of the SPINOR code (STOPRO) to solve the radiative transfer equation in the MURaM cube analyzed by \citet{Siu2018}, which was obtained with non-gray radiative transfer.
Figure \ref{fig:mhd_stokesa} shows the emergent synthetic spectra from a vertical column   located close to the inner edge of the simulated penumbra and which, at the $\log(\tau)=0$ level, intersects with the tail of a CEF-carrying filament and contains  supersonic downflows at that height. 
Similarly to the observed Stokes profiles for the pair of Fe I lines from the LFP in AR 10930, the synthetic Stokes  profiles  from the MHD simulations are highly asymmetric, display large redshifts, and show multi-lobed Stokes $V$ profiles.

\begin{figure*}[htp!]
 \centering

\begin{subfigure}[b]{0.45\textwidth}
\includegraphics[width=\textwidth]{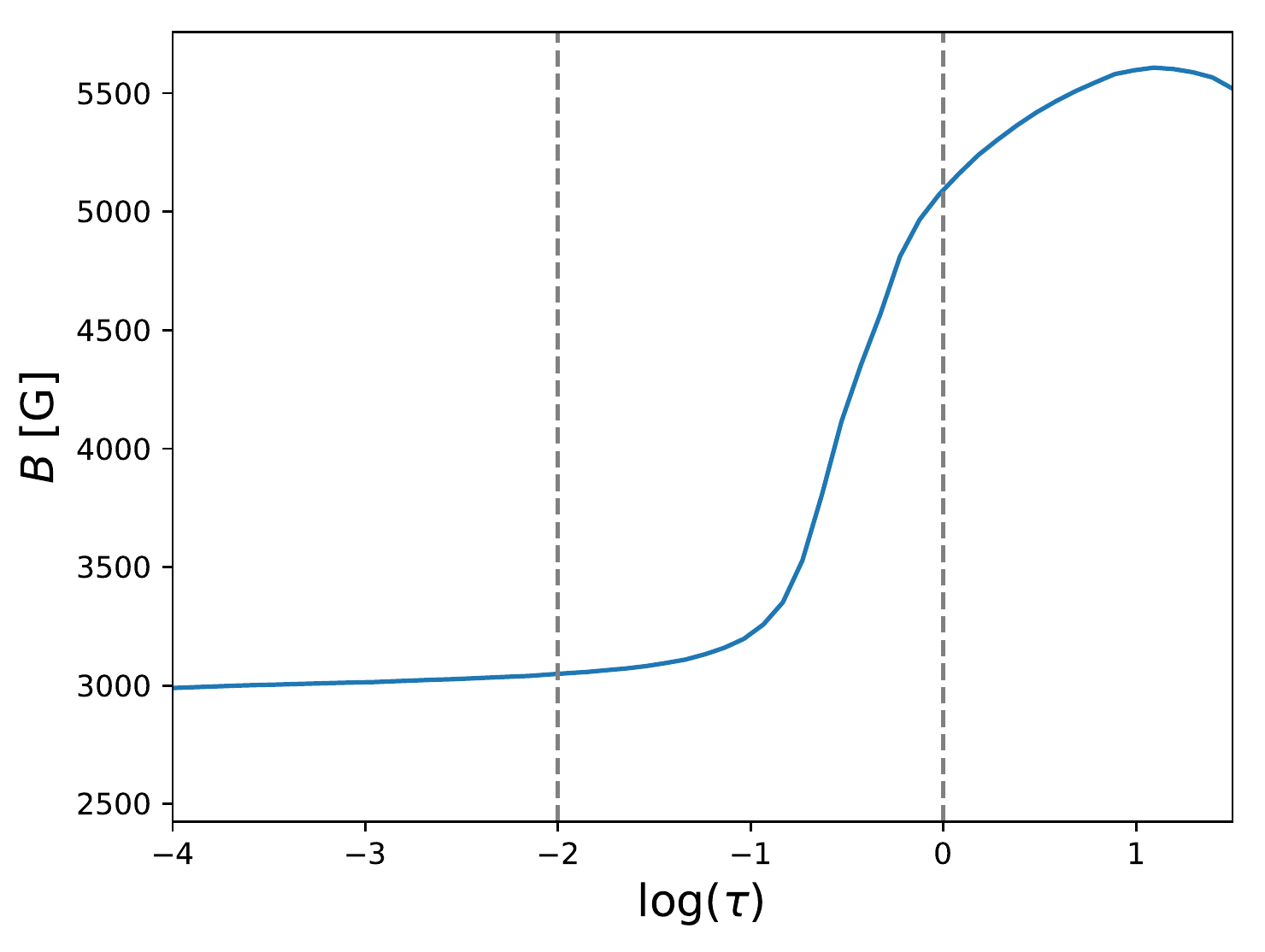}
 \caption{}
\label{fig:mhd_stokes2a}
 \end{subfigure}
~
\begin{subfigure}[b]{0.45\textwidth}
\includegraphics[width=\textwidth]{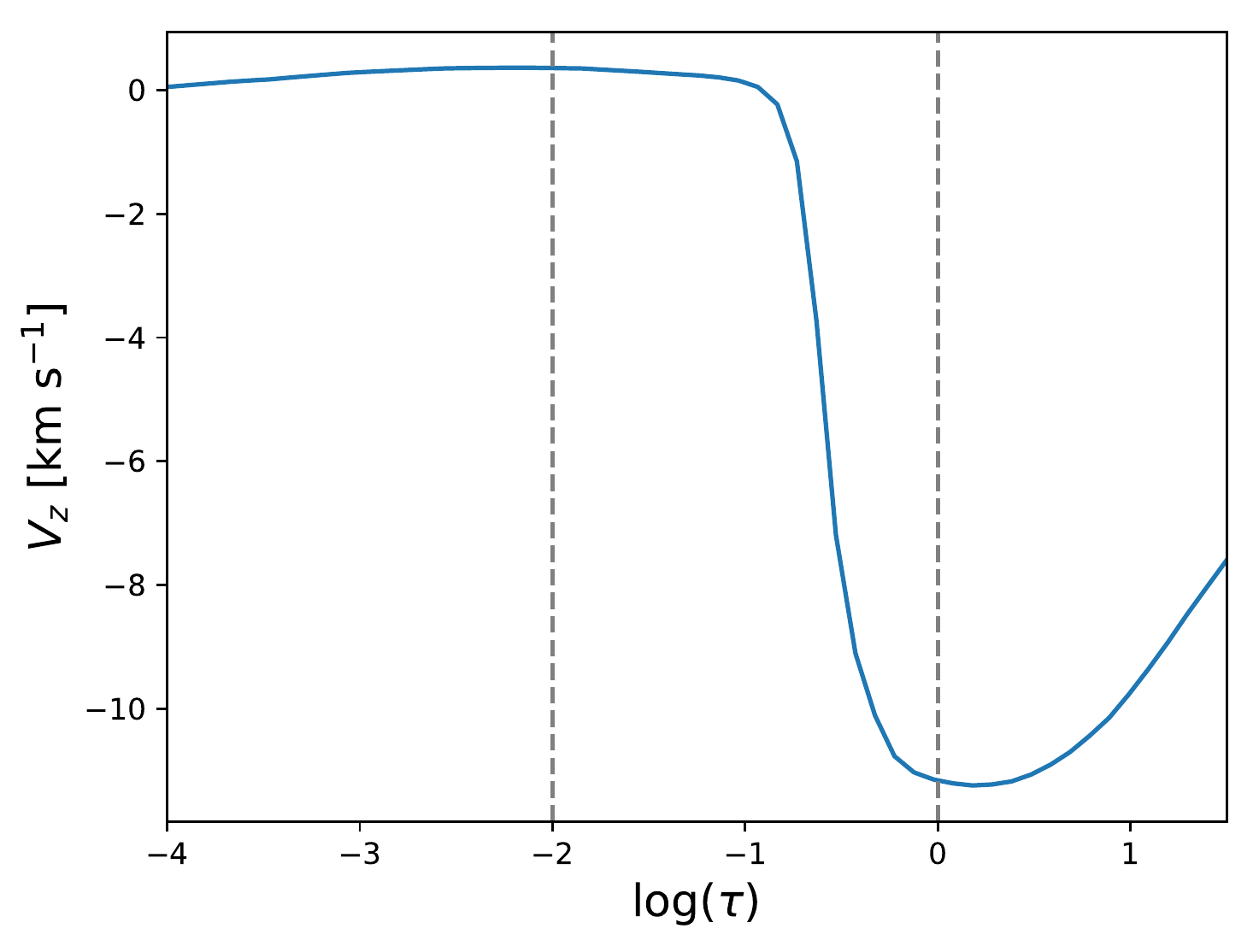}
\caption{}
\label{fig:mhd_stokes2b}
 \end{subfigure}

\centering
\begin{subfigure}[b]{0.45\textwidth}
\includegraphics[width=\textwidth]{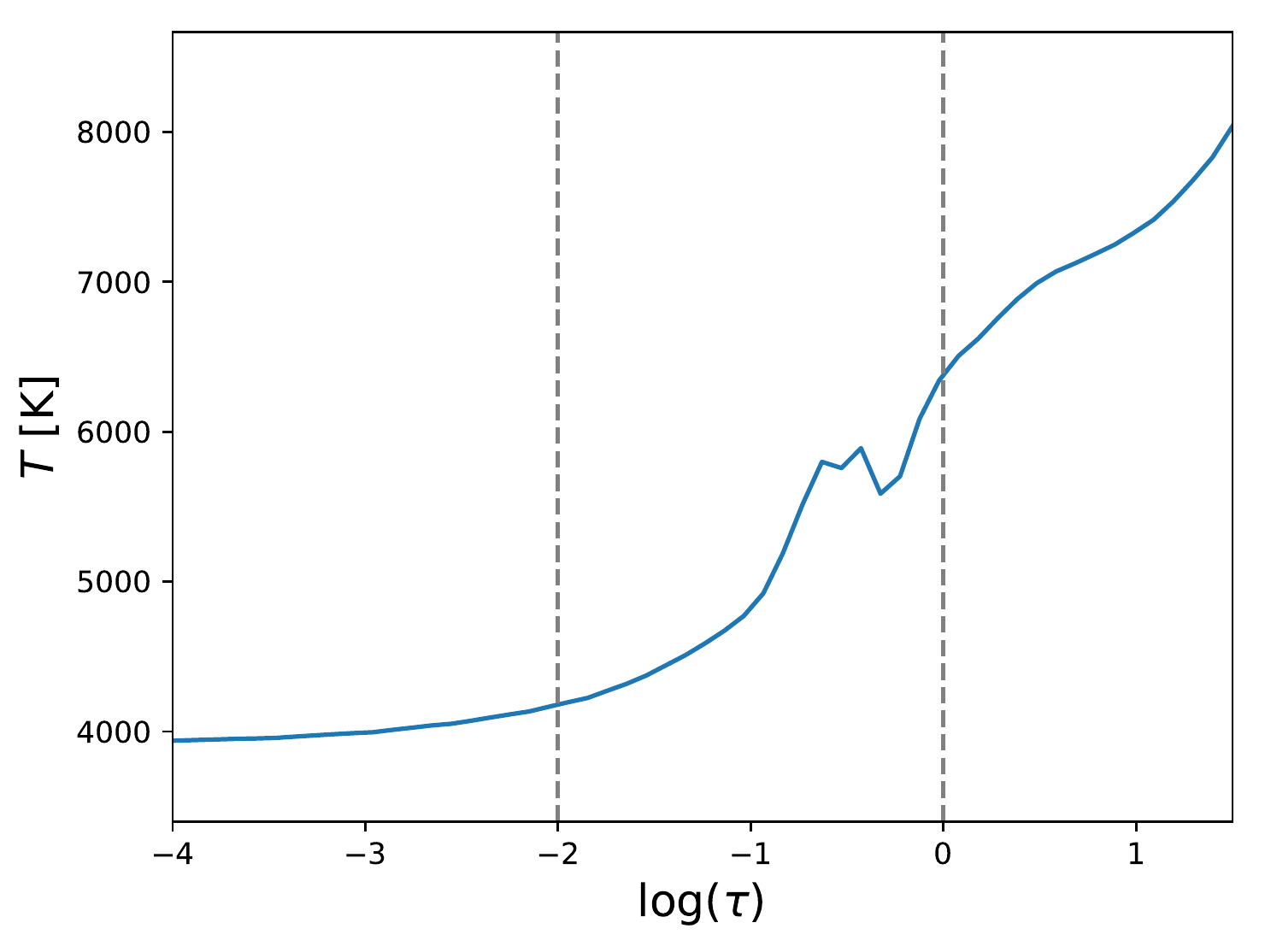}
\caption{}
\label{fig:mhd_stokes2c}
 \end{subfigure}
~
\begin{subfigure}[b]{0.45\textwidth}
\includegraphics[width=\textwidth]{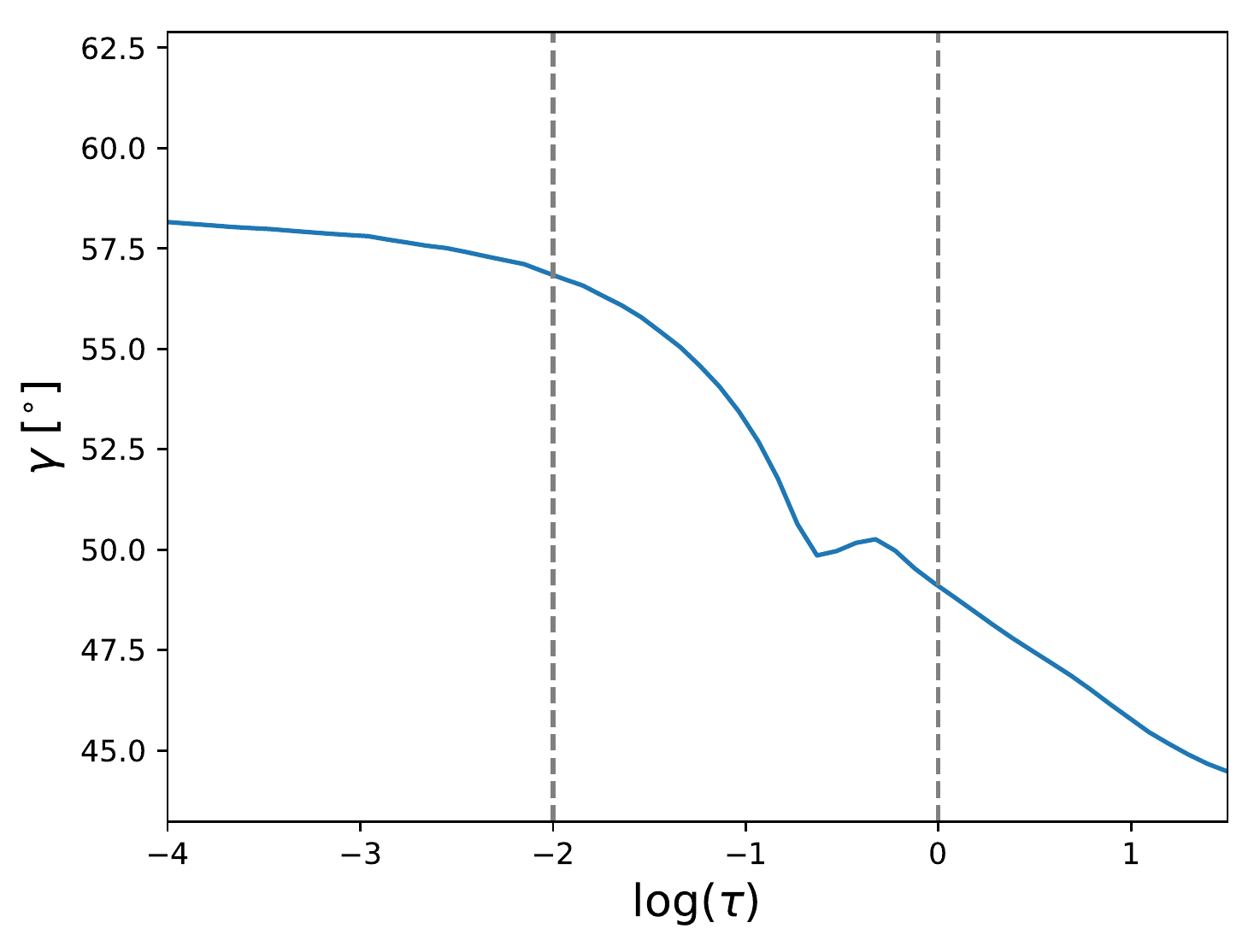}
\caption{}
\label{fig:mhd_stokes2d}
 \end{subfigure}

\centering
\begin{subfigure}[b]{0.45\textwidth}
\includegraphics[width=\textwidth]{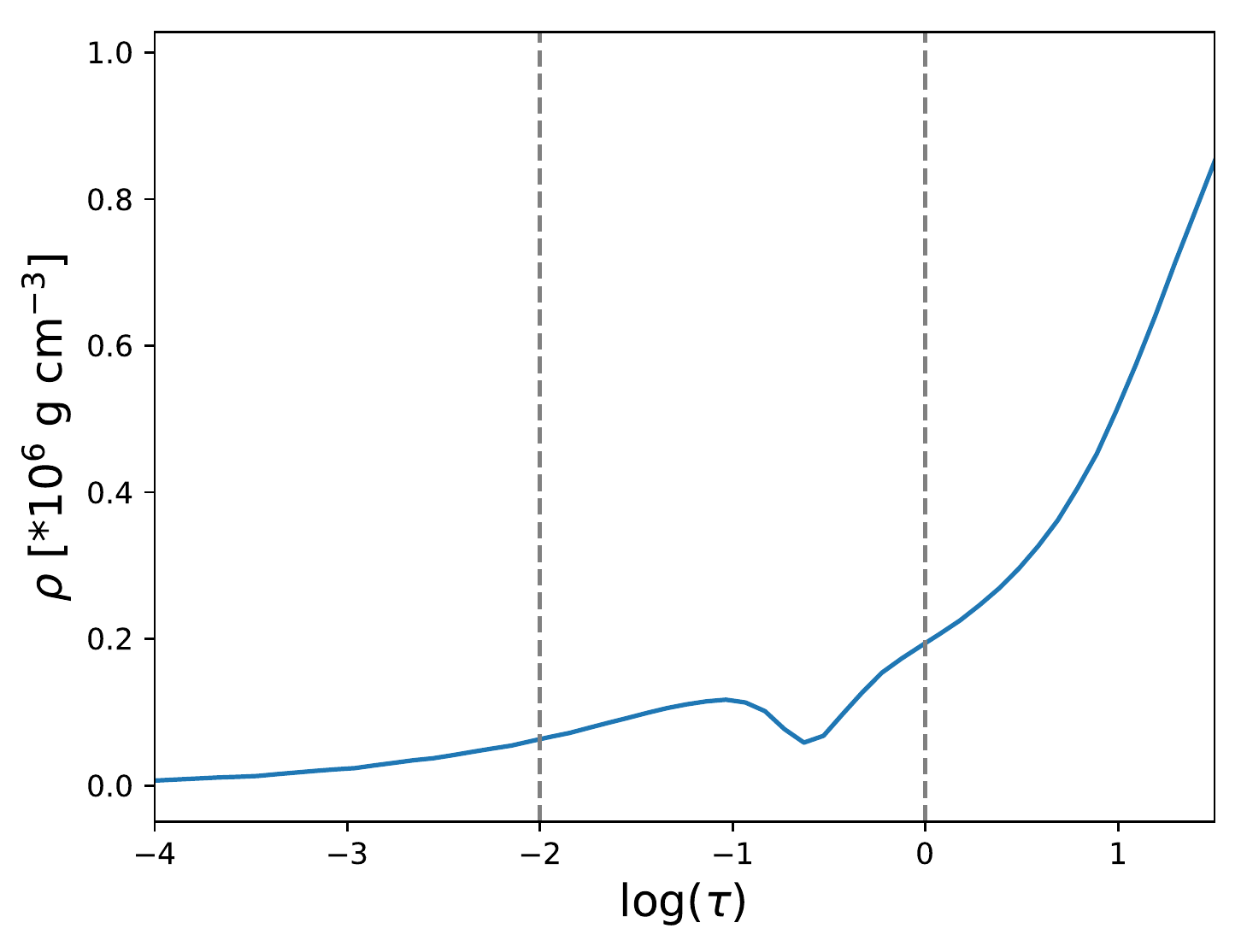}
\caption{}
\label{fig:mhd_stokes2e}
 \end{subfigure}
~
\begin{subfigure}[b]{0.45\textwidth}
\includegraphics[width=\textwidth]{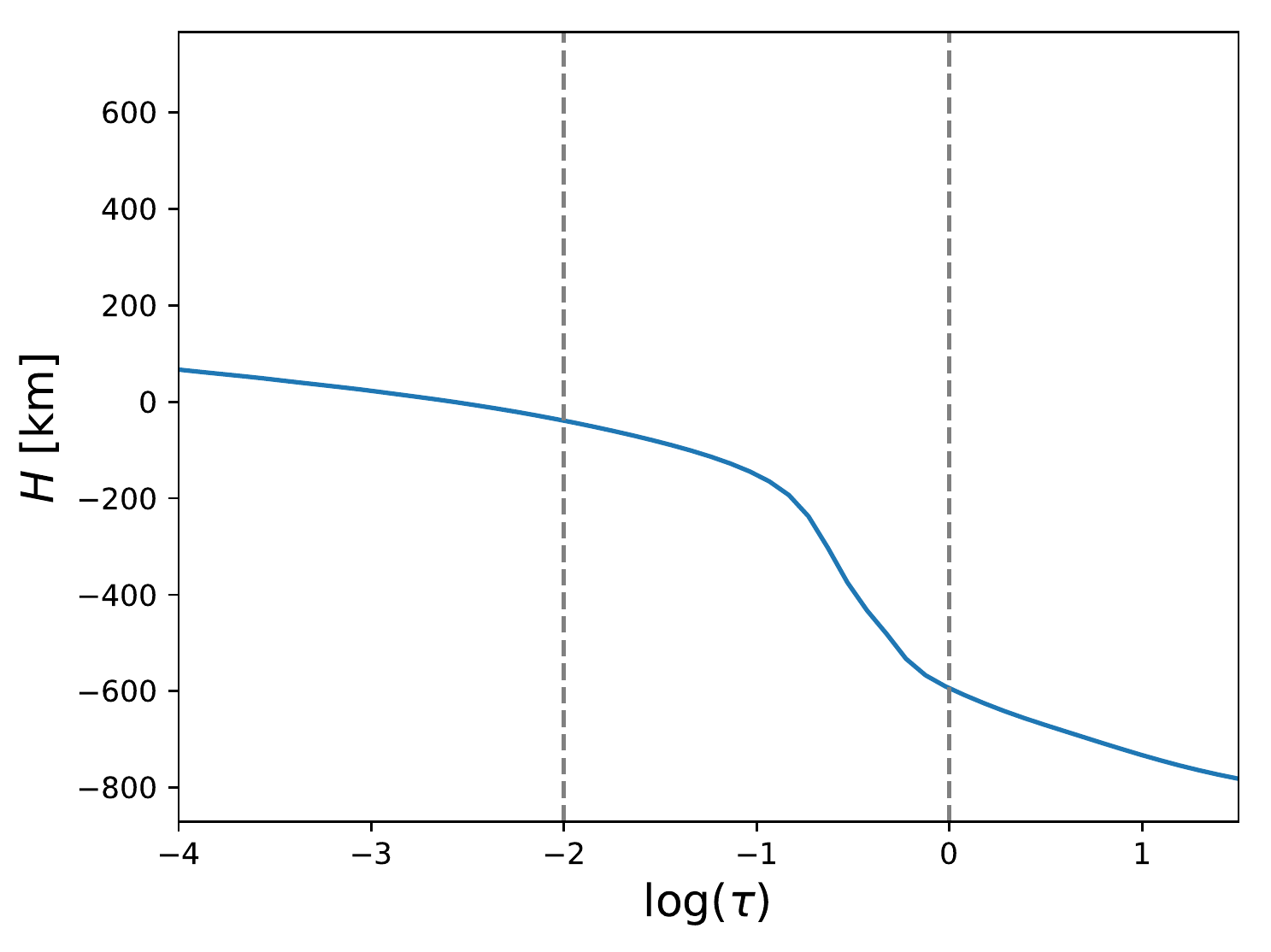}
      \caption{}
\label{fig:mhd_stokes2f}
 \end{subfigure}

    \caption[Vertical profiles of the MHD physical parameters leading to the
 emergent synthetic Stokes profiles shown in Fig. \ref{fig:mhd_stokes1}, at the location of a supersonic downflow in the tail of a CEF-carrying filament] {Vertical profiles, in a $\log(\tau)$-scale, of the MHD physical parameters leading to the
 emergent synthetic Stokes profiles shown in Fig. \ref{fig:mhd_stokes1}, at the location of a supersonic downflow in the tail of a CEF-carrying filament: (a) magnetic field strength, (b)  vertical flow velocity (negative values indicate downflows), (c) temperature, (d)  field inclination, (e)  gas density, and (f) geometrical height at the location of the grid-cell containing the supersonic downflows from the MURaM simulation. Vertical dashed lines  delimit the approximate $\tau-$range where the lines show a significant response, i.e., between $\log(\tau)=-2$ and $0$.}
\label{fig:mhd_stokes2}
\end{figure*}

In Figure  \ref{fig:mhd_stokesb}, we display degraded Stokes profiles which were obtained by convolving the synthetic spectra in Fig. \ref{fig:mhd_stokesa} with an effective PSF=0.16$^{\prime\prime}$ \citep{Danilovic2008} constructed from the pupil
function of the 50-cm Hinode SOT \citep[e. g.,][]{Suematsu2008}.
The degraded profiles also show the main characteristics described above and resemble the observed Stokes profiles presented in the previous sections, which are even more extreme.

\begin{figure*}[htp!]
    \centering

\includegraphics[width=0.45\textwidth]{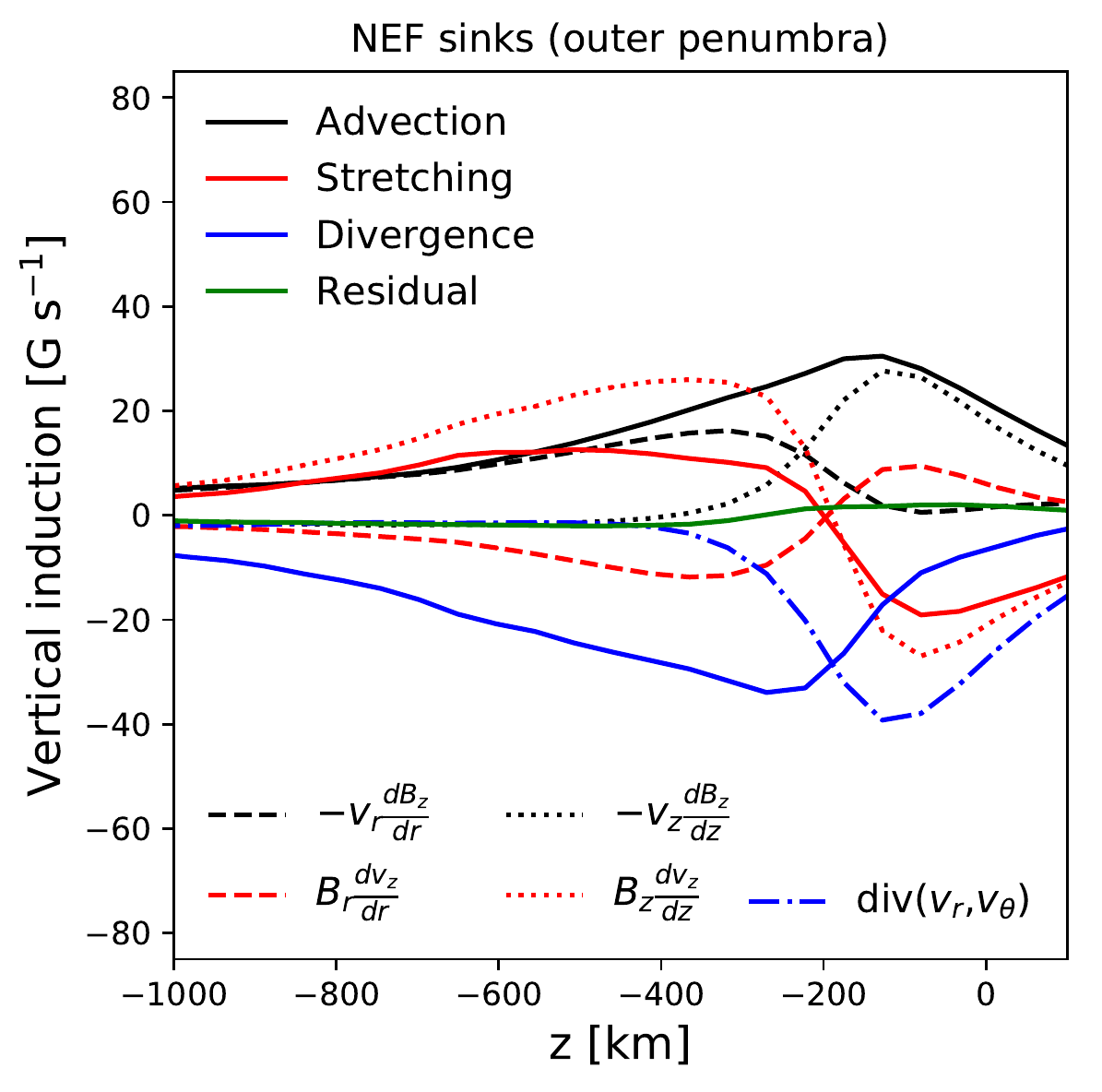} 
\includegraphics[width=0.45\textwidth]{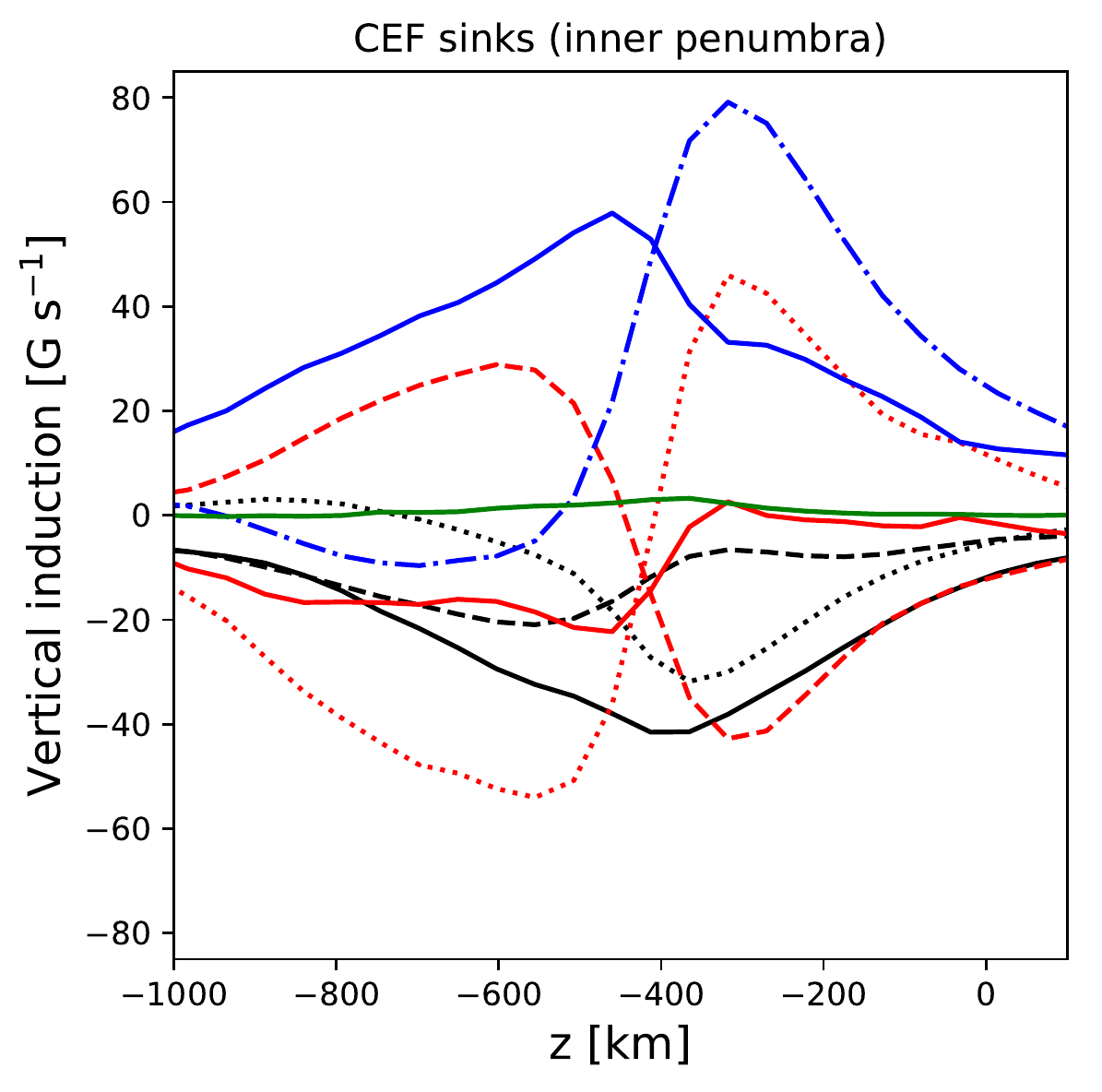} 
    \caption[]{Average contributions to the induction equation from advection (solid black), field-line stretching (solid red), flow divergence (solid blue) and a residual term (solid green) to the vertical field component at the sinks of the NEF (left) and at the sinks of the CEF (right).
The averages have been focused on localized regions that have fairly strong fields ($B_z<-2$ kG for the sinks of the NEF and $B_z>2$ kG for the sinks of the CEF) and supersonic downflows at the local $\log(\tau)=0$ level.
 The residual term represents the numerical magnetic diffusivity as an approximated magnitude that is calculated by the negative sum of the advective, stretching and divergent terms; but it also 
contains potential contributions from time variation. The dashed and dotted lines represent approximated terms that  have a major contribution to the advective term (black) and to the stretching term (red),  whose general expressions are indicated in the lower labels. The dash-dotted blue line represents the contribution to the divergence term  due to flows perpendicular to the vertical field component, i.e. radial and azimuthal flow components.
}\label{fig:12*}
\end{figure*}

Such complex shapes in the MHD case are mainly the result of the large vertical gradients in all the atmospheric quantities and the magnetic field structure. In particular, the  stratification of the field strength (Figure \ref{fig:mhd_stokes2a}), the  flow velocity (Figure \ref{fig:mhd_stokes2b}), and field inclination (Figure \ref{fig:mhd_stokes2c}), qualitatively resemble the SPINOR 2D results in the LFPs (cf. Table \ref{tab:1}), i.e., while the field strength and downflow speeds increase with depth, the field inclination increases with height. 
Another important aspect that might contribute to the complexity of the emergent spectra is the presence of a shock which is seen as the sudden transition of the downflow speed from subsonic to supersonic near $\log(\tau)=-1$, and as the sudden variation of all the physical parameters shown in Figure \ref{fig:mhd_stokes2}.

The presence of strong magnetic fields at the local $\log(\tau)=0$ level (in excess of 5 kG)  in the simulations are mainly due to the influence of the neighboring umbral field and the highly depressed surfaces of constant optical depth as reported by \citet{Siu2018}. Such surface depression is of the order of $400-600$ km in the analyzed case (see Figure \ref{fig:mhd_stokes2f}), and is related to a strong decrease of the gas density beneath $\log(\tau)=-1$ (see Figure \ref{fig:mhd_stokes2e}).

Thus, given the similarity of the physical scenarios that the observations and the simulations present, the determination of the physical processes involved in the maintenance and amplification of the field in the simulations can give us insight into the possible origin of super-strong photospheric magnetic fields observed in sunspot penumbrae.

\subsection{Induction equation}
In order to study how the vertical field is maintained in the penumbra, we evaluate the different terms of the induction equation in the simulation box:

\begin{equation}\label{eq:9}
\frac{\partial \vec{B}}{\partial t}=\underbrace{-(\vec{v} \cdot \nabla)\vec{B}}_ \text{Advection}+ \underbrace{(\vec{B} \cdot \nabla)\vec{v} }_ \text{Stretching}\underbrace{- \vec{B}(\nabla \cdot \vec{v})}_\text{Divergence}.
\end{equation}

We analyze the different terms in eq. \ref{eq:9} during the time-period from
60 to 70 solar hours (the range of time during which the counter-EF are found in the
simulations) of the total of 100 hours simulations run (see \citet{Siu2018} for details), using the transformation to cylindrical coordinates to separate the direction along and perpendicular to the penumbral filaments, i.e. $r$, $\theta$, and $z$ coordinates; and by separating outflows ($v_r$>0) from inflows ($v_r$<0), and upflows ($v_z>0$) from downflows ($v_z<0$) in the simulated penumbra.

\begin{figure*}[htp!]
    \centering

\includegraphics[width=0.45\textwidth]{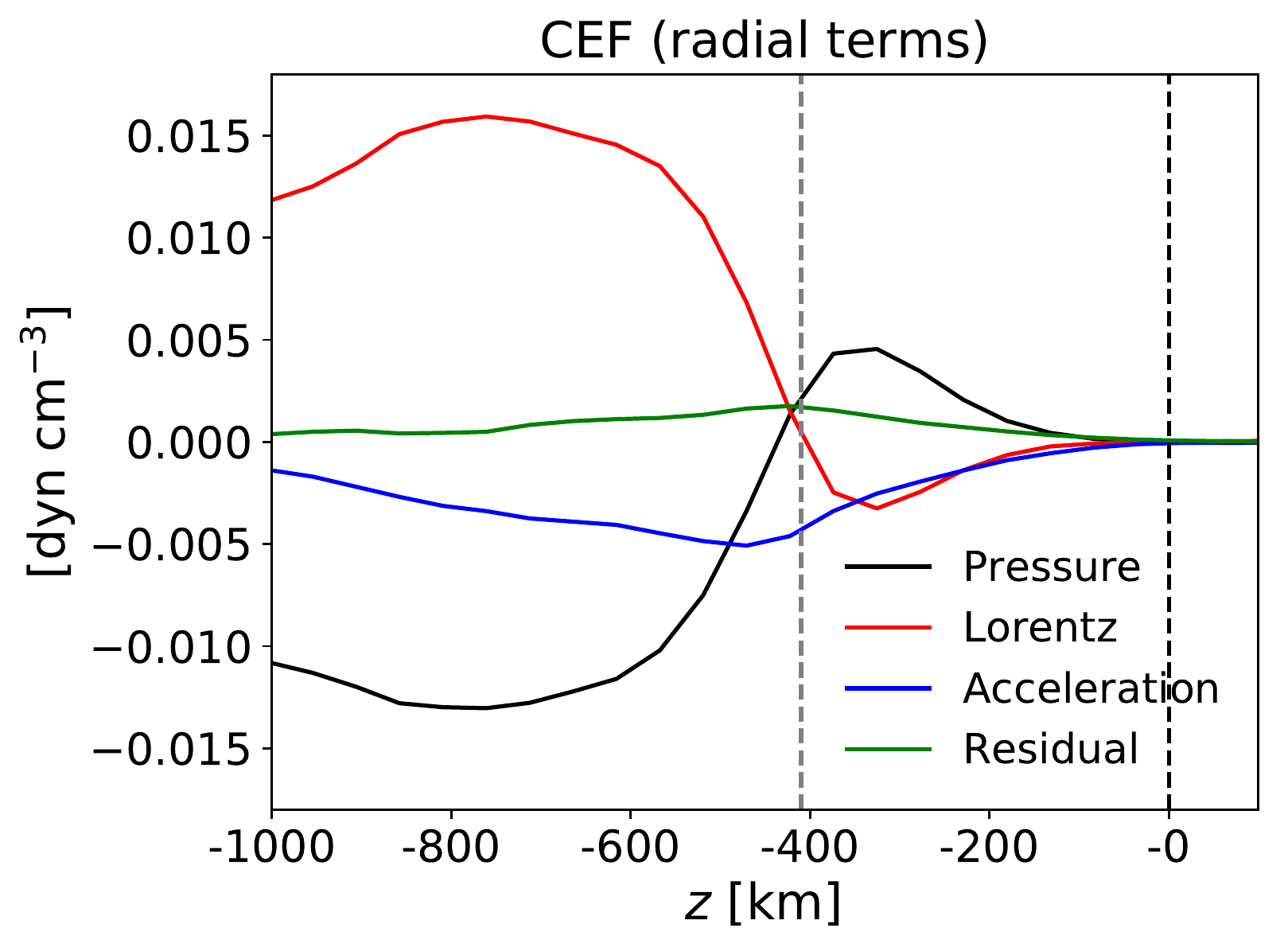} 
\includegraphics[width=0.45\textwidth]{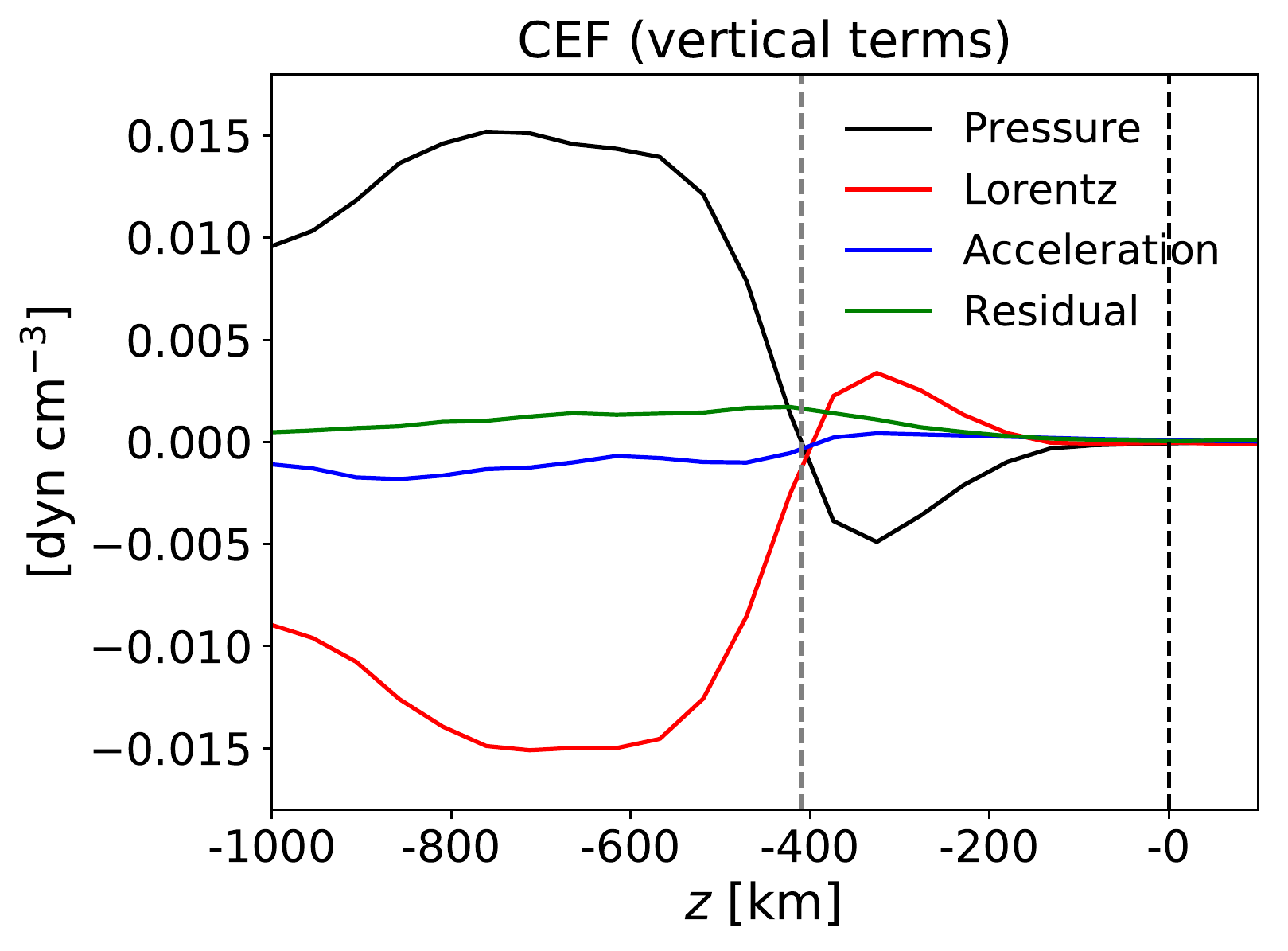}
    \caption[]{Force balance in the radial (left) and vertical (right) directions. The force terms have been horizontally and temporally averaged over the places with strongest fields and supersonic downflows of the CEF in the inner penumbra. Black: pressure forces, red: Lorentz forces, blue: acceleration forces,
green: residual forces. The vertical gray dashed lines are placed at the average height of the $\log(\tau)=0$ level in the selected regions  and are located nearly 400 km beneath the average height of the $\log(\tau)=0$ surface in the quiet sun (i.e. $z=0$ km, black dashed line).
}\label{fig:13}
\end{figure*}

As reported by \citet{Siu2018}, the vertical field component is noticeably enhanced at the tails of the penumbral filaments, i.e. at the filament end-points hosting sinks in both, those filaments carrying a normal-EF (NEF, i.e. outflows) and those carrying a counter-EF (CEF, i.e. inflows), which are mainly located close to the outer and inner penumbral boundary, respectively (see e.g. Fig. 3 in \citet{Siu2018}). Therefore, we explore the mechanisms
that can lead to the amplification of the magnetic field strength at those places. We use different masks in order to separate the sinks (downflows) that occur at the tails of the NEF-carrying filaments (regions where $v_z<0$ and $v_r>0$ in the outer penumbra) from those sinks that happen at the tails of the CEF-carrying filaments (regions where $v_z<0$ and $v_r<0$ in the inner penumbra).

Figure \ref{fig:12*} displays the contributions of each term in equation \ref{eq:9} to the vertical field component, horizontally averaged
over the sinks of the NEF (left plot) and over the sinks of the CEF (right plot), and focusing the averages on localized regions that have fairly strong fields ($B_z<-2$ kG for the sinks of the NEF and $B_z>2$ kG for the sinks of the CEF) as well as supersonic downflows at $\log(\tau)=0$.
In both cases, NEF and CEF sinks, the contributions
from stretching, advection and divergence are mostly in balance, implying that the residual terms, which have potential contributions from the numerical magnetic diffusivity and  from time variations (green lines), do not play a significant role in shaping the magnetic structure of the penumbra in these simulations during the analyzed time period. 

The roles of advection and divergence in the vertical induction (black and blue solid lines respectively) are opposed to each other in both penumbral regions. Besides, they appear with a sign swap in the outer penumbra (left panel) compared to the inner penumbra (right panel). However, the roles of advection and divergence for maintaining the vertical field component are the same in both regions given that  $B_z<0$ at 
the sinks of the NEF and $B_z>0$ at the sinks of the CEF, which causes the sign swap in the vertical induction.

Thus, at the sinks of the NEF (outer penumbra, left plot), there is an opposed contribution from the
 advective term to the maintenance of the (negative) vertical field component at all heights of the analyzed $z-$range (where $z=0$ corresponds to the average height of the $\log(\tau)=0$ surface in quiet sun). 
In contrast, the stretching term behaves as a source for the (negative) vertical field component in the outer penumbra, above $z\sim-200$ km. 
The major contribution to this term comes from vertical stretching (red dotted line), i.e. $B_z\frac{\partial v_z}{\partial z}<0$ due to a strong downward transition of the downflow speeds (from subsonic to supersonic)  that leads to the steepening of $\frac{\partial v_z}{\partial z}$ near $z\sim-200$ km (generally, the supersonic NEF sinks are shallower than in the CEF case, for an example see the white contours in the $v_z$ panel of Fig. 3 in \citet{Siu2018} which enclose regions where $v_z<-8$ km s$^{-1}$). 
Deeper down, below $z\sim-200$ km, there is a radial shear profile  (red dashed line) due to a radial outward increase of the downflow speeds, $B_r\frac{\partial v_z}{\partial r}<0$, that also contributes to the maintenance of the (negative) vertical field component in the outer penumbra.
However, the major source comes from the divergence term (blue line) due to the converging aspect of the downflows, i.e. $\nabla \cdot \vec{v}<0$. The dominant contribution to this term in the near-surface layers is given by flows that are perpendicular to the vertical field component, $v_r$ and $v_{\theta}$, i.e. by a horizontal convergence of the downflowing material (dash-dotted blue line). The remainder is due to the term $-B_z\frac{\partial v_z}{\partial z}$, which becomes negative below $z\sim-200$ km.

Similarly, advection plays an opposite role for the maintenance of the (positive) vertical field component at the sinks of the CEF in the inner penumbra (right plot in Fig. \ref{fig:12*}). However, the role of stretching is not significant above $z\sim-400$ km in this case (height at which most of the CEF sinks become supersonic,  see an example in Figs. \ref{fig:mhd_stokes2b} and \ref{fig:mhd_stokes2f}). Notwithstanding, due to the strong downward acceleration of the gas at the sinks of the CEF, vertical stretching acts as a source for the (positive) vertical field above $z\sim-400$ km, i.e. $B_z\frac{\partial v_z}{\partial z}>0$. Likewise, a radial inward increase of the downflow speed at the sinks of the CEF leads to a radial shear term that contributes positively below $z\sim-400$ km, i.e. $B_r\frac{\partial v_z}{\partial r}>0$.
 The dominant positive contribution to the (positive) vertical field component is given by the divergence term (i.e. $\nabla \cdot \vec{v}<0$, since $B_z$ is positive in the inner penumbra), which means that, similarly to the NEF sinks, the CEF sinks are also convergent downflows, which implies compression and amplification of the magnetic field.

\subsection{Force balance}

In order to investigate how the strongest fields in the penumbra are balanced, we follow the analysis performed by \citet{Rempel2011} and by \citet{Siu2018} and use the following force balance equation which is derived from the momentum equation by assuming stationarity:

\begin{equation}
\underbrace{\rho \vec{g}-\nabla{p}}_\text{Pressure}+\underbrace{\vec{j}\times\vec{B}}_\text{Lorentz}\underbrace{-\rho (\vec{v}\cdot \nabla)\vec{v}}_\text{Acceleration}+\underbrace{\vec{F_{visc}}}_\text{Viscosity}=0
\label{eq:1}
\end{equation}

\begin{figure}[htp!]
    \centering
\includegraphics[width=0.45\textwidth]{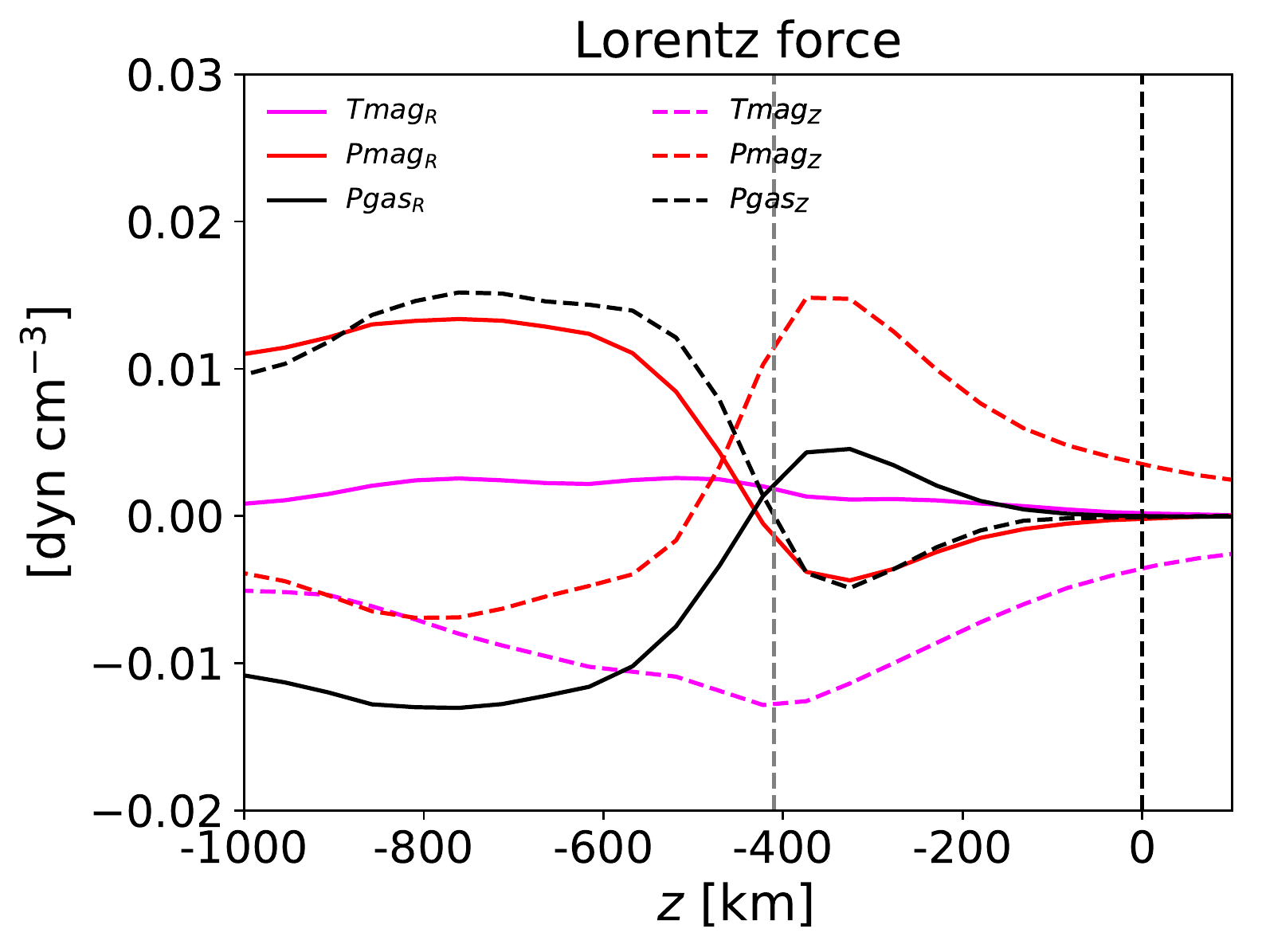} 
    \caption[]{Radial (solid) and vertical (dashed) Lorentz force terms separated for the magnetic pressure (red) and the magnetic tension (magenta) forces at the places with strongest fields and supersonic downflows of the CEF in the inner penumbra. The radial (black solid) and vertical (black dashed) gas pressure forces are overplotted for comparison.  Same format as in Fig. \ref{fig:13}.
}\label{fig:13*}
\end{figure}

Each term in the above equation is then separated into a radial and a vertical component; and a residual force term is introduced as the negative sum of the pressure, the acceleration, and the Lorentz force contributions in the corresponding direction given that the viscosity force terms are not explicitly calculated.

Figure \ref{fig:13} shows the average radial and vertical force balance at the strong field regions of the inner penumbra with supersonic downflows of the CEFs. In both directions the forces are mostly in balance and the residual forces are nearly zero. These plots show that acceleration forces (blue lines) are very significant in the radial direction but almost negligible in the vertical one, which means that the system is close to magneto-hydrostatic in the vertical direction given that the Lorentz force (red line) nearly completely balances with the pressure forces (black line).

The average height of the $\log(\tau)=0$ level over the selected regions lies nearly 400 km below its average height in the quiet sun. Such average height is considerably deeper than in the whole penumbra (approximately at $z=-200$ km) and even deeper than  in the inner penumbra ($z\sim-250$ km).

Figure \ref{fig:13*} displays the Lorentz force separately for the magnetic pressure term ($-\nabla[B^2/8\pi]$, red lines) and the magnetic tension term ($[\vec{B}\cdot\nabla]\vec{B}/4\pi$, magenta lines) in the radial (solid) and vertical (dashed) directions, which have been averaged over the strong field regions of the inner penumbra with supersonic downflows (same mask as in Fig. \ref{fig:13}).  
Looking at the individual components of the Lorentz force, we see in the vertical direction that the strongest fields in the simulation are less force-free in the deeper domain, where the gas pressure (black dashed line)  is strong enough to balance, but they become mostly force-free near the observable photosphere and in the layers above. In contrast, in the radial direction, the magnetic pressure force dominates over the magnetic tension force at most heights. However, the gas pressure force term (black solid line) is also large enough in the radial direction to considerably contribute to keep the balance. In addition, we have seen that the contribution of the radial acceleration forces (blue line in Fig. \ref{fig:13}) plays an important role for maintaining the force balance in the radial direction.

\section{Discussion and conclusion}

Inversion techniques are currently the most powerful tools to infer the physical properties of the solar atmosphere from polarization line profiles, being able to provide reliable and robust results from many types of Stokes profiles according to  numerical tests \citep[e.g.,][]{Ruiz2007}. There are several different inversion techniques in the literature, each of them with its own advantages and shortcomings, which largely depend on the addressed problem. Certainly, after any Stokes inversion the results need always to be validated and one needs to be aware that the resulting model atmosphere is not necessarily the real one since the solution might not be unique or the model underlying the inversion not appropriate to the actual solar situation.
 Nonetheless, as stated by \citet{Sabatier2000}, by means of Stokes inversions it is generally possible to retrieve as much information as possible for a model which is proposed to represent the system in the real world.
 
The existence of $B>7$ kG in the inner  penumbra of a sunspot would require an unusually deep Wilson depression 
to be consistent with idealized magnetohydrostatic models of sunspots \citep{Livingston2006}.
However,  besides the non negligible dynamical effects of the studied penumbra, 
the magnetic field does not have to be in maximally non-force-free state as usually assumed for the photosphere. We have found in the MHD simulations that the strongest fields  in the penumbra ($\sim5$ kG)  are vertically close to force-free in the observable photosphere and the gas pressure  is sufficient to reach a force balance in the deeper layers. Although these fields are less force-free in the radial direction, the radial gas pressure force provides a sufficient balance to keep the system in near equilibrium, but is only with the contribution of radial acceleration forces that an almost complete  balance can be reached. In this sense, the existence of 7 kG magnetic fields in the photosphere would be possible in a highly dynamical environment, such as that inferred by the SPINOR 2D inversions, i.e. with supersonic counter-Evershed flows sinking supersonically in the inner penumbra.  
Such superstrong field concentrations would likely fan out significantly with height and remain close to potential in the observable photosphere. 

Field strengths larger than $7$ kG in penumbral environments have previously been reported by \citet{vannoort2013}. They obtain such large field strengths in supersonic downflow regions in the peripheral penumbra of a sunspot, also by using SPINOR 2D inversions. Nonetheless, they observe $B>7$ kG only in the deepest layer ($\log(\tau)=0$) and obtain a good agreement between the inversion results and a MURaM simulation of a sunspot by \citet{Rempel2012b}. They propose a scenario in which the high magnetic field values are the result of a field intensification in the deep photosphere due to the interaction of the supersonic downflows with an external magnetic barrier (e.g., with a plage region). 
Such a scenario could also be valid for the present observations, with the umbral field playing the role of the magnetic barrier.

According to the SPINOR 2D results (e.g. Fig. \ref{fig:LFP2} and Table \ref{tab:1}), most of the magnetic fields whose strength is above $7$ kG (LFPs) are nearly vertical and have the same polarity as the umbra, which is negative. In addition, they are associated to supersonic downflows, even exceeding $v_{LOS}=20$ km s$^{-1}$ at $\log(\tau)=0$ in some LFPs. 
Moreover, the regions in the sunspot where $B>5$ kG are also associated with supersonic LOS flow velocities (see, e.g., Fig. 5a in \citet{siu2017}). 
As displayed in Figure \ref{fig:LFPa}, the LFPs with  $B>7$ kG (yellow  pixels) are  surrounded by those weaker fields, which are still in excess of 5 kG (red pixels), in a  supersonic flow environment (see also Fig.  \ref{fig:LFP}).
A very low density of the supersonic downflowing material could also 
explain the  observation of unusually strong magnetic fields in the  penumbra, since it would cause the optical depth layers to be strongly depressed. In addition, their close vicinity to the umbral field also plays a role.

This is in agreement with MHD simulations of counter-Evershed flows \citep[][and Section 6]{Siu2018}, which show a $\sim 400$ km depressed  $\tau=1$ surface in average  (and up to 600 km, with respect to its average height in the quiet sun) at the tails of the CEF-carrying filaments, where the supersonic downflowing material becomes a very low density gas (see also Fig. \ref{fig:mhd_stokes2e}). As a consequence, deep-lying field strengths of the  order of 5 kG in the inner penumbra (near the umbra) are visible at the $\tau=1$ level in those simulations.

The resultant synthetic Stokes profiles associated to photospheric fields of the order of 5 kG and downflow speeds of $10-12$ km s$^{-1}$  in the simulated penumbra display large asymmetries, redshifts and multiple lobes in their Stokes $V$ profiles, in agreement with the observed spectra in the LFPs. This result supports the possibility that the observed Stokes profiles associated with the LFPs in AR10930, which display even more extreme characteristics, are produced by larger fields and stronger downflows as inferred by the SPINOR 2D inversions (Fig. \ref{fig:LFP2}).

The strong penumbral magnetic fields in the simulations (nearly 5 kG at the local $\tau=1$ level) are mainly due to the influence of the neighboring umbral field, the highly depressed surfaces of constant optical depth, and the formation of shocks by the transition of the downflow speeds from subsonic to supersonic; but there is also a local  intensification of the field that can be associated to
the converging aspect of the supersonic downflows, which lead to a compression and intensification of the vertical magnetic field component in the inner penumbra. 

A similar mechanism amplifies the negative vertical field component at the sinks of the NEF in the outer penumbra in the simulations, where the field bends over and the downflows are convergent. There, the negative vertical field component can additionally be  intensified by vertical stretching due to the strong downward acceleration of the gas
up to supersonic speeds.
These results are in agreement with the proposed scenario by \citet{vannoort2013} to explain the possible observation of superstrong fields associated with peripheral downflows in the penumbra of the leading spot of NOAA AR 10933.

According to the Bayesian Information Criterion \citep[BIC, ][]{Schwarz1978}, which assesses  a best fit based on the balance between the quality of the fit and the number of free parameters considered by  the model,  the preferred height-dependent model atmospheres for the LFPs are those provided by the SPINOR 2D inversions.
Furthermore, given the extreme characteristics of the observed Stokes profiles associated with the sinks of the CEF in the LFPs and their resemblance to the synthetic spectra derived from the MHD simulations, we finally cannot easily discard the possibility that we are dealing with actual observations of $B\sim7$ kG and even larger in regions of supersonic downflowing material (up to $v_{LOS}\sim 35$ km s$^{-1}$) with very low densities, 
where similar mechanisms to those occurring in the analyzed simulations might explain their origin.

The major strength of SPINOR 2D lies on its simultaneous coupled inversion of all the pixels to self-consistently take into account the  influence  of  straylight  from  neighboring  pixels. This approach is able to reproduce complex multi-lobed profiles with a simple, one-component atmosphere per pixel, keeping an acceptable number of free parameters, which  significantly enhance the reliability and the robustness of the inversion results.
However, we have seen that the highly complex observed Stokes profiles cannot be perfectly reproduced with any of the presented inversion techniques 
without almost doubling the number of free parameters. 
The inherent complexity of these profiles may involve physical aspects that are not considered within the assumptions and approximations made by the inversion codes, which could lead to significant errors in the returned values.
These assumptions include hydrostatic equilibrium, which is  unlikely to be satisfied in such a dynamic environment.

Finally, the field estimations performed by means of the Zeeman splitting and the COG method (methods 1, 2, and 3) provide results that are not entirely consistent among each other.
Unfortunately, all these methods are unable to take into account the errors from the instrumental effects.
 Moreover,
methods 1 and 2 are reliable for ideal cases only,  as when they are applied to single-component profiles produced by homogeneous magnetic fields, which is clearly not the case for the LFPs with highly asymmetric and multi-lobed Stokes profiles. 
As a consequence, the possibility of two (horizontally) unresolved structures  with a large velocity difference cannot be ruled out either, in spite of the shortcomings of the 2-component height-dependent and height-independent inversions. Unusually strong penumbral magnetic fields (5-6 kG) also show up as the more plausible physical solution in some of the 2 components models.
However, the existence of two Doppler shifted components (one carrying nearly 15 km s$^{-1}$ and another one with gas almost at rest) coexisting over several neighboring pixels would require an unresolved fine structure with sub-resolution canals of two types over an extended area. These extreme gradients in velocity required to be present in the one resolution element to produce the observed spectra are considerably less plausible. A solution where the pixels are smoothly connected (as in the 2D inversions), with a height gradient within the line forming region could represent a more plausible scenario and is in agreement with MHD simulations.

For future work, it would be interesting to investigate how the presence of superstrong magnetic fields in the photosphere affects the shape of the observed Stokes profiles in the pair of Fe I lines when considering the Paschen-Back effect. This will be useful to determine whether or not such effects need to be taken into account by the inversion codes when dealing with photospheric field strengths of the order of 7 kG or larger.

\begin{table*}[htp!]
 \caption[Results of different height-dependent inversions applied to the observed and deconvolved Stokes profiles in two LFPs]{Parameters  resulting from 4 different height dependent inversions which were applied to the two sets of observed Stokes profiles ((a) and (b))  and to their corresponding deconvolved Stokes profiles ((c) and (d)) from the two selected LFPs.  From left to right:    number of free parameters $n$, pixel identification number P, atmospheric component C,  optical depth node $\log(\tau)$,  field strength $B$, field inclination $\gamma$,  LOS velocity $v_{LOS}$,  merit function $\chi^2$,  filling factor $\alpha$ for each component, and Bayesian Information Criterion  (BIC) value.} 
\begin{center}
\begin{tabular}{m{3.3cm} m{0.5cm} m{0.5cm}m{0.5cm}  m{0.85cm}m{0.8cm}m{0.65cm} m{1.5cm}m{0.5cm} m{0.7cm} m{0.7cm}}
  \multicolumn{11}{c}{\textbf{Height-dependent inversions}}\\
\hline
  inversion technique &$n$
  &P&C &$\log(\tau)$ &$B$ [kG]&$\gamma$ $[^{\circ}]$&$v_{LOS}$ [km s$^{-1}$]&$\chi^2$&$\alpha$&BIC\\ 
\hline
    \multirow{6}{3cm}{(a) SPINOR 2D\\(observed)}&\multirow{6}{0.2cm}{18}& \multirow{3}{*}{1}&\multirow{3}{*}{1} &$-2.0$&6.4&141&5.6&&&\multirow{3}{*}{98}\\ 
&&& & $-0.8$&8.0&148&9.3&14&1&\\
&& & &$0.0$ &8.3&145&8.3&&&\\
\cline{3-11}
 &&  \multirow{3}{*}{2}& \multirow{3}{*}{1} &$-2.0$&7.0&139&7.9&&&\multirow{3}{*}{119}\\ 
&& & & $-0.8$&8.0&146&9.3&35&1&\\
&& & &$0.0$ &7.4&173&12.1 &&&\\ \hline

\multirow{12}{3.2cm}{(b) SPINOR 1D\\(observed)}&\multirow{12}{0.2cm}{37}&\multirow{6}{*}{1}&\multirow{3}{*}{1}&$-2.0$& 3.6&166&0.2&\multirow{6}{*}{8} &&\multirow{6}{*}{181}\\
& && &$-0.8$&4.0&155&0.4& &0.48&\\ 
&& & &$0.0$&4.2&146&0.5& &&\\ 
\cline{4-8} 
\cline{10-10}
&& &\multirow{3}{*}{2}&$-2.0$&2.9&163&15.5& &&\\ 
&& &&$-0.8$&3.9&177&17.5& &0.52&\\ 
&& & &$0.0$&4.4&179&17.4& &&\\ 
\cline{3-11}
&& \multirow{6}{*}{2}&\multirow{3}{*}{1}&$-2.0$&3.6&159&0.1&\multirow{6}{*}{12}  &&\multirow{6}{*}{185}\\
 && &&$-0.8$&4.0&154&0.3& &0.35&\\ 
&& & &$0.0$&4.2&144&1.1& &&\\ 
\cline{4-8} 
\cline{10-10}
&& &\multirow{3}{*}{2} &$-2.0$&2.3&152&14.3& &&\\ 
&& & &$-0.8$&4.0&169&17.0& &0.65&\\ 
&&  & &$0.0$&5.1&179&16.0& &&\\

\hline
  \multirow{6}{3cm}{(c) SPINOR 1D\\ (deconvolved)}&\multirow{6}{0.2cm}{18}& \multirow{3}{*}{1}&\multirow{3}{*}{1} &$-2.0$&2.5&100&9.8&&&\multirow{3}{*}{128}\\ 
&& & &$-0.8$&7.4&179&12.7&44&1&\\
&& & &$0.0$ &7.2&169& 4.7&&&\\
\cline{3-11}
 &&  \multirow{3}{*}{2}& \multirow{3}{*}{1} &$-2.0$&2.9&67&8.5&&&\multirow{3}{*}{131}\\ 
&&& &$-0.8$&5.2&178&11.4&47&1&\\
&& & &$0.0$ &6.6&132&6.7&&&\\ 

\hline 

\multirow{12}{3cm}{(d) SPINOR  1D\\(deconvolved)}&\multirow{12}{0.2cm}{37}&\multirow{6}{*}{1}&\multirow{3}{*}{1}&-2.0 &3.0&159&0.0&\multirow{6}{*}{14}  &&\multirow{6}{*}{187}\\
&&&&-0.8&3.5&170&0.5& &0.42&\\
& & &&0.0& 4.0&178&1.0& &&\\
\cline{4-8}
\cline{10-10}
 &&&\multirow{3}{*}{2}&-2.0& 2.8&144&14.4& &&\\
&& & &-0.8&4.0&171&16.5& &0.58&\\ 
& &&&0.0& 4.9&179&17.7& &&\\
\cline{3-11}
 && \multirow{6}{*}{2}&\multirow{3}{*}{1}&-2.0&3.2&157&0.2&\multirow{6}{*}{18} &&\multirow{6}{*}{191}\\
&& & &-0.8&3.8&157&0.6&&0.39&\\ 
 && &&0.0 &4.2&156&1.1& &&\\
\cline{4-8}
\cline{10-10}
 && &\multirow{3}{*}{2}&-2.0&1.7&165&15.6& &&\\
&& & &-0.8&4.0&172&17.1& &0.61&\\ 
 && &&0.0 &5.8&176&16.4& &&\\
\hline 

\end{tabular}
\end{center}

\label{tab:1}
\end{table*}

\begin{table*}[htp!]
 \caption[Results of different height-independent inversions applied to the observed and deconvolved Stokes profiles in two LFPs]{Parameters  resulting from two height independent inversions:  (e) two-components Milne-Eddington inversions applied to the two sets of observed Stokes profiles (black dashed lines in Fig. \ref{fig:3**})  and, (f) two-components SPINOR 1D inversions applied to the corresponding deconvolved Stokes profiles (black dashed lines in Fig. \ref{fig:3***}) of the two selected LFPs (blue markers on Fig. \ref{fig:LFPa}).  Columns are in the same format as in Table \ref{tab:1}.} 
\begin{center}
\begin{tabular}{m{3.3cm} m{0.5cm} m{0.5cm}m{0.5cm}  m{0.85cm}m{0.8cm}m{0.65cm} m{1.5cm}m{0.5cm} m{0.7cm} m{0.7cm}}

\multicolumn{10}{c}{}\\
   \multicolumn{10}{c}{\textbf{Height-independent inversions}}\\  
\hline
  inversion technique &$n$
  &P&C & &$B$ [kG]&$\gamma$ $[^{\circ}]$&$v_{LOS}$ [km s$^{-1}$]&$\chi^2$&$\alpha$&BIC\\ 
\hline
\multirow{4}{1.8cm}{(e) ME\\(observed)}&\multirow{4}{0.2cm}{15}
& \multirow{2}{*}{1}&1&& 3.4&169&0.0&\multirow{2}{*}{22}  &0.66&\multirow{2}{*}{92}\\
 \cline{4-8}
\cline{10-10}
 
&& &2 &&4.1&162&15.5& &0.34&\\ 
 
\cline{3-11}
 
&& \multirow{2}{*}{2}&1 &&3.4&165&0.5& \multirow{2}{*}{34} &0.55&\multirow{2}{*}{104}\\ 
 
\cline{4-8}
\cline{10-10}
 
&& &2 &&3.9&161&15.3& &0.45&\\ 
 
\hline

\multirow{4}{3cm}{(f) SPINOR 1D\\(deconvolved) }&\multirow{4}{0.2cm}{17}
&\multirow{2}{*}{1}&1&& 3.5&173&0.0&\multirow{2}{*}{15} &0.32&\multirow{2}{*}{95}\\
 
\cline{4-8}
\cline{10-10}
 
&&&2 &&3.9&165&15.9&&0.68&
\\ 

\cline{3-11}

&&\multirow{2}{*}{2}&1 &&3.5&171&0.5& \multirow{2}{*}{18}&0.35&\multirow{2}{*}{98}\\ 

\cline{4-8}
\cline{10-10}
 
&& &2 &&4.0&166&16.6&&0.65&\\ 
\hline

\end{tabular}
\end{center}

\label{tab:2}
\end{table*}

\begin{acknowledgements}
This work was carried out in the framework of
the International Max Planck Research School (IMPRS) for
Solar System Science at the Max Planck Institute for Solar
System Research (MPS) supported by the Max Planck
Society. The National Center for
Atmospheric Research is sponsored by the National Science
Foundation.
\end{acknowledgements}

\bibliography{aabib}
\bibliographystyle{aa}
%
%
%
%
%
%
%

\end{document}